\documentclass[rmp,aps,lengthcheck,floatfix]{revtex4-1}
\usepackage{bm}
\usepackage{amsfonts}
\usepackage{amssymb}
\usepackage{amsmath}
\usepackage{pdfsync}
\usepackage{graphicx}

\def\lbar{\lambda\hskip-5pt\vrule height4.7pt depth-4.0pt width6pt}

\newcommand{\greeksym}[1]{{\usefont{U}{psy}{m}{n}#1}}
\newcommand{{\rmssmu}}{\mbox{\scriptsize{\greeksym{m}}}}

\newcommand{\rT}{^{\rm T}}
\newcommand{\rL}{^{\rm L}}
\newcommand{\rM}{_{\rm M}}
\newcommand{\rN}{{_{\rm N}}}
\newcommand{\rc}{_{{\rm c}}}
\newcommand{\rd}{{\rm d}}
\newcommand{\re}{{\rm e}}

\newcommand{\ri}{{\rm i}}
\newcommand{\rp}{{_{{\rm p}}}}

\newcommand{\rs}{{_{{\rm s}}}}
\newcommand{\iA}{{\it \Lambda}}

\newcommand{\iO}{{\it \Omega}}
\newcommand{\iP}{{\it \Psi}}

\newcommand{\iX}{{\it \Xi}}
\newcommand{\iPi}{{\it \Pi}}

\newcommand{\cD}{{\cal D}}

\newcommand{\cI}{{\cal I}}

\newcommand{\cV}{{\cal V}}

\newcommand{\0}{{\bm 0}}
\def\fr#1#2{{\textstyle{\frac{#1}{#2}}}}
\def\ft#1#2{{\frac{\textstyle #1}{\textstyle #2}}}

\def\msp{\vbox to 15 pt {}}
\fontfamily{lmss}\selectfont
\begin{document}

\title{Magnetic moments in the Poynting theorem,\\Maxwell
equations, Dirac equation, and QED}

\newcommand{\addrGaithersburg}{National Institute of Standards and Technology,
Gaithersburg, MD 20899-8420, USA}

\author{Peter J.~Mohr}
\affiliation{\addrGaithersburg}

\date{\today}


\begin{abstract}

This paper examines the theory of electron magnetic dipole moment
interactions with magnetic fields or other electrons in classical and
quantum  electrodynamics.  We show that these interactions may be
described by a version of the Poynting theorem that is extended to take
into account energetics of the interaction of magnetic dipole moments
with inhomogeneous magnetic fields.  This extension of the Poynting
theorem is linked to an extension of the Maxwell equations that takes
into account magnetic dipole moment sources.  We provide detailed
descriptions of the interactions based on both the extended Poynting
theorem and on conventional quantum electrodynamics expressed in terms
of electromagnetic fields and show that these apparently different
formulations can give consistent results.  In both cases, we express the
interactions in terms of electromagnetic fields only, without the use of
potentials.  The main focus is on magnetic dipole interactions, and
magnetic monopole interactions are not considered.


\end{abstract}
%
%
\maketitle
%

\tableofcontents

\section{Introduction}
\label{sec:intro}

As is well known, an electron interacts with external
electromagnetic fields through its charge and magnetic moment.
Moreover, there is a Coulomb monopole electric field associated
with the electron's charge and a magnetic dipole field
associated with its magnetic moment.  Possible magnetic monopole
\cite{1990067,2004317} or electric dipole \cite{acme,nistedm}
moments and the corresponding fields for the electron are
experimentally consistent with zero.  Therefore, there is no
consideration of those moments in this work.  Higher-multipole
fields are excluded for a spin one-half particle such as the
electron.

The electron magnetic dipole moment is accurately measured and
calculated, and the comparison provides a test of the Standard
Model.  A recent overview of both theory and experiment is given
by \citet{codata}.  The CODATA recommended value of the electron
magnetic moment is $\mu_{\rm e}=g_{\rm e}\,\mu_{\rm B}/2$, where
$\mu_{\rm B}=e\hbar/(2\,m_{\rm e})$ is the Bohr magneton and
$g_{\rm e}$ is the $g$-factor, currently given by $g_{\rm e} =
-2.002\,319\,304\,360\,92(36)$, with a relative uncertainty of
about 2 parts in $10^{13}$.

The magnetic dipole moment of the electron is the source of a
magnetic field and can be described as a current loop or as two
opposite polarity magnetic monopoles.  These are not realistic
models, but they suggest methods of calculating the associated
magnetic fields.  The current loop model gives a transverse
magnetic field and the dual magnetic monopole model gives a
longitudinal magnetic field.  This latter model has a
resemblance to the quark model of hadrons, because in both cases
the particles are mathematically modeled as having constituents,
quarks in the one case and magnetic monopoles in the other, that
do not appear separately in nature.  The preferred model for the
electron is the loop model, because when associated with quantum
electrodynamics (QED), it gives the correct prediction for the
hyperfine interaction, as discussed by~\citet{CERN}.
However, in Sec.~\ref{ssec:hfs} it is shown that the dual
monopole model also can give the correct hyperfine structure.

The magnetic fields associated with the two models for the
electron magnetic moment are similar in one respect, but
different in another.  Classically, for $|\bm x|>0$, where $\bm
x$ is the location of the electron, they are equal, but they
differ by a delta function at $\bm x=0$.  One aspect of this is
that $\bm \nabla\bm\cdot\bm B^{\rm T}=0$ while $\bm \nabla\bm\cdot\bm
B^{\rm L}\neq0$, where T denotes transverse and L denotes
longitudinal.

In this paper, we consider and compare both forms of the magnetic
moment.  There is interest in the dual monopole model even though the
current loop model and QED give predictions with 14 figure accuracy.
The reason is that QED itself is not entirely satisfactory, so it is
worthwhile to explore alternative formulations.  It is based on a
formalism that is not mathematically well defined and leads to
expressions that require a prescription to remove infinities and thereby
get finite results that may be compared to experiment.  Besides being
mathematically problematic, the infinities result in the calculations to
obtain physical predictions being more difficult than they might
otherwise need to be, based on first-hand
experience~\cite{1974002,1974003}.  Another reason to eliminate the
infinities is that it might allow nonperturbative calculations to be
done.  The order-by-order removal of infinities by renormalization
prevents this.  It is therefore important to seek modifications of QED
that may not have the infinities.

Classically, the energetics of electromagnetic fields and their
interactions with particles are described by the Poynting
theorem, which follows from the Maxwell
equations~\cite{max,poy,1998165}.  Here we examine the
interaction of particles with electromagnetic fields from this
perspective.  The corresponding interactions in quantum
electrodynamics are also considered.

Although the Poynting theorem is a statement of conservation of
energy when energy is exchanged between charged particles and
electromagnetic fields, the conventional form of the theorem
does not take into account the interaction of the magnetic
moment of a particle, such as an electron, with an inhomogeneous
magnetic field.  To remedy this, in Sec~\ref{sec:mfpi} we
suggest a model-independent extension of the Poynting theorem to
take such an interaction into account.  Because the theorem is a
consequence of the Maxwell equations, such a change of the
theorem is not consistent with those equations.  One way to deal
with this is to add appropriate terms to the Maxwell equations
to have consistency with the extended Poynting theorem.  The
added terms are a magnetic dipole source and a magnetic dipole
current that replace the two zero sources in the equations, as
discussed in Sec.~\ref{sec:eme}.  The zeros are sometimes
replaced by magnetic monopole sources, but not in the present
paper.  In Sec.~\ref{sec:li}, we show that the added dipole
terms are consistent with relativistic invariance of the
equations.  In fact, the added magnetic current source term is
shown to also be a consequence of special relativity in
Sec.~\ref{sec:av}, independent of the Poynting theorem.

Magnetic moment sources in the Maxwell equations are routinely
considered in works that derive the equations for macroscopic
media.  See \citet{1998165, FII, mm, bj}, for example.  These
works are based on the current loop model of the magnetic
dipole.  The present work differs in that it is model
independent and links the microscopic source terms for a
particle in the Maxwell equations to conservation of energy as
expressed in the Poynting theorem.  This provides a motivation
for including those terms in the microscopic theory.

Consequences of the extension of the Maxwell equations and the
Poynting theorem must be closely examined.  Foremost is the fact
that a magnetic moment source means that $\bm\nabla\bm\cdot\bm
B\ne0$ instead of $\bm\nabla\bm\cdot\bm B=0$ in the Maxwell
equation.  This could be problematic, because $\bm B =
\bm\nabla\bm\times\bm A$, where $\bm A$ is the vector potential,
implies $\bm\nabla\bm\cdot\bm B=0$.  This raises the question of
whether the vector potential essential in the Dirac equation and
QED?

Feynman points out that the problem of infinities in QED could
be that the assumptions behind it produce an overdetermined
set~\cite{FII}.  These assumptions include quantum mechanics,
special relativity, local interactions, probabilities adding up
to 1, positive energies, causality, and possibly others that we
are not aware of.  The problem may be the assumption that
interactions need to be local.  If they are not local, it could
mean that potentials, which provide local interactions may not
be needed.

We address this question, because vector potentials are
important in quantum mechanics, particularly when considering
the Aharonov-Bohm effect~\cite{1959009}.  But as \citet{2001380}
point out, the vector potential is not necessary to explain this
effect if locality is not imposed.  There are a number of other
reasons why potentials may be necessary.  One is the fact that
external field interactions in the Schr\"odinger and Dirac
equations are implemented via the ``principle of minimal
coupling'', which is the replacement: $(E,\bm p)\rightarrow
(E+e\phi_{\rm ex},\bm p+e\bm A_{\rm ex})$, where $\phi_{\rm ex}$
and $\bm A_{\rm ex}$ are the scalar and vector potentials
associated with the external fields, respectively, and $-e$ is
the charge of the electron.  It has been suggested that this
demonstrates the necessity of potentials~\cite{1959009,FII}.
However, external field interactions may be introduced into the
Schr\"odinger and Dirac equations by using the Poynting theorem
rather than the minimal coupling principle, as shown in
Sec.~\ref{sec:ptde}.  To further address this question, in
Sec.~\ref{ssec:ope} we show that the QED expression for
one-photon exchange may be given in terms of electric and
magnetic fields alone.  In fact, this was already known to be
the case for the electron self energy by \citet{weisskopf}.

How could the Poynting theorem and thus the Maxwell equations
omit a magnetic moment source term that should be included for
160 years?  A possible reason is that these equations were
firmly embedded in the culture and textbooks for over 55 years
by the time it was realized that particles such as the electron
had magnetic moments.  At the same time, potentials played an
important role in electrodynamics.  For example,
\citet{einstein} used potentials in his proof of the
relativistic invariance of the Maxwell equations.  Moreover,
gauge theories such as QED and QCD are viewed as being
fundamental.  However, conservation of energy in the Poynting
theorem is a compelling argument and the connection to the
Maxwell equations is straightforward algebra.  It is worthwhile
to consider the consequences of including the magnetic source
terms with emphasis on QED as a simple example of a gauge
theory.

The form in fields that the interaction takes in QED is
proportional to $|\bm E|^2-|c\bm B|^2$, while according to the
Poynting theorem, the interaction energy is proportional to
$|\bm E|^2+|c\bm B|^2$.  This apparent discrepancy is linked to
the way the magnetic dipole field is treated.  In QED, the
magnetic field of a dipole source is a transverse field
(current-loop model), whereas in the extended Poynting-Maxwell
case, the magnetic field is a longitudinal field (dual magnetic
monopole model).  This a subtle difference, because these
properties of the fields differ only by a delta function at the
location of the source, as explained in this paper.

The QED expression for the interaction energy in terms of fields is
curious, because it ascribes a negative value to the energy of the
magnetic field.  Besides being counterintuitive, it incorrectly predicts
the behavior of macroscopic bar magnets.  Another curiosity is that the
QED interaction energy is a Lorentz scalar, while one would expect it to
transform as an energy, that is, as the zero component of an
energy-momentum four vector.

The extended Poynting-Maxwell interaction energy has neither of these
curious properties.  The magnetic energy is positive, giving the proper
behavior of bar magnets, and $|\bm E|^2+|c\bm B|^2$ indeed transforms
as the zero component of an energy-momentum four vector.  Whether this
version of electrodynamics can be the basis for an alternative
formulation of QED is an interesting question, but outside of the scope
of this paper.

\section{Particle-field interactions}
\label{sec:pfi}

If a particle is in an external electromagnetic field that applies a
force $\bm F(\bm x)$ to it, then motion of the particle opposing the
force will require work done on the particle.  This has the effect of
increasing the energy of the combined particle field and external field.
The increase in energy of the fields will be the work done to move the
particle against the force, or the negative of the force on the particle
times the distance moved.  If $U(\bm x_0)$ is the energy of the combined
fields for a particle at $\bm x_0$, then the change in energy when the
particle moves incrementally by $\rd\bm x_0$ is
\begin{eqnarray}
\rd U(\bm x_0) = -\bm F(\bm x_0)\bm\cdot\rd\bm x_0  ,
\end{eqnarray}
and the force on the particle is
\begin{eqnarray}
\bm F(\bm x_0) = -\bm\nabla_0\,U(\bm x_0)  .
\end{eqnarray}
If the particle moves with a velocity $\bm v$, the rate of change of the
energy of the fields is 
\begin{eqnarray}
\frac{\rd}{\rd t}\,U(\bm x_0) = -\bm F(\bm x_0)\bm\cdot
\frac{\rd\bm x_0}{\rd t} = -\bm F(\bm x_0)\bm\cdot\bm v  .
\label{eq:eex}
\end{eqnarray}
The key element is the force on the particle due to its interaction with
the external fields.  This is examined in the following sections.

\section{Poynting theorem}
\label{sec:pt}

The Poynting theorem describes the energetics of electromagnetic fields,
$\bm E$ and $\bm B$, and their interactions with charged particles.  It
is conventionally given in vacuum by \cite{poy,1998165}
\begin{eqnarray}
\frac{\partial u}{\partial t} + \bm\nabla\bm\cdot\bm S =
-\bm J\bm\cdot\bm E  ,
\label{eq:pt}
\end{eqnarray}
where
\begin{eqnarray}
u = \frac{\epsilon_0}{2}\left(\big|\bm E\big|^2 
+ \big|c\bm B\big|^2\right)
\label{eq:ed}
\end{eqnarray}
is the energy density of the electromagnetic fields, and
\begin{eqnarray}
\bm S = \frac{1}{\mu_0}\,\bm E\bm\times\bm B
\label{eq:pv}
\end{eqnarray}
is the Poynting vector, which is the energy flow per unit area of the
fields.  In Eqs.~(\ref{eq:pt})-(\ref{eq:pv}), $\bm J$ is the electric
current density, $\epsilon_0$ is the vacuum electric permittivity,
$\mu_0$ is the vacuum magnetic permeability, and $c$ is the speed of
light, with $\epsilon_0\mu_0c^2 = 1$.

The integral of Eq.~(\ref{eq:pt}) over a volume V is
\begin{eqnarray}
\int_{\rm V}\rd\bm x \,\frac{\partial u}{\partial t} + \int_{\rm S}\rd
A\,\bm{\hat n}\bm\cdot \bm S &=& -\int_{\rm V}\rd\bm x\,\bm J\bm\cdot\bm E .
\label{eq:cpt}
\end{eqnarray}
where the volume integral of the divergence of $\bm S$ in the second
term has been replaced by an integral over the normal to the surface
S according to the Gauss-Ostrogradsky theorem.  In words,
Eq.~(\ref{eq:cpt}) states that the electromagnetic field energy in a
volume decreases by the outward flow of energy through the surface of
the volume and by the work done by the fields on charged particles in
the volume.  Conversely, the field energy in the volume will increase by
the energy flow into the volume and by the work done against the forces
of the fields on the particles by their motion which is provided by an
independent source.  This is a statement of conservation of energy,
which follows from the Maxwell equations.

Since the Poynting theorem follows from the Maxwell equations, which are
consistent with special relativity, we expect the theorem to also show
such a consistency.  However, to provide a qualitative description of
various aspects of the theorem, here we restrict attention to
nonrelativistic motion of the relevant particles, which means neglecting
all but the leading terms in powers of $|\bm v|/c$, where $\bm v$ is the
velocity of a particle.  We also assume that the external fields are
slowly varying.  Higher-order terms are properly accounted for in the
relativistic formulation of the theorem in Sec.~\ref{sec:rpt}.

If we take the current density for a particle with charge $q$ at the
point $\bm x_0$ moving with the velocity $\bm v$ inside the volume to be
\begin{eqnarray}
\bm J(\bm x) = \rho(\bm x)\,\bm v = q\,\delta(\bm x-\bm x_0)\bm v \, ,
\label{eq:chcurrent}
\end{eqnarray}
to lowest order in $|\bm v|/c$, then the rate of change of the energy of
the fields $U_q^{\,\rm I}(\bm x)$ due to the motion of the particle in
the external field $\bm E_{\rm ex}(\bm x)$ is~\cite{1998165}
\begin{eqnarray} 
\frac{\rd}{\rd t}\,U_q^{\,\rm I}(\bm x_0) &=& 
-\bm v\bm\cdot\bm F_q(\bm x_0) = 
-q\,\bm v\bm\cdot\bm E_{\rm ex}(\bm x_0)
\nonumber\\[10 pt]&=&
-\int_{\rm V}\rd\bm x \, \bm J(\bm x)\bm\cdot 
\bm E_{\rm ex}(\bm x)  ,
\label{eq:jeint}
\end{eqnarray}
as appears in Eq.~(\ref{eq:cpt}).

A particle with a magnetic moment in an inhomogeneous magnetic
flux density also experiences a force.  Independent of any
model, a particle with a magnetic moment interacts with an
inhomogeneous magnetic field, with examples being the
Stern-Gerlach experiment \cite{GS0} and magnetic neutron
scattering \cite{JS,HJ,ns1,ns2}.  Moreover, the interaction may
involve energy exchange between the field and the particle, as
shown by the deceleration of hydrogen atoms in the triplet
hyperfine ground state in an inhomogeneous magnetic
field~\cite{mz}.  However, Eq.~(\ref{eq:cpt}) does not include a
magnetic interaction between an inhomogeneous magnetic field and
the magnetic moment of a particle, needed to account for
conservation of energy.  In the following section, we consider
such an interaction.

\section{Magnetic field-particle interactions}
\label{sec:mfpi}

Here, we examine the interactions of a particle with a magnetic moment
$\bm m$ with an inhomogeneous magnetic flux density $\bm B_{\rm ex}(\bm
x)$.  For this purpose, it is useful to consider separately the
transverse (T) and longitudinal (L) components of the external field
(see Appendix \ref{app:tl})
\begin{eqnarray}
\bm B_{\rm ex}(\bm x) &=& \bm B_{\rm ex}^{\rm T}(\bm x) 
+ \bm B_{\rm ex}^{\rm
L}(\bm x)  .
\end{eqnarray}
For a magnetic field resulting from a steady-state current $\bm J_{\rm
ex}(\bm x)$, we have from Eq.~(5.16) of \citet{1998165}
\begin{eqnarray}
\bm B_{\rm ex}^{\rm T}(\bm x) &=& \frac{\mu_0}{4\pi}\,\bm\nabla\bm\times
\int\rd\bm x^\prime\,\frac{\bm J_{\rm ex}(\bm x^\prime)}
{|\bm x - \bm x^\prime|}  ,
\label{eq:btj}
\end{eqnarray}
which is transverse because
\begin{eqnarray}
\bm\nabla\bm\cdot\bm B_{\rm ex}^{\rm T}(\bm x) &=& 0  .
\label{eq:divbt}
\end{eqnarray}
Thus following \citet{1998165} up to Eq.~(5.68) in that text, the force
from the external current is
\begin{eqnarray}
\bm F_{\bm m}^{\rm T}(\bm x_0) &=& \left(\bm m\bm\times\bm \nabla_0\right)
\bm\times\bm B_{\rm ex}^{\rm T}(\bm x_0)
\nonumber\\[5 pt]
&=& \bm\nabla_0\,\bm m\bm\cdot \bm B_{\rm ex}^{\rm T}(\bm x_0)
- \bm m\,\bm\nabla_0\bm\cdot\bm B_{\rm ex}^{\rm T}(\bm x_0)
\nonumber\\[5 pt]
&=&\bm\nabla_0\,\bm m\bm\cdot \bm B_{\rm ex}^{\rm T}(\bm x_0)  ,
\label{eq:f12}
\end{eqnarray}
where the last line follows from Eq.~(\ref{eq:divbt}).  This is the
force from the current in Eq.~(\ref{eq:btj}) that could be the current in
a solenoid, for example.  However, this provides no information about
the interaction of the particle with a possible longitudinal component
of the field $\bm B_{\rm ex}^{\rm L}(\bm x)$.  To address this, we
consider the interaction of the magnetic moment of a particle with an
external longitudinal magnetic field from the perspective of the
Poynting theorem.

A magnetic dipole moment $\bm m$ located at $\bm x = 0$ is the source of
a magnetic field given for $|\bm x|>0$ by \cite{1998165}
\begin{eqnarray}
\bm B_{\bm m}(\bm x) &=& \frac{\mu_0}{ 4 \pi}  \,
\frac{3 \bm{\hat x}(\bm{\hat x}\bm\cdot \bm m) - \bm m }
{ |\bm x|^3}  . 
\label{eq:bexl}
\end{eqnarray}
An equivalent way of expressing this for $|\bm x|>0$ is
\begin{eqnarray}
\bm B_{\bm m}(\bm x) &=&
\frac{\mu_0}{4 \pi} \, \bm m\bm\cdot\bm\nabla\,\bm\nabla\
\frac{1}{|\bm x|}  ,
\label{eq:gsder}
\end{eqnarray}
where the derivatives reproduce Eq.~(\ref{eq:bexl}).  If
Eq.~(\ref{eq:gsder}) is extended to $\bm x = 0$, then it includes a delta
function at the origin as shown by taking the angular average, which
gives
\begin{eqnarray}
\frac{1}{4\pi}\int\rd\iO_{\bm x}\, \bm B_{\bm m}(\bm x) &=&
\frac{\mu_0}{12\pi} \,
\bm m\,\bm\nabla^2\,\frac{1}{|\bm x|} 
\nonumber\\[10 pt]&=&
 - \frac{\mu_0}{3}\,\bm m\,\delta(\bm x)  .
\label{eq:bav}
\end{eqnarray}
We also have
\begin{eqnarray}
\bm\nabla\bm\cdot\bm B_{\bm m}(\bm x) &=& 
\frac{\mu_0}{4 \pi}\,
\bm m\bm\cdot\bm\nabla\,\bm\nabla^2\, \frac{1}{|\bm x|}
\nonumber\\[10 pt]&=&
 -\mu_0 \, \bm m\bm\cdot\bm\nabla \, \delta(\bm x).
\label{eq:divb1}
\end{eqnarray}
The derivative of the delta function on the right-hand-side is
meaningful as a distribution or generalized function~\cite{ls1,1964013}.
The field given by Eq.~(\ref{eq:gsder}) is longitudinal, i.e.,
\begin{eqnarray}
\bm \nabla \bm\times \bm B_{\bm m}(\bm x) &=& 0  .
\end{eqnarray}

If a particle with a magnetic dipole moment $\bm m$ is located at $\bm
x_0$, the field at $\bm x$ is
\begin{eqnarray}
\bm B_{\bm m}^{\rm L}(\bm x,\bm x_0) &=&
\frac{\mu_0}{4 \pi} \, \bm m\bm\cdot\bm\nabla\,\bm\nabla\
\frac{1}{|\bm x - \bm x_0|}  .
\label{eq:gsx0}
\end{eqnarray}
The energy of the combined fields of the particle and the external field
$\bm B_{\rm ex}^{\rm L}(\bm x)$ is
\begin{eqnarray}
U_{\bm m}^{\rm L}(\bm x_0) &=& \frac{\epsilon_0c^2}{2}\int\rd\bm x
\left(\big|\bm B^{\rm L}_{\bm m}(\bm x,\bm x_0) 
+ \bm B_{\rm ex}^{\rm L}(\bm x)\big|^2\right)
\nonumber\\[5 pt]&=&
 \frac{1}{2\mu_0}\int\rd\bm x \Big(\big|\bm B^{\rm L}_{\bm m}(\bm x,\bm
 x_0)\big|^2 
 \nonumber\\[10 pt]&&
 + 2\bm B^{\rm L}_{\bm m}(\bm x,\bm x_0)\bm\cdot\bm B_{\rm
 ex}^{\rm L}(\bm x) + \big|\bm B_{\rm ex}^{\rm L}(\bm
 x)\big|^2\Big)
  . \qquad
\end{eqnarray}
The first term is divergent for a point source, but it is finite for a
finite magnetic moment distribution.  It is the magnetic self energy of
the particle, which is independent of the external field.  The third
term is independent of $\bm x_0$, so the dependence of the interaction
energy on $\bm x_0$ is confined to the second term, which gives
\begin{eqnarray}
U_{\bm m}^{\rm L,I}(\bm x_0) &=&
\frac{1}{\mu_0}\int\rd\bm x\,
\bm B^{\rm L}_{\bm m}(\bm x,\bm x_0) \bm\cdot \bm B_{\rm ex}^{\rm L}(\bm x)
\nonumber\\[3 pt]&=&
\frac{1}{4\pi}\int\rd\bm x \left[\bm m\bm\cdot\bm\nabla\,\bm\nabla\,
\frac{1}{|\bm x-\bm x_0|}
\right] \bm\cdot \bm B_{\rm ex}^{\rm L}(\bm x)
\nonumber\\[3 pt]&=&
\frac{1}{4\pi}\,\bm m\bm\cdot\bm\nabla_0
\int\rd\bm x\, \frac{1}{|\bm x-\bm x_0|} \, 
\bm\nabla \bm\cdot \bm B_{\rm ex}^{\rm L}(\bm x)
\nonumber\\[3 pt]&=& - \bm m\bm\cdot \bm B_{\rm ex}^{\rm L}(\bm x_0) ,
\label{eq:ulmi}
\end{eqnarray}
and so [see Eq.~(\ref{eq:long})]
\begin{eqnarray}
\bm F_{\bm m}^{\rm L}(\bm x_0) 
= -\bm\nabla_0\,U_{\bm m}^{\rm L,I}(\bm x_0)
= \bm\nabla_0\,\bm m\bm\cdot\bm B_{\rm ex}^{\rm L}(\bm x_0)  .
\label{eq:fpt}
\end{eqnarray}
In Eq.~(\ref{eq:ulmi}) the gradient operators in square brackets act
only on the function within the square brackets.  The total force on the
particle is thus
\begin{eqnarray}
\bm F_{\bm m}(\bm x_0) 
&=&\bm F_{\bm m}^{\rm T}(\bm x) + \bm F_{\bm m}^{\rm L}(\bm x)
\nonumber\\[5 pt]
&=& \bm \nabla_0 \,\bm m \bm\cdot\left[\bm B_{\rm ex}^{\rm T}(\bm x_0)
+ \bm B_{\rm ex}^{\rm L}(\bm x_0) \right]
\nonumber\\[5 pt]
&=& \bm \nabla_0 \,\bm m \bm\cdot\bm B_{\rm ex}(\bm x_0)  .
\end{eqnarray}
Evidently, the expression for the force is the same for both the
transverse and longitudinal external fields.  Thus, the rate of
change of the energy in the field due to the interaction with the
particle is
\begin{eqnarray}
\frac{\rd}{\rd t}\,U_{\bm m}^{\,\rm I}(\bm x_0) &=&
-\bm v\bm\cdot\bm F_{\bm m}(\bm x_0) 
\nonumber\\[10 pt]&=&
 -\bm v\bm\cdot\bm\nabla_0\,\bm
m\bm\cdot\bm B_{\rm ex}(\bm x_0)  .
\label{eq:kbint}
\end{eqnarray}

Let a magnetic current density be defined as
\begin{eqnarray}
\bm K(\bm x) = -
(\bm v/c)\,\bm m\bm\cdot\bm\nabla\,\delta(\bm x-\bm x_0)  ,
\label{eq:kdef}
\end{eqnarray}
and consider the integral 
\begin{eqnarray}
\int\rd\bm x \, \bm K(\bm x)\bm\cdot
c\bm B_{\rm ex}(\bm x) &=&
\bm m\bm\cdot\bm\nabla_0\,\bm v\bm\cdot\bm B_{\rm ex}(\bm x_0) 
\nonumber\\&=&
\bm v\bm\cdot\bm\nabla_0\,\bm m\bm\cdot\bm B_{\rm ex}(\bm x_0) 
\nonumber\\[10 pt]&&
-
\bm v\bm\cdot\bm m\bm\times\left[\bm\nabla_0
\bm\times\bm B_{\rm ex}(\bm x_0)\right]  . 
\nonumber\\
\end{eqnarray}
For the longitudinal component of $\bm B_{\rm ex}(\bm x_0)$ we have
$\bm\nabla_0 \bm\times\bm B_{\rm ex}^{\rm L}(\bm x_0) = 0$, as noted above.
For the transverse component, we have the Maxwell equation
\begin{eqnarray}
\bm \nabla\bm\times\bm B_{\rm ex}^{\rm T}(\bm x_0) &=& 
\frac{1}{c^2}\,\frac{\partial \bm E^{\rm T}(\bm
x_0)} {\partial t} + \mu_0\,\bm J^{\rm T}(\bm x_0)  .
\end{eqnarray}
In the limit of a slowly varying or zero electric field and assuming
that the external transverse charge current vanishes at $\bm x_0$, the
location of the particle, we have $\bm\nabla_0 \bm\times\bm B_{\rm ex}^{\rm
T}(\bm x_0) = 0$.  Then
\begin{eqnarray}
\frac{\rd}{\rd t}\,U_{\bm m}^{\,\rm I}(\bm x_0) &=&
-\int\rd\bm x \, \bm K(\bm x)\bm\cdot
c\bm B_{\rm ex}(\bm x)  .
\end{eqnarray}
Thus, if the magnetic interaction is included in the energy exchange,
then Eq.~(\ref{eq:pt}) is replaced by
\begin{eqnarray}
\frac{\partial u}{\partial t} + \bm\nabla\bm\cdot\bm S =
-\bm J\bm\cdot\bm E - \bm K\bm\cdot c\bm B \,
\label{eq:ept}
\end{eqnarray}
as an extended form of the Poynting theorem.

The assumptions mentioned above are not particularly
restrictive.  For example, they would apply to the interaction
of a particle with a magnetic moment in the magnetic field of a
solenoid, provided only that the particle is not embedded in the
coil windings that produce the magnetic field.  This treatment
also applies exactly to the magnetic interactions of a particle
with other particles with magnetic moments, assuming they are
longitudinal interactions.  Relativistic effects will change
things.  A relativistic formulation of the extended Poynting
theorem with no assumptions is given in Sec.~\ref{sec:rpt}.

\section{Extended Maxwell equations}
\label{sec:eme}

The Poynting theorem, with the extension in Eq.~(\ref{eq:ept}), accounts
for conservation of energy for the particle-field interactions,
including a magnetic moment interaction with a magnetic flux density.
However, since the Poynting theorem without the magnetic interaction
follows from the conventional Maxwell equations~\cite{1998165}, the
extended Poynting theorem is not consistent with those equations.  In
this section, we consider a way to resolve this inconsistency.

The vacuum Maxwell equations are given (in SI units) by~\cite{1998165}
\begin{eqnarray}
\bm \nabla\bm\cdot\bm E &=& \frac{\rho }{\epsilon_0}
\label{eq:me1}  ,\\[5 pt]
\bm \nabla\bm\times\bm B - \frac{1}{c^2}\,\frac{\partial \bm E}
{\partial t} &=& \mu_0\,\bm J
\label{eq:me2}  , \\[8 pt]
\bm \nabla\bm\times\bm E + \frac{\partial \bm B}
{\partial t} &=& 0 
\label{eq:me3}  , \\[5 pt]
\bm \nabla\bm\cdot\bm B &=& 0  . 
\label{eq:me4}
\end{eqnarray}

Following \citet{1998165}, multiplication of Eq.~(\ref{eq:me2}) by $\bm
E$ gives
\begin{eqnarray}
\bm E\bm\cdot\bm\nabla\bm\times\bm B 
- \frac{1}{c^2}\,\bm E\bm\cdot
\frac{\partial\bm E}{\partial t} 
&=&
\mu_0\,\bm E\bm\cdot\bm J  ,
\end{eqnarray}
where
\begin{eqnarray}
\bm \nabla \bm\cdot \bm E \bm\times \bm B &=& 
\bm B\bm\cdot\bm \nabla \bm\times \bm E
-\bm E \bm\cdot \bm \nabla \bm\times \bm B  ,
\end{eqnarray}
so that
\begin{eqnarray}
-\bm\nabla\bm\cdot\bm E\bm\times \bm B 
+\bm B\bm\cdot\bm \nabla\bm\times\bm E
- \frac{1}{c^2}\,\bm E\bm\cdot
\frac{\partial\bm E}{\partial t}
&=&
\mu_0\,\bm J\bm\cdot\bm E . \qquad
\end{eqnarray}
The replacement from Eq.~(\ref{eq:me3}),
\begin{eqnarray}
\bm\nabla\bm\times \bm E&\rightarrow& 
-\frac{\partial \bm B}
{\partial t}  ,
\label{eq:convrep}
\end{eqnarray}
yields
\begin{eqnarray}
&&-\bm\nabla\bm\cdot\bm E\bm\times \bm B 
- \frac{1}{c^2}\,\bm E\bm\cdot
\frac{\partial\bm E}{\partial t}
- \bm B\bm\cdot\frac{\partial\bm B}{\partial t}
\nonumber\\[10 pt]&&\qquad=
-\bm\nabla\bm\cdot\bm E\bm\times \bm B 
- \frac{1}{2c^2}\,\frac{\partial}{\partial t}
\left(\left|\bm E\right|^2 
+ \left|c\bm B\right|^2\right) 
\nonumber\\[10 pt]&&\qquad=
\mu_0\,\bm J\bm\cdot\bm E  ,
\label{eq:subs}
\end{eqnarray}
which is equivalent to Eq.~(\ref{eq:pt}).  However, the magnetic
contribution in Eq.~(\ref{eq:ept}) must be included to have conservation
of energy.  If instead of the replacement made in
Eq.~(\ref{eq:convrep}), the replacement 
\begin{eqnarray}
\bm\nabla\bm\times\bm E \rightarrow -\frac{\partial\bm B}{\partial t}
- c\mu_0\bm K
\label{eq:replace}
\end{eqnarray}
is made, then we have
\begin{eqnarray}
&&-\bm\nabla\bm\cdot\bm E\bm\times \bm B 
- \frac{1}{c^2}\,\bm E\bm\cdot\frac{\partial\bm E}{\partial t}
- \bm B\bm\cdot\frac{\partial\bm B}{\partial t}
\nonumber\\[10 pt]
&&\qquad=
\mu_0\,\bm J\bm\cdot\bm E + \mu_0\,\bm K\bm\cdot c\bm B
\end{eqnarray}
or
\begin{eqnarray}
\frac{\partial u}{\partial t} + \bm\nabla\bm\cdot\bm S =
-\bm J\bm\cdot\bm E - \bm K\bm\cdot c\bm B  ,
\end{eqnarray}
which is the desired result.  The replacement shown in
Eq.~(\ref{eq:replace}) corresponds to a modification of the third
Maxwell equation, Eq.~(\ref{eq:me3}), to be
\begin{eqnarray}
\bm\nabla\bm\times\bm E +\frac{\partial\bm B}{\partial t}
&=& -c\mu_0\bm K  .
\label{eq:rme3}
\end{eqnarray}

To consider such an extension of the Maxwell equations, it is necessary
to examine possible conflicts it may cause.  In particular, the
extension must be consistent with relativistic invariance of the Maxwell
equations.  This question is addressed in Sec.~\ref{sec:li}, where it is
shown that Eq.~(\ref{eq:rme3}) is consistent with Lorentz invariance of
the Maxwell equations, provided a corresponding source term is added to
Eq.~(\ref{eq:me4}) to give
\begin{eqnarray}
\bm \nabla\bm\cdot \bm B
&=& c\,\mu_0\,\sigma
\label{eq:rme4}
\end{eqnarray}
where $c\,\sigma$ and $\bm K$ are components of a four-vector, just as
$c\,\rho$ and $\bm J$ are.  The source $\sigma$ is the magnetic moment
density associated with a particle with a magnetic moment.  If the
particle at rest is given a velocity boost $\bm v$, then to lowest order
in $|\bm v|/c$, there will be a resulting current $\sigma\,\bm v$,
according to the lower component of Eq.~(\ref{eq:transk}).  From
Eq.~(\ref{eq:kdef}), we have
\begin{eqnarray}
c\,\sigma(\bm x) &=& -\bm m\bm\cdot\bm\nabla\delta(\bm x-\bm x_0)
\end{eqnarray}
in agreement with Eq.~(\ref{eq:divb1}).

We thus have extended Maxwell equations in vacuum as
\begin{eqnarray}
\bm \nabla\bm\cdot\bm E
&=& \frac{\rho }{\epsilon_0} \label{eq:xme1}  , \\[5 pt]
\frac{\partial \bm E}{\partial ct}
-\bm \nabla\bm\times c\bm B
&=& -\frac{1}{c\epsilon_0}\,\bm J \label{eq:xme2} ,\\[8 pt]
\frac{\partial c \bm B} {\partial ct} 
+\bm \nabla\bm\times\bm E
&=& - c\mu_0\,\bm K \label{eq:xme3} ,\\[5 pt]
\bm \nabla\bm\cdot c \bm B(x) 
&=& c^2\mu_0\,\sigma \label{eq:xme4} \, .
\end{eqnarray}
Alternatively, Eqs.~(\ref{eq:xme3}) and (\ref{eq:xme4}) may be
written as
\begin{eqnarray}
\frac{\partial c \bm B} {\partial ct} 
+\bm \nabla\bm\times\bm E
&=& - \frac{1}{c\epsilon_0}\,\bm K
\label{eq:rgme3} ,\\[5 pt]
\bm \nabla\bm\cdot c \bm B 
&=& \frac{\sigma}{\epsilon_0} \label{eq:rgme4}  .
\end{eqnarray}
The electric continuity equation
\begin{eqnarray}
\frac{\partial\rho}{\partial t} + \bm\nabla\bm\cdot\bm J = 0 \, ,
\label{eq:conj}
\end{eqnarray}
which follows from Eqs.~(\ref{eq:xme1}) and (\ref{eq:xme2}), is matched
by a corresponding magnetic continuity equation
\begin{eqnarray}
\frac{\partial\sigma}{\partial t} 
+ \bm\nabla\bm\cdot\bm K = 0  ,
\label{eq:conk}
\end{eqnarray}
which follows from Eqs.~(\ref{eq:xme3}) and (\ref{eq:xme4}).  The
continuity equations only give information about the longitudinal
components of $\bm J$ and $\bm K$.

\section{Perspective on the extended Maxwell equations}
\label{sec:peme}

The extended Maxwell equations are a departure from the traditional
Maxwell equations, so some further remarks are included here.  The most
straightforward observation is that even though magnetic moment currents
do not explicitly exist in the conventional Maxwell equations, they do
exist in nature.  Some examples are polarized electron
beams~\cite{ALLEY19951}, possible triplet Cooper pair currents of
electrons in superconductivity~\cite{supercond}, polarized neutron
beams~\cite{RMP.89.045004}, and atomic beams of hydrogen atoms in the
triplet hyperfine state~\cite{mz}.  The equations with magnetic sources
are similar to the equations that are sometimes considered to include
magnetic monopoles, where the sources $\sigma$ and $\bm K$ would
describe the density and current of hypothetical
monopoles~\cite{1998165}.  However, the dipole source and current are
independent of such considerations and the monopole sources are not
considered here.

Evidently, Eq.~(\ref{eq:xme4}) is in conflict with the conventional
expression $\bm B=\bm\nabla\bm\times\bm A$, where $\bm A$ is a vector
potential, which implies $\bm\nabla\bm\cdot\bm B = 0$.  The non-zero
divergence of $\bm B$ can be traced back to Eq.~(\ref{eq:gsder}), which
is a longitudinal field from a magnetic dipole moment of a particle.
Although, we do not assume any model for the dipole moment in this work,
the longitudinal field  corresponds to the dual magnetic monopole model,
which differs from the conventional current loop model for the
source~\cite{FII,1998165}.  The latter field is transverse as is
evident from Eq.~(\ref{eq:bt}).

Conventional QED is based on transverse magnetic fields for the
interactions, and with this restriction, one has $\bm\nabla\bm\cdot\bm
B^{\rm T} = 0$, so the vector potential is not ruled out.  Thus it
appears that both the extended Maxwell equations and the associated
extended Poynting theorem can coexist with conventional QED, which
simply does not deal with longitudinal magnetic fields, even though it
does have longitudinal electric fields.  All magnetic fields are taken
to be transverse, including the magnetic dipole moment of the electron.
On the other hand, from this perspective the $|\bm B|^2$ term in the
conventional Poynting theorem has no explicit relation to the
interactions of inhomogeneous magnetic fields with magnetic moments.

\section{Alternative approach to the extended Maxwell equations}
\label{sec:av}

The form of the Maxwell equations described in Sec.~\ref{sec:eme} is
arrived at by extending the Poynting theorem to be consistent with
energy conservation and seeing that a modification of the Maxwell
equations can be made to arrive at this result.  In this section, we
take a different tack to check the consistency of this result.  Here we
consider the example of the field of a moving particle with a magnetic
dipole moment to show that the extension in Eq.~(\ref{eq:xme3}) is
consistent with the conventional Lorentz transformation of the field.

Consider a particle with a magnetic dipole moment at the location $\bm
x_0$ in its rest frame that is moving with a constant velocity $\bm v$
relative to the lab frame.  The magnetic field in the rest frame of the
particle is
\begin{eqnarray}
\bm B^\prime &=& \frac{\mu_0}{4\pi}\,\bm
m\bm\cdot\bm\nabla\,\bm\nabla\,\frac{1}{|\bm x - \bm x_0|}  ,
\end{eqnarray}
and in the lab frame, to lowest order in $|\bm v|/c$, there is an
electric field given by [see p. 558 of \citet{1998165} and
Eq.~(\ref{eq:ltem})]
\begin{eqnarray}
\bm E &=& -\bm v \bm\times \bm B^\prime
= -\frac{\mu_0}{4\pi}\,
\bm v \bm\times \bm\nabla\,
\bm m\bm\cdot\bm\nabla\,
\frac{1}{|\bm x - \bm x_0(t)|}  . \qquad
\end{eqnarray}
Thus
\begin{eqnarray}
\bm\nabla\bm\times\bm E &=&
 -\frac{\mu_0}{4\pi}\,
\bm\nabla\bm\times(\bm v \bm\times \bm\nabla)\,
\bm m\bm\cdot\bm\nabla\,
\frac{1}{|\bm x - \bm x_0(t)|}  \, ,
\nonumber\\[10 pt]
&=& \frac{\mu_0}{4\pi}\left[\bm v\bm\cdot\bm \nabla \, \bm\nabla
-\bm v \bm\nabla^2
\right]
\bm m\bm\cdot\bm\nabla\,
\frac{1}{|\bm x - \bm x_0(t)|} \, ,
\nonumber\\[10 pt]
&=& \bm v\bm\cdot\bm\nabla\bm B
+\mu_0\,\bm v\,\bm m\bm\cdot\bm\nabla\,\delta\big(\bm x - \bm x_0(t)\big)
\, .
\end{eqnarray}
For the first term,
\begin{eqnarray}
\frac{\partial}{\partial t} \,
\frac{1}{|\bm x - \bm x_0(t)|} &=&
\sum_{i=1}^3
\left[\frac{\partial}{\partial t} \, x_0^i(t)\right] 
\frac{\partial}{\partial x_0^i(t)} 
\frac{1}{|\bm x - \bm x_0(t)|} \qquad
\nonumber\\[10 pt] &=&
-\bm v\bm\cdot\bm\nabla \,
\frac{1}{|\bm x - \bm x_0(t)|} \, ,
\end{eqnarray}
so that to lowest order in $|\bm v|/c$
\begin{eqnarray}
\bm v\bm\cdot\bm\nabla\,\bm B^\prime = 
-\frac{\partial}{\partial ct}\,c\bm B^\prime
= -\frac{\partial}{\partial ct}\,c\bm B,
\end{eqnarray}
and for the second term, from Eq.~(\ref{eq:kdef})
\begin{eqnarray}
\bm v\,\bm m\bm\cdot\bm\nabla\,\delta\big(\bm x - \bm x_0(t)\big)
&=& -c\bm K  ,
\end{eqnarray}
which gives
\begin{eqnarray}
\bm\nabla\bm\times\bm E &=& -\frac{\partial c\bm B}
{\partial ct} -c\mu_0 \bm K  ,
\end{eqnarray}
in agreement with Eq.~(\ref{eq:xme3}).

\section{Matrix form of the Maxwell equations}
\label{sec:mme}

As already mentioned, it is necessary to confirm the relativistic
invariance of the extended Maxwell equations.  To do this, it is useful
to write the equations in a matrix form that provides a compact notation
for the otherwise complicated algebraic equations.  In this section, a
brief review of this approach provides the basic tools.  See also
\citet{2010043} and \citet{ja} for additional information.

We can express a three-vector $\bm a$ with Cartesian coordinates
$a^1,a^2,a^3$ as the matrix
\begin{eqnarray}
\bm a\rc &=&
\left(\begin{array}{c} a^1 \\ a^2 \\ a^3 \end{array}\right)  ,
\end{eqnarray}
and in a spherical basis, we have
\begin{eqnarray}
\bm a\rs
=\bm M \bm a\rc 
&=& \frac{1}{\sqrt{2}}\left(\begin{array}{ccc}
 -1 & \ri & 0 \\
 0 &   0  & \sqrt{2} \\
 1 &  \ri & 0 \end{array}\right)
 \left(\begin{array}{c} a^1 \\ a^2 \\ a^3 \end{array}\right)
\nonumber\\[10 pt]
&=& \left(\begin{array}{c}-\frac{1}{\sqrt{2}}(a^1 - {\rm i}\,a^2) \\
a^3 \\
\frac{1}{\sqrt{2}}(a^1 + {\rm i}\,a^2)\end{array}\right)  .
\label{eq:asph}
\end{eqnarray}
The dot product of two vectors is given by
\begin{eqnarray}
\bm a\bm\cdot\bm b &=& \bm a\rc^\dagger\bm b\rc
= \bm a\rs^\dagger\bm b\rs
= a^{i*}b^i  .
\label{eq:dot}
\end{eqnarray}
Three Hermitian $(\tau^{i\dagger} = \tau^i)$ matrices are defined
as
\begin{eqnarray}
\tau^1 = 
\frac{1}{\sqrt{2}} \left(\begin{array}{c@{\quad}c@{\quad}c} 
0   & 1  &     0
\\ 1 &   0     &  1 \\ 0   & 1  &     0
\end{array}\right)\!; \,
&&\tau^2 = 
\frac{\ri}{\sqrt{2}} \left(\begin{array}{c@{\quad}c@{\quad}c} 
0    & \! -1  &
0   \\ 1 &    0     & \! -1 \\ 0    & 1  &     0
\end{array}\right)\!; \quad
\nonumber\\[10 pt]
\tau^3 &=& \left(\begin{array}{c@{\quad}c@{\quad}c} 
1    &    0     &     0
\\ 0    &    0     &     0   \\ 0    &    0     & \!\! -1
\end{array}\right) . 
\end{eqnarray}
Similar matrices have been given by \citet{1931004}, by
Majorana \cite{1974037}, and others.  We use this form for the matrices,
because they are direct analogs of the Pauli spin matrices.  The dot
product with a vector $\bm a$ is 
\begin{eqnarray}
&&\bm \tau \bm\cdot \bm a = \tau^i\,a^i
\\[10 pt]\nonumber
&=& \left(\begin{array}{c@{\quad}c@{\quad}c} 
a^3   & \frac{1}{\sqrt{2}}(a^1 - \ri\,a^2)  &     0
\\ \frac{1}{\sqrt{2}}(a^1 + \ri\,a^2) 
&   0     &  \frac{1}{\sqrt{2}}(a^1 - \ri\,a^2)
\\ 0   & \frac{1}{\sqrt{2}}(a^1 + \ri\,a^2)  &    -a^3 
\end{array}\right) .\quad 
\end{eqnarray}
These matrices have the property that
\begin{eqnarray}
\bm\tau\bm\cdot\bm a \, \bm b\rs &=& \ri\,(\bm a \bm\times \bm b)\rs  ,
\label{eq:fund}
\end{eqnarray}
where $(\bm a \bm\times \bm b)\rs$ is the ordinary vector cross product
expressed in the spherical basis.

The Maxwell equations are conventionally written in 3-dimensional vector
notation, but for the purposes of this paper, it is convenient to also
use a $6\bm\times6$ matrix notation.  This matrix version of the Maxwell
equations is the direct analog of the $4\bm\times4$ matrix Dirac equation,
in which the $2\bm\times2$ Pauli (sigma) matrices are replaced by
$3\bm\times3$ (tau) matrices.  Taking into account the corresponding
relations
\begin{eqnarray}
\bm \tau \bm\cdot \bm \nabla \, \bm B\rs &=&
\ri\,(\bm\nabla\bm\times\bm B)\rs  ,
\\[10 pt]
\bm \tau \bm\cdot \bm \nabla \, \bm E\rs &=& \ri\,(\bm\nabla\bm\times\bm
E)\rs  ,
\end{eqnarray}
we can write the two source-free Maxwell equations in vacuum
\begin{eqnarray}
\frac{\partial \bm E(x)}{\partial ct}
-\bm \nabla\bm\times c\bm B(x)
&=& 0 \label{eq:gme2}\, ,\\[8 pt]
\frac{\partial c \bm B(x)} {\partial ct} 
+\bm \nabla\bm\times\bm E(x)
&=& 0 \label{eq:gme3}\, ,
\end{eqnarray}
as
\begin{eqnarray}
\left(\begin{array}{ccc}
\bm I\,\ft{\partial}{\partial ct} && \bm\tau\bm\cdot\bm\nabla \\
-\bm\tau\bm\cdot\bm\nabla && -\bm I\,\ft{\partial}{\partial ct}
\end{array}\right) \left(\begin{array}{c}
\bm E\rs(x) \\ \ri \, c \bm B\rs(x) \msp \end{array}\right) = 
0 ,
\label{eq:max6}
\end{eqnarray}
where $\bm I$ is the $3\bm\times3$ identity matrix, and the four-vector $x$
is defined by Eq.~(\ref{eq:xdef}).

Employing the analogy with the Dirac equation, we define
$6\bm\times6$ gamma matrices, which are analogs of the Dirac $4\bm\times4$
gamma matrices, by 
\begin{eqnarray}
\gamma^0 = \left(\begin{array}{ccc} \bm I && \0 \\
\0 && -\bm I \end{array}\right)\! ; \
\gamma^i = \left(\begin{array}{ccc} \0 && \tau^i \\
-\tau^i && \0 \end{array}\right)  , \ i = 1,2,3 \, ,
\label{eq:ggammas}
\\ \nonumber
\end{eqnarray}
where $\bm 0$ is the $3\bm\times3$ matrix of zeros.  With the derivatives
\begin{eqnarray}
\partial_0 = \frac{\partial}{\partial ct}; \quad
\partial_i = \frac{\partial}{\partial x^i} \, , \quad i=1,2,3
\label{eq:partials}
\end{eqnarray}
we have
\begin{eqnarray}
\left(\begin{array}{ccc}
\bm I\,\ft{\partial}{\partial ct} && \bm\tau\bm\cdot\bm\nabla \\
-\bm\tau\bm\cdot\bm\nabla && -\bm I\,\ft{\partial}{\partial ct}
\end{array}\right) &=&
\gamma^\mu\,\partial_\mu  .
\end{eqnarray}
If a six-row matrix containing the electric and magnetic fields is
written as
\begin{eqnarray}
\left(\begin{array}{c}
\bm E\rs(x) \\[5 pt] 
\ri c \bm B\rs(x)
\end{array}\right) &=& {\it \Psi}(x)  ,
\label{eq:psi}
\end{eqnarray}
then
\begin{eqnarray}
\gamma^\mu\partial_\mu \iP(x) &=& 0  ,
\label{eq:dmax}
\end{eqnarray}
which has the same form as the Dirac equation (with zero mass), with the
exception of the dimensionality of the tau matrices.  The other two
source-free Maxwell equations
\begin{eqnarray}
\bm \nabla\bm\cdot\bm E(x) &=& 0  , \\[5 pt]
\bm \nabla\bm\cdot c \bm B(x) &=& 0  ,
\end{eqnarray}
may be written as
\begin{eqnarray}
{\cal D}\,{\it \Psi}(x) = 0  ,
\label{eq:leq}
\end{eqnarray}
where
\begin{eqnarray}
{\cal D} &=& \left(\begin{array}{cc}
-\bm\nabla\rs^\dagger & \bm 0
\\[10 pt]
\bm 0 & \bm\nabla\rs^\dagger
\end{array}\right)  ,
\end{eqnarray}
and $\bm 0$ is a row of 3 zeros.

If sources are present, we have
\begin{eqnarray}
\gamma^\mu\partial_\mu \iP(x) &=& \iX(x)  ,
\label{eq:mmeq}
\\[5 pt]
{\cal D}\,\iP(x) &=& {\cal X}(x)  ,
\label{eq:other}
\end{eqnarray}
where $\iX(x)$ and ${\cal X}(x)$ are source terms given by
\begin{eqnarray}
\iX(x) &=&
\left(\begin{array}{c}
-\ft{1}{c\epsilon_0}\bm J_{\rm s}(x) \\[5 pt]
\ri\,c\mu_0\,\bm K_{\rm s}(x)
 \end{array} \right) 
=
 \frac{1}{c\epsilon_0}\left(\begin{array}{c}
-\bm J_{\rm s}(x) \\[5 pt]
 \ri\bm K_{\rm s}(x)
 \end{array} \right) 
 \nonumber\\[10 pt]
&=&
 c\mu_0\left(\begin{array}{c}
-\bm J_{\rm s}(x) \\[5 pt]
 \ri\bm K_{\rm s}(x)
 \end{array} \right) 
\label{eq:sources}
\end{eqnarray}
and
\begin{eqnarray}
{\cal X}(x) &=& 
\frac{1}{\epsilon_0}
\left(\begin{array}{c} -\rho(x) \\[10 pt]
\ri\,\sigma(x)\end{array}\right)  .
\label{eq:lsource}
\end{eqnarray}
We also have
\begin{eqnarray}
\overline{\iP}(x)\overleftarrow{\partial}_{\!\mu}\gamma^\mu
\label{eq:adjeq}
=\overline{\iX}(x) ,
\\[10 pt]
\overline{\iP}(x)\overleftarrow{\cal D}^\dagger
=\overline{\cal X}(x) ,
\label{eq:tmmeq}
\end{eqnarray}
where
\begin{eqnarray}
\overline{\iP}(x) = \iP^\dagger(x)\gamma^0 ;\
\overline{\iX}(x) = \iX^\dagger(x)\gamma^0 ;\
\overline{\cal X}(x) = {\cal X}^\dagger(x)\gamma^0  ,
\nonumber\\
\end{eqnarray}
and
\begin{eqnarray}
\gamma^{\mu\,\dagger} = \gamma^0\gamma^\mu\gamma^0  .
\end{eqnarray}
Although the matrix formulation simplifies complicated calculations, we
shall in general use ordinary vector notation.

\section{Lorentz invariance}
\label{sec:li}

The Maxwell equations are consistent with special relativity, but it is
necessary to show that the equations with the added magnetic source
terms are also consistent with special relativity.  Despite the symmetry
between the electric and magnetic sources, this is not obvious because
the magnetic current is a three-vector.  Here we extend the method of
showing Lorentz invariance of the conventional Maxwell equations given
by~\citet{2010043} to include the magnetic source terms.

To establish the invariance, it is sufficient to restrict our attention
to the homogeneous Lorentz transformations. The rotation, velocity, and
discrete transformations in this subset may be considered individually.
These transformations leave the four-vector scalar product $x\bm\cdot x$
invariant, where
\begin{eqnarray}
x = \left(\begin{array}{c} ct \\ \bm x\rc
\end{array}\right)
= \left(\begin{array}{c} x^0 \\ x^1 \\ x^2 \\ x^3
\end{array}\right) 
\label{eq:xdef}
\end{eqnarray}
and
\begin{eqnarray}
x\bm\cdot x &=& 
x^\top g \, x =
(ct)^2 - \bm x^2 
\, ,
\label{eq:sprod}
\end{eqnarray}
where $\top$ denotes the matrix transpose.  The $4\bm\times4$ metric tensor
$g$ is given by
\begin{eqnarray}
g = \left(\begin{array}{ccc}
1 && \0 \\
\0 && -\bm I \msp \end{array}\right)  ,
\label{eq:gc}
\end{eqnarray}
where $\bm 0$ signifies a $1\bm\times3$ array
of zeros in the upper-right position and a $3\bm\times1$ array in the
lower-left position.

\subsection{Rotations}

Invariance under rotations is self-evident since the spatial dependence
of the four-vector scalar product is  $\bm x^2$ and the Maxwell
equations transform as either scalars or 3-vectors under rotations.  In
particular, the magnetic terms are in this class.  The source $\sigma$
is the scalar product of two quantities that transform like vectors
under rotations, namely the gradient operator and the magnetic moment
vector.  Moreover, the magnetic current $\bm K$ is a three-vector that
transforms like an ordinary vector under rotations.  However, invariance
under velocity transformations requires closer examination.

\subsection{Velocity transformations}
\label{ssec:vtpsi}

Lorentz invariance of the extended Maxwell equations is established by
showing that if $\iP(x)$ is a solution of Eqs.~(\ref{eq:mmeq}) and
(\ref{eq:other}), then \cite{bd}
\begin{eqnarray}
\gamma^\mu\partial_\mu^{\,\prime} \iP^\prime(x^\prime) &=&
\iX^\prime(x^\prime) ,
\label{eq:transheq}
\\[10 pt]
{\cal D}^\prime\iP^\prime(x^\prime) 
&=& {\cal X}^\prime(x^\prime) \, ,
\label{eq:longeq}
\end{eqnarray}
where the primes denote Lorentz transformed quantities.

The velocity transformation of the four-vector coordinate is given by
$x^\prime = V(\bm v)x$, where $V(\bm v)$ is the $4\bm\times4$ matrix
\cite{2010043}
\begin{eqnarray}
V(\bm v) &=& \re^{\zeta \iA(\bm{\hat v})}
\nonumber\\[10 pt]
&=&\left(\begin{array}{ccc}
\cosh{\zeta} && \bm{\hat v}\rc^\top\sinh{\zeta} \\
\bm{\hat v}\rc\sinh{\zeta} && \bm I
+\bm{\hat v}\rc\bm{\hat v}\rc^\top\left(\cosh{\zeta}-1\right)
\msp\end{array}\right)  , \qquad
\label{eq:cvt}
\end{eqnarray}
$\bm v = c\tanh{\zeta} \,\bm{\hat v}$ is the velocity of the
transformation, and
\begin{eqnarray}
\iA(\bm{\hat v}) &=& \left(\begin{array}{cc}
0 & \bm{\hat v}\rc^\top \\
\bm{\hat v}\rc & \0 \end{array}\right)  .
\label{eq:K}
\end{eqnarray}
In Eq.~(\ref{eq:K}) and in the following, $\bm 0$ denotes the
appropriate array of zeros to fill out the unoccupied spaces.  This
transformation leaves the scalar product invariant, because
\begin{eqnarray}
x^\prime\bm\cdot x^\prime 
&=& x^\top V^\top(\bm v)gV(\bm v) x
\nonumber\\[10 pt]
&=& x^\top gV^{-1}(\bm v)V(\bm v) x
=x\bm\cdot x \, ,
\end{eqnarray}
and $V^\top(\bm v) = V(\bm v) = gV^{-1}(\bm v)g$.  The infinitesimal
transformation
\begin{eqnarray}
x^\prime = x + \zeta \iA(\bm{\hat v}) x + \dots
= \left(\begin{array}{c} ct + \bm v\bm\cdot\bm x/c +\dots\\
\bm x_c + \bm v_c t + \dots \end{array}\right)
\end{eqnarray}
shows that the transformed coordinate has the appropriate form, i.e.,
the boosted space coordinate is increasing with the velocity $\bm v$ and
$x^\prime\bm\cdot x^\prime = x\bm\cdot x + {\cal O}\left(\bm v\bm\cdot\bm
x\,t\right)$.  We use the convention that the transformations are
applied to the properties of the physical system, rather than to the
observers coordinates.

For the derivatives
$\partial_\mu$, we have $x = V^{-1}(\bm v) x^\prime = V(-\bm v)
x^\prime$ or $x^\nu=V_{\nu\mu}(-\bm v)x^{\prime\mu}$ so that
\begin{eqnarray}
\partial_\mu^{\,\prime} &=&
\frac{\partial }{\partial x^{\prime\,\mu}} = 
\frac{\partial x^\nu}{\partial x^{\prime\,\mu}}
\,\frac{\partial}{\partial x^\nu}
\nonumber\\[10 pt]
&=& V_{\nu\mu}(-\bm v) \, \partial_\nu 
= V_{\mu\nu}(-\bm v) \, \partial_\nu  \, .
\end{eqnarray}
Thus
\begin{eqnarray}
&&\left(\begin{array}{c} \ft{\partial}{\partial ct^\prime} \\[5
pt]
\bm \nabla^\prime\rc\end{array}\right)
=V(-\bm v)
\left(\begin{array}{c} \ft{\partial}{\partial ct} \\[5 pt]
\bm \nabla\rc\end{array}\right)
\nonumber\\[10 pt]&&
=
\left(\begin{array}{ccc}
\ft{\partial}{\partial ct}
\cosh{\zeta} -\bm{\hat v}\bm\cdot\bm\nabla\sinh{\zeta} \\
\bm \nabla\rc
 +\bm{\hat v}\rc\bm{\hat v}\bm\cdot \bm \nabla\left(\cosh{\zeta}-1\right)
-\bm{\hat v}\rc \ft{\partial}{\partial ct}\sinh{\zeta} 
\msp\end{array}\right)  . \qquad
\label{eq:tderivs}
\end{eqnarray}

The source currents are the space components of the electric and
magnetic four-currents, which include the sources in
Eqs.~(\ref{eq:xme1}) and (\ref{eq:xme4}) as the timelike components.
Explicitly, we have
\begin{eqnarray}
J\rs(x) &=& \left(\begin{array}{c}
c\rho(x) \\[5 pt] 
\bm J\rs(x) \end{array}\right)  ; \quad
K\rs(x) = \left(\begin{array}{c}
c\sigma(x) \\[5 pt] 
\bm K\rs(x) \end{array}\right)  , \qquad
\end{eqnarray}
both of which transform as four-vectors.  In this case, the
transformation matrix is the spherical version of $V$
\begin{eqnarray}
V\rs(\bm v) &=&
\left(\begin{array}{cc} 1 & \0 \\ \0 & \bm M \end{array}\right)
V(\bm v)
\left(\begin{array}{cc} 1 & \0 \\ \0 & \bm M^\dagger \end{array}\right)
\nonumber\\[10 pt]
&=& \left(\begin{array}{ccc}
\cosh{\zeta} && \bm{\hat v}\rs^\dagger\sinh{\zeta} \\
\bm{\hat v}\rs\sinh{\zeta} && \bm I
+\bm{\hat v}\rs\bm{\hat
v}\rs^\dagger\left(\cosh{\zeta}-1\right)\msp\end{array}\right) ,
\qquad
\label{eq:sav}
\end{eqnarray}
which gives
\begin{widetext}
\begin{eqnarray}
J\rs^\prime(x^\prime) &=& V\rs(\bm v) J\rs(x) = 
 \left(\begin{array}{c}
c\rho(x)\cosh{\zeta} + \bm{\hat v}\bm\cdot\bm J(x)\sinh{\zeta} \\
\bm J\rs(x) +\bm{\hat v}\rs \, \bm{\hat v}\bm\cdot\bm J(x)
\left(\cosh{\zeta}-1\right) +
\bm{\hat v}\rs \,c\rho(x) \sinh{\zeta} \end{array}\right)  ,
\label{eq:transj}
\\[10 pt]
K\rs^\prime(x^\prime) &=& V\rs(\bm v) K\rs(x)
= \left(\begin{array}{c}
c\sigma(x)\cosh{\zeta} 
+ \bm{\hat v}\bm\cdot\bm K(x)\sinh{\zeta} \\
\bm K\rs(x) +\bm{\hat v}\rs \, \bm{\hat v}\bm\cdot\bm K(x)
\left(\cosh{\zeta}-1\right) +
\bm{\hat v}\rs \,c\sigma(x) \sinh{\zeta} \end{array}\right)  .
\label{eq:transk}
\end{eqnarray}

The function $\iP^\prime(x^\prime)$ in Eq.~(\ref{eq:transheq}) is
\begin{eqnarray}
\iP^\prime(x^\prime) &=& \cV(\bm v)\iP\!\big(x\big),
\label{eq:sol}
\end{eqnarray}
or
\begin{eqnarray}
\iP^\prime(x) &=& \cV(\bm v)\iP\!\big(V^{-1}(\bm v)\,x\big),
\label{eq:solp}
\end{eqnarray}
where $\cV(\bm v)$ is a $6\bm\times6$ matrix that gives the linear
transformation of $\iP(x)$.  It can be written as \cite{2010043}
\begin{eqnarray}
\cV(\bm v) = \re^{\zeta\bm \iA \bm\cdot \bm{\hat v}}
&=& \left(\begin{array}{cc}
\bm I + (\bm \tau \bm\cdot \bm{\hat v})^2 \,(\cosh{\zeta}-1) 
& \bm \tau \bm\cdot \bm{\hat v} \,\sinh{\zeta} \\
\bm \tau \bm\cdot \bm{\hat v}\,\sinh{\zeta} &
\bm I + \left(\bm \tau \bm\cdot \bm{\hat v}\right)^2(\cosh{\zeta}-1)
\end{array}\right)
\nonumber\\[10 pt]
&=& \left(\begin{array}{cc}
\bm I\cosh{\zeta} -
\bm{\hat v}\rs\bm{\hat v}\rs^\dagger \,(\cosh{\zeta}-1) 
& \bm \tau \bm\cdot \bm{\hat v} \,\sinh{\zeta} \\
\bm \tau \bm\cdot \bm{\hat v}\,\sinh{\zeta} &
\bm I\cosh{\zeta} -
\bm{\hat v}\rs\bm{\hat v}\rs^\dagger \,(\cosh{\zeta}-1) 
\end{array}\right)  ,
\label{eq:vdef}
\end{eqnarray}
where
\begin{eqnarray}
\bm \iA = \left(\begin{array}{cc} \0 & \bm \tau \\ \bm \tau & \0
\end{array}\right) .
\label{eq:kdefx}
\end{eqnarray}
Equation~(\ref{eq:vdef}) follows from the series expansion of the
exponential function together with the identities
\begin{eqnarray}
(\bm\tau\bm\cdot\bm{\hat v})^2 &=& \bm I -
\bm{\hat v}\rs\bm{\hat v}\rs^\dagger  ,
\\[0 pt]
(\bm\tau\bm\cdot\bm{\hat v})^3 &=& \bm\tau\bm\cdot\bm{\hat v}  .
\end{eqnarray}
The transformed fields are thus\footnote{Our convention
differs from \citet{1998165} by the sign of $\bm v$.}
\begin{eqnarray}
\cV(\bm v)\iP(x) &=& 
\left(\begin{array}{c}
\bm E\rs(x)\cosh{\zeta} - \bm{\hat v}\rs 
\bm{\hat v}\bm\cdot \bm E(x)
(\cosh{\zeta}-1) + \ri \, \bm \tau \bm\cdot \bm{\hat v} 
\, c \bm B\rs(x) \sinh{\zeta}
\\ \ri \left[c\bm B\rs(x)\cosh{\zeta} - \bm{\hat v}\rs
\bm{\hat v}\bm\cdot
c\bm B\rs(x) (\cosh{\zeta}-1) 
- \ri \, \bm \tau \bm\cdot \bm{\hat v} \, \bm E\rs(x)
\sinh{\zeta}\,\right] 
\msp \end{array}\right). \qquad
\label{eq:ltem}
\end{eqnarray}
\end{widetext}
Incidentally, the relations $\cV^\dagger \gamma^0 \cV = \gamma^0$ and
$\cV^\dagger \gamma^0 \eta \cV = \gamma^0 \eta$, where
\begin{eqnarray}
\eta = \left(\begin{array}{ccc} \0 && \bm I \\
\bm I && \0 \end{array}\right),
\label{eq:eta}
\end{eqnarray}
yield
\begin{eqnarray}
\overline{\iP}^{\,\prime}(x^\prime)\iP^{\,\prime}(x^\prime) &=& 
\overline{\iP}(x)\iP(x)
\\
\overline{\iP}^{\,\prime}(x^\prime)\eta\iP^{\,\prime}(x^\prime) &=& 
\overline{\iP}(x)\eta\iP(x) \, ,
\end{eqnarray}
or in vector notation
\begin{eqnarray}
|\bm E^\prime(x^\prime)|^2 - c^2|\bm B^\prime(x^\prime)|^2 &=&
|\bm E(x)|^2 - c^2|\bm B(x)|^2 \, ,
\\
{\rm Re} \, \bm E^\prime(x^\prime)\bm\cdot \bm B^\prime(x^\prime) &=&
{\rm Re} \, \bm E(x)\bm\cdot \bm B(x) \, ,
\end{eqnarray}
which are the conventional invariants of electromagnetism.

To specify our convention for the transformations, the physical system
is the combination of electric and magnetic fields along with the source
terms that are boosted by the velocity $\bm v$.  However, a particle
moving with a velocity $\bm v$ relative to a reference (laboratory)
frame will observe fields transformed by a velocity $-\bm v$ relative to
its reference frame, and will be subjected to the corresponding forces.
In this case, the relevant transformation for small $|\bm v|/c$ is
\begin{eqnarray}
&&\cV(-\bm v)\iP(x) = 
\cV^{-1}(\bm v)\iP(x) 
\nonumber\\[10 pt]
&&\qquad= 
\left(\begin{array}{c}
\bm E\rs(x) + \left(\bm v \bm\times \bm B(x)\right)\rs + \dots\\[10 pt]
\ri \left[\, c \bm B\rs(x) - \left(\bm v \bm\times \bm E(x)\right)\rs 
+ \dots \right]
\end{array}\right) . \qquad
\end{eqnarray}

To calculate $\gamma^\mu\partial_\mu^{\,\prime} \iP^\prime(x^\prime)$,
we start with 
\begin{eqnarray}
\gamma^\mu \partial^{\,\prime}_\mu
&=& \left(\begin{array}{ccc}
\bm I\,\frac{\textstyle \partial}{\textstyle \partial ct^\prime} 
&& \bm\tau\bm\cdot\bm\nabla^\prime
\\
-\bm\tau\bm\cdot\bm\nabla^\prime && -\bm I\,\frac{\textstyle
\partial}{\textstyle \partial ct^\prime}
\end{array}\right) ,
\label{eq:tdiffop}
\\ \nonumber
\end{eqnarray}
where the derivatives are given in Eq.~(\ref{eq:tderivs}).  The product
of Eq.~(\ref{eq:tdiffop}) and $\cV(\bm v)$ is (see Appendix
\ref{app:lti1})
\begin{widetext}
\begin{eqnarray}
\gamma^\mu \partial^{\,\prime}_\mu \, {\cal V}(\bm v)
&=&
\left(\begin{array}{ccc}
\bm I+\bm{\hat v}_{\rm s}\bm{\hat v}_{\rm s}^\dagger
(\cosh{\zeta}-1)
&& \0 \\
\0 &&
\bm I+\bm{\hat v}_{\rm s}\bm{\hat v}_{\rm s}^\dagger
(\cosh{\zeta}-1)
 \end{array} \right)
\gamma^\mu \, \partial_\mu 
 +
 \left(\begin{array}{ccc}
 \bm{\hat v}_{\rm s}\sinh{\zeta} &&
 \0 \\
 \0 && \bm{\hat v}_{\rm s}\sinh{\zeta}
 \end{array}\right){\cal D}  .
 \label{eq:firstid}
\end{eqnarray}
We thus have
\begin{eqnarray}
\gamma^\mu\partial_\mu^{\,\prime} \iP^\prime(x^\prime) &=& 
\gamma^\mu\partial_\mu^{\,\prime} \cV(\bm v)\iP(x) 
\nonumber\\[10 pt] &=& 
\left(\begin{array}{ccc}
\bm I+\bm{\hat v}_{\rm s}\bm{\hat v}_{\rm s}^\dagger
(\cosh{\zeta}-1) && \0 \\
\0 &&
\bm I+\bm{\hat v}_{\rm s}\bm{\hat v}_{\rm s}^\dagger
(\cosh{\zeta}-1)
 \end{array} \right)
\iX(x)
+\left(\begin{array}{ccc}
 \bm{\hat v}_{\rm s}\sinh{\zeta} &&
 \0 \\
 \0 && \bm{\hat v}_{\rm s}\sinh{\zeta}
 \end{array}\right){\cal X}(x)
 \nonumber\\[10 pt] &=& 
\left(\begin{array}{c}
-\ft{1}{c\epsilon_0}\left[\bm J_{\rm s}(x)+\bm{\hat v}_{\rm s}
\bm{\hat v} \bm\cdot \bm J(x)
(\cosh{\zeta}-1)
+ \bm{\hat v}_{\rm s} c \rho(x)\sinh{\zeta}\,\right] \\
\ri\,c\mu_0\left[
\bm K_{\rm s}(x)+\bm{\hat v}_{\rm s}
\bm{\hat v} \bm\cdot \bm K(x)
(\cosh{\zeta}-1)
+ \bm{\hat v}_{\rm s} c \sigma(x)\sinh{\zeta}\,
\right]
 \end{array} \right)  .
 \label{eq:transprod}
 \end{eqnarray}
The transformed source term is obtained directly from the bottom
lines of Eqs.~(\ref{eq:transj}) and (\ref{eq:transk}), which give
\begin{eqnarray}
\iX^\prime(x^\prime) &=&
\left(\begin{array}{c}
-\ft{1}{c\epsilon_0}\bm J_{\rm s}^\prime(x^\prime) \\
\ri\,c\mu_0\,\bm K_{\rm s}^\prime(x^\prime)
 \end{array} \right) 
=\left(\begin{array}{c}
-\ft{1}{c\epsilon_0}\left[\bm J_{\rm s}(x)+\bm{\hat v}_{\rm s}
\bm{\hat v} \bm\cdot \bm J(x)
(\cosh{\zeta}-1)
+ \bm{\hat v}_{\rm s} c \rho(x)\sinh{\zeta}\,\right] \\
\ri\,c\mu_0\left[
\bm K_{\rm s}(x)+\bm{\hat v}_{\rm s}
\bm{\hat v} \bm\cdot \bm K(x)
(\cosh{\zeta}-1)
+ \bm{\hat v}_{\rm s} c \sigma(x)\sinh{\zeta}\,
\right]
 \end{array} \right)  ,
 \label{eq:ltrans}
\end{eqnarray}
\end{widetext}
in agreement with Eq.~(\ref{eq:transprod}), which confirms
Eq.~(\ref{eq:transheq}).

For Eq.~(\ref{eq:longeq}), to calculate ${\cal
D}^\prime\iP^\prime(x^\prime)$, we first write (see
Appendix~\ref{app:lti2})
\begin{eqnarray}
{\cal D}^\prime\cV(\bm v) &=&
 \left(\begin{array}{c@{\quad}c}
\bm I\cosh\zeta & \bm{0}
\\[10 pt]
\bm{0}&
\bm I\cosh\zeta
\end{array}\right) {\cal D}
\nonumber\\[10 pt]&&+
 \left(\begin{array}{c@{\quad}c}
\bm{\hat v}\rs^\dagger \sinh{\zeta}
&\bm{0} \\[10 pt]
\bm{0} &
\bm{\hat v}\rs^\dagger \sinh{\zeta}
\end{array}\right) \gamma^\mu\partial_\mu . \qquad
\label{eq:secondid}
\end{eqnarray}
The product ${\cal D}^\prime\Psi^\prime(x^\prime)={\cal D}^\prime\cV(\bm
v)\Psi(x)$ is thus
\begin{eqnarray}
&&{\cal D}^\prime\cV(\bm v)\Psi(x) =
\left(\begin{array}{c@{\quad}c}
\bm I\cosh\zeta & \bm{0}
\\[10 pt]
\bm{0}&
\bm I\cosh\zeta
\end{array}\right) {\cal X}(x)
\nonumber\\[10 pt]&&\qquad\qquad\qquad+
\left(\begin{array}{c@{\quad}c}
\bm{\hat v}\rs^\dagger \sinh{\zeta}
&\bm{0} \\[10 pt]
\bm{0} &
\bm{\hat v}\rs^\dagger \sinh{\zeta}
\end{array}\right) \iX(x)
\nonumber\\[10 pt]&&=
\frac{1}{\epsilon_0} \left(\begin{array}{c}
-\rho(x) \cosh\zeta
-\ft{1}{c}\,\bm{\hat v}\bm\cdot\bm J(x)
\sinh{\zeta} \\[10 pt]
 \ri\,\sigma(x) \cosh\zeta
+ \ri\,\ft{1}{c}\,
\bm{\hat v}\bm\cdot\bm K(x)
\sinh{\zeta}
\end{array}\right) . \qquad
\label{eq:longtrans}
\end{eqnarray}
The transformed source term may be directly read from the top lines of
Eqs.~(\ref{eq:transj}) and (\ref{eq:transk}) to be
\begin{eqnarray}
&&{\cal X}^\prime(x^\prime) = 
\frac{1}{\epsilon_0}
\left(\begin{array}{c} -\rho^\prime(x^\prime) \\[10 pt]
\ri\,\sigma^\prime(x^\prime)\end{array}\right)
\nonumber\\[10 pt]
&&\quad = \frac{1}{\epsilon_0}
\left(\begin{array}{c} -\rho(x)\cosh\zeta
- \ft{1}{c}\,\bm{\hat v}\bm\cdot\bm J(x)\sinh\zeta\\[10 pt]
\ri\,\sigma(x)\cosh\zeta
+ \ri\,\ft{1}{c}\,\bm{\hat v}\bm\cdot\bm K(x)\sinh\zeta\\[10 pt]
\end{array}\right) , \qquad
\end{eqnarray}
in agreement with Eq.~(\ref{eq:longtrans}), which confirms
Eq.~(\ref{eq:longeq}).

\subsection{Parity and time-reversal}
\label{ssec:ptpsi}

If magnetic source and current terms are included in the Maxwell
equations, they must be consistent with parity inversion and time
reversal transformations.  There is consistency for the equations
without these additions, so it is only necessary to consider the effects
of the additional terms.  The properties of various quantities and the
Maxwell equations under these transformations are summarized in
Table~\ref{tab:pt}.
\begin{table}
\caption{Properties of various quantities and the Maxwell equations
under space inversion or time reversal.  The symbols $\bm+$ and $\bm-$
in the second and third columns indicate evenness or oddness under the
corresponding transformation.  In the first column, $q$ represents
charge which by convention does not change under either transformation.
\label{tab:pt}}
$$
\begin{array}{rcl@{\qquad}c@{\quad}c}
\hline
\hline
\multicolumn{3}{c}{\mbox{Quantity/Equation}} &\mbox{Space}
&\mbox{Time}\\[0 pt]
&&& \mbox{inversion} & \mbox{reversal} \\[0 pt]
\hline
\bm x,\bm\nabla &&& \bm{-} & \bm{+} \\[0 pt]
t &&& \bm{+} & \bm{-} \\[0 pt]
q&&& \bm{+} & \bm{+} \\[0 pt]
\bm m &&& \bm{+} & \bm{-} \\[5 pt]
 \bm E(x) &&
 & \bm{-} & \bm{+} \\[5 pt]
 \bm B(x) &&
 & \bm{+} & \bm{-} \\[5 pt]
\bm \nabla\bm\cdot\bm E(x)
&=& \ft{\rho(x) }{\epsilon_0} & \bm{+} & \bm{+} \\[5 pt]
\ft{\partial \bm E(x)}{\partial ct}
-\bm \nabla\bm\times c\bm B(x)
&=& -\ft{1}{c\epsilon_0}\,\bm J(x) & \bm{-} & \bm{-} \\[5 pt]
\ft{\partial c \bm B(x)} {\partial ct}
+\bm \nabla\bm\times\bm E(x)
&=& - c\mu_0\,\bm K(x) & \bm{+} & \bm{+} \\[5 pt]
\bm \nabla\bm\cdot c \bm B(x)
&=& c^2\mu_0\,\sigma(x) & \bm{-} & \bm{-} \\
\hline
\end{array}
$$
\end{table}

The dipole source term $\sigma(x) = -\bm m\bm\cdot \bm\nabla \delta(\bm
x)/c$ is odd under a parity reversal.  If the moment is viewed as the
result of a current loop, or more specifically to a single charge on a
circular path, then the parity transformation transports the particle to
the opposite side of the circle and also reverses the direction of the
motion, so there is no net change in the current.  It is conventional to
require that charge does not change under a parity reversal.
Alternatively, one may use the fact that the magnetic moment of an
electron is proportional to its spin angular momentum and angular
momentum does not change sign under a parity reversal.  On the other
hand, if the magnetic moment is considered as two opposite polarity
magnetic monopoles, the locations of the monopoles are interchanged, so
the monopole polarity must change sign under a parity transformation.
The gradient operator $\bm\nabla$ is odd under the parity reversal, so
the combined result is that $\sigma$ is odd under the inversion.  On the
left-hand side of the fourth Maxwell equation, the magnetic flux density
is even under the parity change, while the gradient operator is odd.
The consequence is that both sides of the fourth Maxwell equation are
odd under space inversion, which is the consistent result.

Under time-reversal, $\sigma$ changes sign because the magnetic moment
is odd and the gradient operator is even.  The odd nature of the
magnetic moment may be visualized as the reversal of the velocity of the
rotating charge in a hypothetical current loop, while the location of
the charge does not change.  Moreover, it is conventional to require
that charge does not change under time reversal.  The gradient operator
is even so the net result is that $\sigma$ is odd under time reversal.
On the left-hand side of the fourth equation, the magnetic flux density
is odd under time reversal and the gradient operator is even.  So both
sides of that equation are odd under time reversal, which is again the
consistent result.

The magnetic current $\bm K(x)$ can be thought of as a source
$\sigma(x)$ in motion.  Velocity is odd under either space inversion or
time reversal, so both parity and time reversal evenness or oddness are
the opposite for $\bm K(x)$ of what they are for $\sigma(x)$.  Thus for
the third Maxwell equation, both sides are even under parity or time
inversion, giving the consistent result.  The opposite properties of
$\sigma(x)$ and $\bm K(x)$ are also necessary for consistency with the
continuity equation, Eq.~(\ref{eq:conk}).

The above considerations apply to electric monopole and magnetic dipole
sources, $\rho$ and $\sigma$.  On the other hand, one may consider more
general sources.  For the electric source, from the definition of the
electric field $\bm E$, the source term must be even under a parity
transformation.  This rules out electric dipole sources, but not
electric quadrupole or higher-moment sources, provided they are even
moments, which are positive under space inversion.  Similarly, the
magnetic source term must be odd under a parity transformation, which
rules out magnetic monopoles, allows magnetic dipoles, rules out
magnetic quadrupoles, but may allow higher moments with odd parity.  The
parity restrictions carry over to the currents associated with the
charges.  This allows monopole electric currents and dipole magnetic
currents and the corresponding higher multipole generalizations.

\section{Relativistic extended Poynting theorem}
\label{sec:rpt}

This section provides a relativistic derivation of the extended Poynting
theorem, with no restriction to the small velocity limit imposed in
Sec~\ref{sec:mfpi}.  The extended Poynting theorem is a consequence of
the extended Maxwell equations, and because the extended Maxwell
equations are Lorentz invariant, it follows that the extended Poynting
theorem is also Lorentz invariant.  In particular, if the fields and
currents in the Poynting theorem are replaced by their Lorentz
transformed counterparts, the theorem will remain valid.

We define an energy-momentum density operator to be
\begin{eqnarray}
p^\mu = \frac{\epsilon_0}{2c}\,\gamma^\mu  ,
\end{eqnarray}
so that
\begin{eqnarray}
\overline{\iP}cp^0\iP &=& \frac{\epsilon_0}{2}
\left(|\bm E|^2 + |c\bm B|^2\right) = u  ,
\label{eq:pt0}
\\
\overline{\iP}\bm p\iP &=& \frac{\ri\epsilon_0}{2}
\left(\bm E\rs^\dagger\bm\tau\bm B\rs
-\bm B\rs^\dagger\bm\tau\bm E\rs\right)
\nonumber\\&=&\frac{1}{c^2\mu_0}\,{\rm Re}\, \bm E\bm\times\bm B^* = \bm g  .
\end{eqnarray}
Equations (\ref{eq:mmeq}) and (\ref{eq:adjeq}) give
\begin{eqnarray}
\partial_\mu\overline{\iP}\gamma^\mu\iP =
\overline{\iX}\iP+\overline{\iP}\iX  ,
\end{eqnarray}
or
\begin{eqnarray}
\frac{\partial u}{\partial t} + \bm\nabla\bm\cdot\bm S =
-{\rm Re}\,\bm J\bm\cdot\bm E - {\rm Re}\,\bm K\bm\cdot c\bm B  ,
\end{eqnarray}
where
\begin{eqnarray}
\bm S = c^2\bm g \, ,
\end{eqnarray}
which is just the relativistic extended Poynting theorem.

\section{Comparison of magnetic dipole moment models}
\label{sec:cmdmm}

Two classical models of the source of the magnetic dipole field
associated with a particle are the dual magnetic monopole model and the
current loop model.  Both of these give the same apparent field away
from the source, but the fields are fundamentally different, as are the
consequences of the difference.  In this section, we examine these
differences and the consequences.

\begin{figure}[t]
\includegraphics[angle=-90,trim=0 15 0
15,clip,width=.45\textwidth]{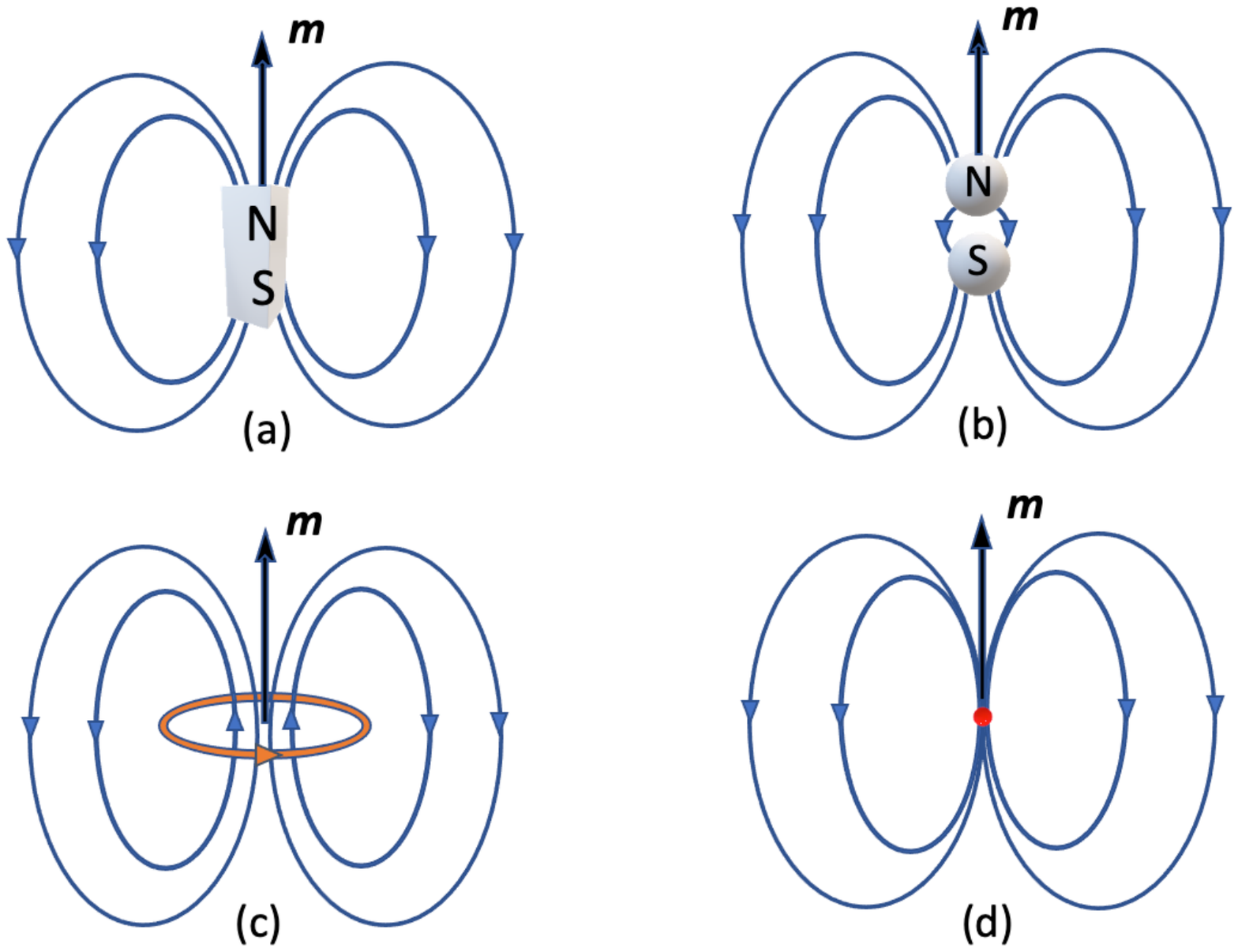}
\caption{Various depictions of a magnetic dipole field source.  (a): A
bar magnet. (b): A dual monopole model which produces a longitudinal
field.  (c): A current loop model which produces a transverse field.
(d) A model-independent point source which produces a field that can be
longitudinal, transverse, or both.
\label{fig:dipoles}}
\end{figure}

\subsection{Longitudinal vs. transverse fields}

One of the differences between the dual magnetic monopole model and the
current loop model for the source of the dipole field is that the former
produces a longitudinal field, while the latter produces a transverse
field.  These are global properties of the fields, although the origin
of the difference is confined to the location of the source.

\subsubsection{Dual monopole model}
\label{sec:dmf}

For a single magnetic monopole, we assume a longitudinal field of the
form
\begin{eqnarray}
\bm{B}_{\rm M}^{\,\rm L}(\bm{x}) &=& 
- \frac{\mu_0m}{4\pi} \,\bm\nabla\,\frac{1}{|\bm x|} \, ,
\end{eqnarray}
with the corresponding dipole field
\begin{eqnarray}
\bm{B}_{\bm m}^{\,\rm L}(\bm{x}) &=& 
\left[\bm{B}_{\rm M}^{\rm L}\left(\bm{x}+\frac{\bm{a}}{2}\right) 
- \bm{B}_{\rm M}^{\rm L}\left(\bm{x}-\frac{\bm{a}}{2}\right)\right]_{\bm
  a\rightarrow 0} , \qquad
\end{eqnarray}
where the magnetic moment is
\begin{eqnarray}
\bm m &=& m\,\bm{a} \, .
\end{eqnarray}
The expansion
\begin{eqnarray}
\frac{1}{\left|\bm{x}\pm\ft{\bm{a}}{2}\right|} &=& 
\ft{1}{|\bm x|} \pm \frac{\bm a}{2}\bm\cdot \bm\nabla\,\frac{1}{|\bm x|}
+ \dots
\end{eqnarray}
yields
\begin{eqnarray}
\bm B_{\bm m}^{\,\rm L}(\bm x) &=&
\frac{\mu_0}{4 \pi} \, \bm m\bm\cdot\bm\nabla\,\bm\nabla\
\frac{1}{|\bm x|} \, ,
\label{eq:newlong}
\end{eqnarray}
which agrees with Eq.~(\ref{eq:gsder}).  This is the longitudinal field
associated with the dual monopole model.

\subsubsection{Current loop model}

\citet{1998165} gives the result for this case.  We paraphrase that
derivation in the following.  For a steady state, i.e., $\partial \bm
E/\partial t = 0$, Eq.~(\ref{eq:me2}) is
\begin{eqnarray}
\bm\nabla\bm\times\bm B = \mu_0 \bm J  ,
\label{eq:dcbeqj}
\end{eqnarray}
where $\bm J$ is transverse, because $\bm\nabla\bm\cdot\bm J = 0$, and the
field $\bm B$ is the transverse component, because $\bm\nabla\bm\times\bm
B^{\rm L} = 0$.  From Eq.~(\ref{eq:trans}), we have
\begin{eqnarray}
\bm B^{\rm T}(\bm x) &=& \frac{1}{4\pi}\int\rd\bm x^\prime \,
\frac{1}{|\bm x - \bm x^\prime|}\,
\bm\nabla^\prime\bm\times\left[\bm\nabla^\prime\bm\times
\bm B(\bm x^\prime)\right] \qquad
\nonumber\\[10 pt]
&=& \frac{\mu_0}{4\pi} \,
\bm\nabla\bm\times\int\rd\bm x^\prime \,
\frac{1}{|\bm x - \bm x^\prime|}\,
\bm J^{\rm T}(\bm x^\prime) .
\label{eq:bt}
\end{eqnarray}
The dipole contribution follows from the expansion
\begin{eqnarray}
\frac{1}{|\bm x - \bm x^\prime|} = \frac{1}{|\bm x|}
+ \frac{\bm x\bm\cdot\bm x^\prime}{|\bm x|^3} + \dots \, ,
\end{eqnarray}
where the first term gives no contribution, because
\begin{eqnarray}
\int\rd\bm x^\prime \,
 J^{\rm T\,i}(\bm x^\prime) &=&
 \int\rd\bm x^\prime \,\big[\bm\nabla^\prime\bm\cdot
\bm J^{\rm T}(\bm x^\prime)x^{\prime\, i}
\nonumber\\[10 pt]&&
-x^{\prime\, i}\bm\nabla^\prime\bm\cdot
\bm J^{\rm T}(\bm x^\prime)\big] =0  .
\end{eqnarray}
The second term gives the dipole contribution
\begin{eqnarray}
\bm B^{\rm T}(\bm x) &=&
\frac{\mu_0}{4\pi} \,
\bm\nabla\bm\times\int\rd\bm x^\prime \,
\frac{\bm x\bm\cdot\bm x^\prime}{|\bm x|^3}\,
\bm J^{\rm T}(\bm x^\prime)  .
\end{eqnarray}
We have
\begin{eqnarray}
\bm x \bm\times\left[\bm x^\prime\bm\times
\bm J^{\rm T}(\bm x^\prime)\right]
&=& \bm x\bm\cdot\bm J^{\rm T}(\bm x^\prime)\,\bm x^\prime 
\nonumber\\[10 pt]&&
- \bm x\bm\cdot \bm x^\prime\,\bm J^{\rm T}(\bm x^\prime)
\end{eqnarray}
or
\begin{eqnarray}
&&\bm x\bm\cdot \bm x^\prime\,\bm J^{\rm T}(\bm x^\prime) =
-\frac{1}{2}\,\bm x \bm\times\left[\bm x^\prime\bm\times
\bm J^{\rm T}(\bm x^\prime)\right]
\nonumber\\[10 pt]&&\qquad
+\frac{1}{2}\left[\bm x\bm\cdot \bm x^\prime\,\bm J^{\rm T}(\bm x^\prime)
+\bm x\bm\cdot\bm J^{\rm T}(\bm x^\prime)\,\bm x^\prime\right] .
\label{eq:Dprods}
\end{eqnarray}
The first term on the right-hand-side of Eq.~(\ref{eq:Dprods}) gives the
dipole field
\begin{eqnarray}
\bm B_{\bm m}^{\rm T}(\bm x) &=& 
-\frac{\mu_0}{4\pi} \, \bm\nabla\bm\times
\left(\frac{\bm x}{|\bm x|^3}\bm\times\bm m\right)
\nonumber\\[10 pt]&=& 
 \frac{\mu_0}{4\pi} \, \bm\nabla\bm\times
\left(\bm\nabla\bm\times\frac{\bm m}{|\bm x|}\right)  ,
\label{eq:res}
\end{eqnarray}
where the magnetic dipole moment is
\begin{eqnarray}
\bm m = \frac{1}{2}\int\rd\bm x^\prime\,\bm x^\prime\bm\times
\bm J^{\rm T}(\bm x^\prime) \, .
\end{eqnarray}
The second term in Eq.~(\ref{eq:Dprods}) makes no contribution, because
\begin{eqnarray}
&&\int\rd\bm x^\prime\left[x^{\prime\,i}\, J^{\rm T\,j}(\bm x^\prime)
+x^{\prime\,j}\, J^{\rm T\,i}(\bm x^\prime)\right]
\nonumber\\[10 pt]
&&\qquad= \int\rd\bm x^\prime\,\nabla^{\prime\,k}\, x^{\prime\,j}\,x^{\prime\,i}
J^{\rm T\,k}(\bm x^\prime)
\nonumber\\[10 pt]&&\qquad\quad
- \int\rd\bm x^\prime\,x^{\prime\,j}\,x^{\prime\,i} \,
\bm\nabla\bm\cdot\bm J^{\rm T}(\bm x^\prime) = 0 \, .
\end{eqnarray}
We can also write Eq.~(\ref{eq:res}) as
\begin{eqnarray}
\bm B_{\bm m}^{\rm T}(\bm x) &=&
\frac{\mu_0}{4\pi}\left(\bm m\bm\cdot\bm\nabla\,\bm\nabla -\bm
m\,\bm\nabla^2\right)\frac{1}{|\bm x|}
\nonumber\\[10 pt] &=&
\bm B_{\bm m}^{\rm L}(\bm x) + \mu_0\,\bm m \, \delta(\bm x) \, .
\label{eq:clm}
\end{eqnarray}
In view of the delta function contained in $\bm B_{\bm m}^{\rm L}(\bm
x)$ according to Eq.~(\ref{eq:bav}), the total delta function
contribution in Eq.~(\ref{eq:clm}), $\fr{2}{3}\,\mu_0\,\bm m\,\delta(\bm
x)$, is in agreement with the corresponding term in Eq.~(5.64) of
\citet{1998165}, based on the current loop model.

\subsection{Comparison of the models}

These two models correspond to two different formulations of classical
electromagnetism and how they deal with particles with a magnetic dipole
moment, such as the electron.

On the one hand, there is the dual monopole model for magnetic dipole
moments, where the associated field is longitudinal.  In this case, the
theoretical framework can be the extended Poynting theorem and the
associated extended Maxwell equations.  Energetics of magnetic dipole
interactions with magnetic field gradients are taken into account by the
extended Poynting theorem, and $\bm \nabla\bm\cdot\bm B \ne 0$ in general.
We note in passing that the dual magnetic monopole model is not
necessary to arrive at this formulation, as shown in Sec.~\ref{sec:mfpi}
which makes no such assumption.

On the other hand, there is the current loop model for the magnetic
dipole moment, where the associated field is transverse.  The
theoretical framework for this model is the classical electrodynamics
associated with conventional quantum electrodynamics.  In it
$\bm\nabla\bm\cdot\bm B = 0$, and a vector potential describes magnetic
interactions between particles and fields.

\section{The Poynting theorem and classical electrodynamics}

The electromagnetic energy considerations described by the Poynting
theorem may be applied to calculate the interactions of particles with
fields, and thereby interactions between particles.

\subsection{Electric interaction between two charged particles}
\label{ssec:efi}

The electric interaction between two charged particles may be obtained
from the Poynting theorem.  The interaction energy, and thus the force
between them, is obtained by calculating the total energy of the
combined electric fields of the two particles~\cite{1998165}.  The
electric field of each particle is
\begin{eqnarray}
\bm E^{\rm L}(\bm x,\bm x_i) = -\frac{q_i}{4\pi\epsilon_0}\,\bm\nabla\,
\frac{1}{|\bm x - \bm x_i|}  \, ;\qquad i = 1,2 \, ,
\end{eqnarray}
which is the field at the point $\bm x$ due to the particle at the point
$\bm x_i$.   These fields are longitudinal because $\bm\nabla\bm\times\bm
E^{\rm L} = 0$.  The energy density is
\begin{eqnarray}
u_{\rm E}(\bm x) &=& \frac{\epsilon_0}{2}\,
|\bm E^{\rm L}(\bm x,\bm x_1)+\bm E^{\rm L}(\bm x,\bm x_2)|^2 
\nonumber\\[10 pt]&=&
\frac{\epsilon_0}{2} \Big[|\bm E^{\rm L}(\bm x,\bm x_1)|^2
\\[10 pt]&& +
2\bm E^{\rm L}(\bm x,\bm x_1)\bm\cdot\bm
E^{\rm L}(\bm x,\bm x_2) + |\bm E^{\rm L}(\bm x,\bm x_2)|^2\Big]
\nonumber
. \qquad
\label{eq:eie}
\end{eqnarray}
The first and third terms on the right-hand-side are the individual
particle electric self-energy densities, and the second term is the
interaction energy density $u_{\rm E}^{\rm I}$.  Thus, the total
interaction energy $U_{\rm E}^{\rm I}$ is
\begin{eqnarray}
U_{E}^{\rm I} &=& 
\frac{q_1q_2}{(4\pi)^2\epsilon_0}
\int\rd\bm x \left[\bm\nabla\,\frac{1}{|\bm x - \bm x_1|}\right]
\bm\cdot\left[\bm\nabla\,\frac{1}{|\bm x - \bm x_2|}\right]
\nonumber\\[10 pt] &=&
-\frac{q_1q_2}{(4\pi)^2\epsilon_0}
\int\rd\bm x \, \frac{1}{|\bm x - \bm x_1|}
\bm\nabla^2\,\frac{1}{|\bm x - \bm x_2|}
\nonumber\\[10 pt] &=&
\frac{q_1q_2}{4\pi\epsilon_0}
\int\rd\bm x \, \frac{1}{|\bm x - \bm x_1|}
\,\delta(\bm x - \bm x_2)
\nonumber\\[10 pt] &=&
\frac{q_1q_2}{4\pi\epsilon_0} \,
\frac{1}{|\bm x_2- \bm x_1|} \, .
\label{eq:eeint}
\end{eqnarray}

For two electrons, this is
\begin{eqnarray}
U_{E}^{\rm I} &=&  \frac{\alpha \hbar c}{|\bm x_2- \bm x_1|}
= \frac{\alpha \lbar_{\rm e}}{|\bm x_2- \bm x_1|} \, m_{\rm e}c^2  ,
\end{eqnarray}
where $\alpha = e^2/4\pi\epsilon_0\hbar c$, $\hbar$ is the Planck
constant, and $\lbar_{\rm e}=\hbar/m_{\rm e}c$ is the reduced Compton
wavelength of the electron.  The energy is just the conventional Coulomb
interaction energy.

\subsection{Interaction between particles with magnetic moments}
\label{ssec:mmi}

The interaction between magnetic dipoles may be considered for both the
longitudinal field model and the transverse field model.

\subsubsection{Longitudinal magnetic dipole interaction}
\label{sssec:lmdi}

For the longitudinal magnetic interaction of two particles, the
extension of the Pointing theorem considered earlier is relevant.  It
takes into account longitudinal magnetic fields, in analogy with the
longitudinal electric fields.

We have
\begin{eqnarray}
\bm B_{\bm m_i}^{\rm L}(\bm x,\bm x_i) &=&
\frac{\mu_0}{4 \pi}  \, \bm m_{i}\bm\cdot\bm\nabla\,\bm\nabla\,
\frac{1}{|\bm x-\bm x_i|} \quad i = 1,2 , \qquad
\label{eq:ball}
\end{eqnarray}
for the longitudinal dipole field at the point $\bm x$ due to the
particle at the point $\bm x_i$.  Such longitudinal magnetic fields are
excluded by the condition $\bm\nabla\bm\cdot\bm B = 0$ associated with
the conventional Maxwell equations.  The energy density is
\begin{eqnarray}
&&u_{B^{\rm L}}(\bm x) = \frac{\epsilon_0}{2}\,
|c\bm B_{\bm m_1}^{\rm L}(\bm x,\bm x_1)
+c\bm B_{\bm m_2}^{\rm L}(\bm x,\bm x_2)|^2
\nonumber\\[10 pt]&&\quad=
 \frac{\epsilon_0}{2} \Big[|c\bm B_{\bm m_1}^{\rm L}(\bm x,\bm x_1)|^2 
 \\[10 pt]&&\quad
 + 2c\bm B_{\bm m_1}^{\rm L}(\bm x,\bm x_1)\bm\cdot 
 c\bm B_{\bm m_2}^{\rm L}(\bm x,\bm x_2)
+ |c\bm B_{\bm m_2}^{\rm L}(\bm x,\bm x_2)|^2\Big] ,
\nonumber
\label{eq:mie}
\end{eqnarray}
where the interaction energy density is
\begin{eqnarray}
u_{B^{\rm L}}^{\rm I}(\bm x) &=& \epsilon_0c^2
\bm B_{\bm m_1}^{\rm L}(\bm x,\bm x_1)\bm\cdot
 \bm B_{\bm m_2}^{\rm L}(\bm x,\bm x_2) \, .
\end{eqnarray}
Thus, the total interaction energy is
\begin{eqnarray}
U_{B^{\rm L}}^{\rm I} &=& 
\frac{\mu_0}{(4\pi)^2} 
\int\rd\bm x \left[
\bm m_{1}\bm\cdot\bm\nabla\,
\bm\nabla\,\frac{1}{|\bm x - \bm x_1|}\right]
\nonumber\\[10 pt]&&\qquad\bm\cdot
\left[\bm m_{2}\bm\cdot\bm\nabla\,\bm\nabla
\frac{1}{|\bm x - \bm x_2|}\right]
\nonumber\\&=&
\frac{\mu_0}{(4\pi)^2} \,
\bm m_{1}\bm\cdot\bm\nabla_1\,
\bm m_{2}\bm\cdot\bm\nabla_2\,
\int\rd\bm x \left[\bm\nabla\,\frac{1}{|\bm x - \bm x_1|}\right]
\nonumber\\[10 pt]&&\qquad\bm\cdot
\left[\bm\nabla\,\frac{1}{|\bm x - \bm x_2|}\right]
\nonumber\\&=& 
\frac{\mu_0}{4\pi}
 \,\bm m_{1}\bm\cdot\bm\nabla_1\,
 \bm m_{2}\bm\cdot\bm\nabla_2\,
\frac{1}{|\bm x_2- \bm x_1|} \, .
\label{eq:bsqint}
\end{eqnarray}
For $|\bm x_2- \bm x_1|>0$, differentiation yields
\begin{eqnarray}
U_{B^{\rm L}}^{\rm I} &=& 
\frac{\mu_0}{4\pi} \,
 \frac{\bm m_1\bm\cdot
 \bm m_2 
 -3\,\bm m_1\bm\cdot\bm{\hat{x}}_{21} \,
\bm m_2\bm\cdot
 \bm{\hat{x}}_{21}}{|\bm x_{21}|^3} , \qquad
\label{eq:magint}
\end{eqnarray}
where $\bm x_{21} = \bm x_2 - \bm x_1$.  

This expression gives the proper form of the interaction of two
classical dipoles.  In particular, the sign is correct as shown by the
following considerations.  If two magnetic moments, or magnets, are side
by side pointing the same direction perpendicular to their separation,
then $\bm m_1\bm\cdot\bm m_2 = m_1m_2 > 0$, $\bm m_i\bm\cdot\bm{\hat
x}_{21}=0$, and
\begin{eqnarray}
U_{B^{\rm L}}^{\rm I} &\rightarrow&
\frac{\mu_0\,m_1m_2}{4\pi\,x_{21}^3}\, , 
\end{eqnarray}
which means that the energy in the field increases if the magnets are
moved closer together, so work is done against a repulsive force,
consistent with experience with magnets.  Similarly, if the moments are
pointing in the same direction and are collinear, then $\bm
m_1\bm\cdot\bm m_2 = m_1m_2$, $\bm m_1\bm\cdot\bm{\hat{x}}_{21} \, \bm
m_2\bm\cdot \bm{\hat{x}}_{21} = m_1m_2$, with $m_1m_2 > 0$, and
\begin{eqnarray}
U_{B^{\rm L}}^{\rm I} &\rightarrow&
-\frac{\mu_0\,m_1m_2}{2\pi\,x_{21}^3}\, , 
\end{eqnarray}
corresponding to an attractive force, as expected.

There is also a contact delta function interaction between the dipole
moments.  The integral over $\bm x_2$ of Eq.~(\ref{eq:bsqint}) for a
sphere of radius $R$, centered at $\bm x_1$ is
\begin{eqnarray}
\frac{\mu_0}{4\pi}\int_R\rd\bm x_2\, U_{B^{\rm L}}^{\rm I}
&=& 
-\frac{\mu_0}{12\pi}\,\bm m_1\bm\cdot\bm m_2
\int_R\rd\bm x_2\,
\bm\nabla_2^2 \,
\frac{1}{|\bm x_2 - \bm x_1|}
\nonumber\\[10 pt]&=&\frac{\mu_0}{3} \,
 \bm m_1\bm\cdot \bm m_2  ,
\end{eqnarray}
so the total is
\begin{eqnarray}
U_{B^{\rm L}}^{\rm I} &=& \frac{\mu_0}{4\pi}
\bigg[
 \frac{\bm m_1\bm\cdot
 \bm m_2 
 -3\,\bm m_1\bm\cdot\bm{\hat{x}}_{21} \,
\bm m_2\bm\cdot
 \bm{\hat{x}}_{21}}{x_{21}^3} 
 \nonumber\\[10 pt]&&
 +\frac{4\pi}{3}\,\bm m_1\bm\cdot
 \bm m_2\,\delta(\bm x_2-\bm x_1)
 \bigg] .
\label{eq:dipint}
\end{eqnarray}

We note that the coefficient of the delta function in
Eq.~(\ref{eq:dipint}) differs by a factor of $-2$ from the corresponding
expression in Eq.~(5.73) given by \citet{1998165}.  The reason is that
the expression in that equation is based on the current loop model for
the magnetic moment source, while Eq.~(\ref{eq:dipint}) is not.  This
equation is not in conflict with the hyperfine interaction in
QED, because the classical dipole contact interaction
is not the source of the delta function in the nonrelativistic hyperfine
Hamiltonian.  Instead, the delta function is a surface term that results
from the nonrelativistic reduction of the Dirac equation, as shown in
Sec.~\ref{ssec:hfs}.

\subsubsection{Transverse magnetic dipole interaction}
\label{sssec:tmdi}

Here, we carry out the same calculation as in the previous section with
transverse magnetic fields rather than with longitudinal
magnetic fields.  We have from Eq.~(\ref{eq:clm})
\begin{eqnarray}
\bm B_{\bm m_i}^{\rm T}(\bm x - \bm x_i) &=&
\bm B_{\bm m_i}^{\rm L}(\bm x - \bm x_i) + 
\mu_0\,\bm m_i \, \delta(\bm x - \bm x_i) , \qquad
\end{eqnarray}
and the term corresponding to the interaction energy density is
\begin{eqnarray}
u_{B^{\rm T}}^{\rm I}(\bm x) &=& 
\epsilon_0c^2 
\bm B_{\bm m_1}^{\rm T}(\bm x,\bm x_1)
 \bm\cdot\bm B_{\bm m_2}^{\rm T}(\bm x,\bm x_2)
\nonumber\\[10 pt]&=& 
\epsilon_0c^2 \Big\{\left[
\bm B_{\bm m_1}^{\rm L}(\bm x,\bm x_1)+
\mu_0\,\bm m_1 \, \delta(\bm x - \bm x_1)\right]
\nonumber\\[10 pt]&&\bm\cdot\left[
 \bm B_{\bm m_2}^{\rm L}(\bm x,\bm x_2)+
 \mu_0\,\bm m_2 \, \delta(\bm x - \bm x_2)
 \right]\Big\}
\nonumber\\[10 pt]&=& 
u_{\rm B^{\rm L}}^{\rm I}(\bm x) +
\bm B_{\bm m_1}^{\rm L}(\bm x,\bm x_1) \bm\cdot
\bm m_2 \, \delta(\bm x - \bm x_2) 
\nonumber\\[10 pt]&&+
 \bm B_{\bm m_2}^{\rm L}(\bm x,\bm x_2) \bm\cdot
 \bm m_1 \, \delta(\bm x - \bm x_1)
\nonumber\\[10 pt] &&+ \mu_0\,\bm m_1\bm\cdot\bm m_2
\, \delta(\bm x - \bm x_1) \, \delta(\bm x - \bm x_2)
 \, . \qquad
\end{eqnarray}
Integration over $\bm x$ yields [see Eqs.~(\ref{eq:ball}) and
(\ref{eq:bsqint})]
\begin{eqnarray}
U_{B^{\rm T}}^{\rm I} &=& U_{B^{\rm L}}^{\rm I}
+\bm B_{\bm m_1}^{\rm L}(\bm x_2,\bm x_1) \bm\cdot \bm m_2
+\bm B_{\bm m_2}^{\rm L}(\bm x_1,\bm x_2) \bm\cdot \bm m_1
\nonumber\\[10 pt]&&
+ \mu_0\,\bm m_1\bm\cdot\bm m_2
\, \delta(\bm x_2 - \bm x_1)
\nonumber\\[10 pt]&=& -U_{B^{\rm L}}^{\rm I}
 + \mu_0\,\bm m_1\bm\cdot\bm m_2
\, \delta(\bm x_2 - \bm x_1)
\, . \qquad
\end{eqnarray}
We thus have
\begin{eqnarray}
U_{B^{\rm T}}^{\rm I}
&=& -\frac{\mu_0}{4\pi}
\Big[
 \frac{\bm m_{1}\bm\cdot \bm m_{2} 
 -3\,\bm m_{1}\bm\cdot\bm{\hat{x}}_{21} \,
\bm m_{2}\bm\cdot
 \bm{\hat{x}}_{21}}{x_{21}^3} 
 \nonumber\\[10 pt]&& \qquad
 -\frac{8\pi}{3}\,\bm m_{1}\bm\cdot
 \bm m_{2}\,\delta(\bm x_2-\bm x_1)
 \Big] .
\end{eqnarray}
This is essentially in agreement with the hyperfine Hamiltonian given by
Eq.~(5.73) in \citet{1998165}, but it has the opposite sign.  Moreover,
the sign of the interaction energy is the opposite of what is observed
with classical magnets, which, in contrast, is given correctly by the
magnetic monopole model in Sec.~\ref{sssec:lmdi}.  This difference in
sign is explained in Sec.~\ref{sec:iiq}.

\subsection{Electric self energy}
\label{ssec:cese}

Considerations of the electric field of an electron as the source of its
mass date back to Thompson in 1881~\cite{thomson}. Subsequent work
examined various dynamical effects that would contribute to such a mass.
Most recently, particle masses are attributed to the Higgs mechanism.
Here we consider the mass equivalent of the electric field energy as
given by the Poynting theorem.  For a point charge, this mass is
infinite due to the singularity at the location of the electron, so we
consider the field energy with a lower cutoff $x_{\rm c}$, as a function
of the cutoff.

The self energy density, as appears in Eq.~(\ref{eq:eie}) for example,
is given by 
\begin{eqnarray} 
u_{\rm ESE}(\bm x) &=& \frac{\epsilon_0}{2}
\, |\bm E(\bm x)|^2  , 
\end{eqnarray} 
where $\bm E(\bm x)$ denotes $\bm E(\bm x,0)$ and
\begin{eqnarray}
\bm E(\bm x) = \frac{e}{4\pi\epsilon_0}\,\bm\nabla\,\frac{1}{|\bm x|}
= -\frac{e}{4\pi\epsilon_0}\,\frac{\bm x}{|\bm x|^3} \, .
\end{eqnarray}
Thus, the electric field energy is
\begin{eqnarray}
U_{\rm ESE}(x_{\rm c}) &=& \frac{\epsilon_0}{2}
\left(\frac{e}{4\pi\epsilon_0}\right)^2
\int_{x > x_{\rm c}}\rd \bm x \, \frac{1}{|\bm x|^4} 
\nonumber\\[10 pt]&=&
\frac{\alpha\lbar_{\rm C}}{2\,x_{\rm c}}\,m_{\rm e}c^2 \, .
\end{eqnarray}
For $x_{\rm c} = r_0 = \alpha\lbar_{\rm C}$, the classical electron
radius, the field-energy mass equivalent is $m_{\rm e}/2$, which is the
mass of the electron up to a factor of 1/2.  However, this radius is too
small compared to the radius of vacuum quantum fluctuations for the
electric field energy to be a plausible source of the electron mass.

At the substantially larger radius $x_{\rm c} = n^2a_0 = n^2\lbar_{\rm
C}/\alpha$, the Bohr radius of an electron in an atomic bound state with
quantum number $n$, the electric field energy is
\begin{eqnarray}
U_{\rm ESE}(n^2a_0) &=& 
\frac{\alpha^2}{2\,n^2}\,m_{\rm e}c^2  .
\end{eqnarray}
This is exactly the nonrelativistic binding energy of an electron in a
static Coulomb field.  A simple classical model the bound electron could
be taken to be a spherical shell of charge with radius $n^2a_0$.  This
would shield the proton's electric field for $x>n^2a_0$.  Removing the
electron would require providing the energy to create the electric
field for $x>n^2a_0$, which is the same as the binding energy of the
electron.

\subsection{Magnetic self energy}
\label{ssec:mse}

Next, we consider the energy in the field of the magnetic dipole moment
of an electron.  As with the electric field of the electron, the
magnetic field energy is infinite for a point source, so only the
contribution for $x > x_{\rm c}$ is considered.  The magnetic energy
density is
\begin{eqnarray} 
u_{\rm BSE}(\bm x) &=& \frac{\epsilon_0}{2}
\, |c\bm B_{\bm m}(\bm x)|^2
\end{eqnarray} 
and
\begin{eqnarray}
\bm B_{\bm m}(\bm x) &=&
\frac{\mu_0}{4 \pi} \, \bm m\bm\cdot\bm\nabla\,\bm\nabla\
\frac{1}{|\bm x|} 
\nonumber\\[10 pt]&=& 
\frac{\mu_0}{ 4 \pi}  \,
\frac{3 \bm{\hat x}(\bm{\hat x}\bm\cdot \bm m) - \bm m }
{ |\bm x|^3}  \, . 
\end{eqnarray}
Since the delta function at the origin is excluded, we have
$\bm\nabla\bm\times\bm B_{\bm m}(\bm x) = 0$, so the magnetic field is
essentially longitudinal.  The cutoff energy is
\begin{eqnarray}
U_{\rm BSE}(x_{\rm c}) &=& \frac{\epsilon_0}{2}
\left(\frac{\mu_0c}{4\pi}\right)^2
\int_{x > x_{\rm c}}\rd \bm x \, 
\frac{3 (\bm{\hat x}\bm\cdot \bm m)^2 + \bm m^2 }{|\bm x|^6} 
\nonumber\\[10 pt]&=&
\frac{\mu_0\,\bm m^2}{12 \pi x_{\rm c}^3}  .
\end{eqnarray}

For an electron, $|\bm m| = (g_{\rm e}/2)\mu_{\rm B}$, where $g_{\rm e}
\approx 2$ is the electron $g$-factor and $\mu_{\rm B} = e\hbar/2m_{\rm
e}$ is the Bohr magneton.  We thus have (assuming $g_{\rm e} = 2$)
\begin{eqnarray}
U_{\rm BSE}(x_{\rm c}) &=& \frac{\alpha}{ 12}
\left(\frac{\lbar_{\rm C}}{ x_{\rm c}}\right)^3 \, m_{\rm e}c^2  ,
\end{eqnarray}
which gives
\begin{eqnarray}
U_{\rm BSE}(a_0) &=& \frac{\alpha^4}{ 12} \, m_{\rm e}c^2  ,
\end{eqnarray}
which is the order of magnitude of magnetic effects on the energy of an
electron bound in a hydrogen atom, and
\begin{eqnarray}
U_{\rm BSE}(\lbar_{\rm C}/12) &\approx&  m_{\rm e}c^2  .
\end{eqnarray}
Evidently, for the electron, seemingly small magnetic field effects may
be larger than the electric field effects due to the stronger field near
the nucleus.  The cutoff of $\lambda_{\rm C}/12$ is plausible, because
nonperturbative vacuum fluctuations can be expected to mitigate the
divergence of the field closer to the location of the nucleus.

It is of interest to consider the corresponding effect for a muon.  In
this case, the mass is about 207 times larger and the magnetic moment is
about 207 times smaller, for a change in the relative magnetic field
energy by a factor of $(1/207)^{3}$.  However, if the cutoff is taken to
be proportional to the Compton wavelength of the muon, which is about
$\lbar_{\rm C}/207$, the $x_{\rm c}^{-3}$ behavior of the energy
compensates for these effects, with the result that
\begin{eqnarray}
U_{\rm BSE}(\lbar_{{{\rmssmu}}}/12) &\approx&  m_{{{\rmssmu}}}c^2  .
\end{eqnarray}

As mentioned above, the mass of elementary particles is currently
considered to be due to the Higgs mechanism.  On the other hand, it is
hard to ignore the energy in the electric and magnetic fields given by
the Poynting theorem.  These are seemingly non-local effects and the
2022 Nobel prize on the violation of Bell inequalities suggests that
quantum mechanics allows such effects.

\section{The Poynting theorem and the Dirac equation}
\label{sec:ptde}

Interactions of electrons with external electromagnetic fields are
generally described by including scalar and/or vector potentials in the
Dirac equation.  This is implemented by the minimal coupling
substitution
\begin{eqnarray}
p^\mu \rightarrow p^\mu + e\,A^\mu  , 
\label{eq:mcs}
\end{eqnarray}
where $-e$ is the charge of the electron and $A^\mu$ is the four-vector
potential of the external fields, with electrostatic component
$A^0={\it\Phi}/c$ and three-vector potential $\bm A$.  However, the
interactions may also be derived by employing only electric and magnetic
fields, as described in this section.

The Dirac equation for a free electron is
\begin{eqnarray}
\left[c\,\bm\alpha\bm\cdot\bm p + \beta \, m_{\rm e} c^2
-E_0\right]\phi(\bm x) = 0  ,
\end{eqnarray}
where $\phi$ is the normalized four-component wave function, $E_0$ is
the energy of the state, $\bm p = -\ri\,\hbar\bm \nabla$, $\bm\alpha$
and $\beta$ are $4\bm\times4$ Dirac matrices
\begin{eqnarray}
\alpha_0 &=& \left(\begin{array}{c@{\quad}c} \bm I & \bm 0 \\
\bm 0 & \bm I \end{array}\right) ; \
\bm\alpha = \left(\begin{array}{c@{\quad}c} \bm 0 & \bm\sigma \\
\bm\sigma & \bm 0 \end{array}\right) ; \
\beta = \left(\begin{array}{c@{\quad}c} \bm I & \bm 0 \\
\bm 0 & -\bm I \end{array}\right) ; \nonumber\\[10 pt]
\bm I &=& \left(\begin{array}{c@{\quad}c} 1 & 0 \\
0 & 1 \end{array}\right) ; \
\bm 0 = \left(\begin{array}{c@{\quad}c} 0 & 0 \\
0 & 0 \end{array}\right)  ,
\label{eq:diracmat}
\end{eqnarray}
and where $\bm\sigma$ denotes a vector of Pauli matrices, with
components given by
\begin{eqnarray}
\sigma^1 = \left(\begin{array}{c@{\quad}c} 0 & 1 \\ 1 & 0 
\end{array}\right) ; \
\sigma^2 = \left(\begin{array}{c@{\quad}c} 0 & -\ri \\ \ri & 0 
\end{array}\right) ; \
\sigma^3 = \left(\begin{array}{cc} 1 & 0 \\ 0 & -1 
\end{array}\right) .
\nonumber\\
\end{eqnarray}
We thus have
\begin{eqnarray}
E_0 &=& \int\rd\bm x \,\phi^\dagger(\bm x)
\left[c\,\bm\alpha\bm\cdot\bm p + \beta \, m_{\rm e} c^2
\right]\phi(\bm x)  .
\label{eq:freee}
\end{eqnarray}

\subsection{External electric field}
\label{ssec:eef}

The charge density $\rho_\phi$ associated with $\phi$ is
\begin{eqnarray}
\rho_\phi(\bm x) = -e \,\phi^\dagger(\bm x)\,\phi(\bm x)  ,
\label{eq:j0}
\end{eqnarray}
which corresponds to a longitudinal electric field from [see
Eq.~(\ref{eq:xme1})]
\begin{eqnarray}
\bm\nabla\bm\cdot\bm E_\phi(\bm x) &=& 
-\frac{e}{\epsilon_0} \,\phi^\dagger(\bm x)\,\phi(\bm x)
\end{eqnarray}
or [see Eq.~(\ref{eq:long})]
\begin{eqnarray}
\bm E_\phi^{\rm L}(\bm x)
&=& \frac{e}{4\pi\epsilon_0}\,\bm\nabla\int\rd\bm x^\prime \,
\frac{1}{|\bm x - \bm x^\prime|}
\,\phi^\dagger(\bm x^\prime)\,\phi(\bm x^\prime) . \quad
\end{eqnarray}
Similarly, a charge density $\rho_{\rm ex}$ is associated with an
external field $\bm E_{\rm ex}$ by the relation
\begin{eqnarray}
\bm\nabla\bm\cdot\bm E_{\rm ex}(\bm x) 
= \frac{1}{\epsilon_0}\,\rho_{\rm ex}(\bm x) 
\end{eqnarray}
or
\begin{eqnarray}
\bm E_{\rm ex}^{\rm L}(\bm x)
&=& -\frac{1}{4\pi\epsilon_0}\,\bm\nabla\int\rd\bm x^\prime \,
\frac{1}{|\bm x - \bm x^\prime|}
\,\rho_{\rm ex}(\bm x^\prime)  .
\end{eqnarray}
There is no contribution to the total energy in Eq.~(\ref{eq:ebl}) from
an external transverse field because of the orthogonality given by
Eq.~(\ref{eq:orth}).

The energy density of the combined fields is
\begin{eqnarray}
u_E(\bm x) &=& \frac{\epsilon_0}{2}\,\left|\bm E_\phi^{\rm L}(\bm x)
+\bm E_{\rm ex}^{\rm L}(\bm x)\right|^2   ,
\end{eqnarray}
the interaction energy density is
\begin{eqnarray}
u_E^{\,\rm I}(\bm x) &=& \epsilon_0 
\bm E_\phi^{\rm L}(\bm x)\bm\cdot\bm E_{\rm ex}^{\rm L}(\bm x)  ,
\end{eqnarray}
and so the total interaction energy is
\begin{eqnarray}
U_E^{\,\rm I} &=& -\frac{e}{(4\pi)^2\epsilon_0}\int\rd\bm x\,
\left[\bm\nabla\int\rd\bm x^\prime \,
\frac{1}{|\bm x - \bm x^\prime|}
\,\phi^\dagger(\bm x^\prime)\,\phi(\bm x^\prime)\right]
\nonumber\\[10 pt]&& \qquad
\bm\cdot\left[\bm\nabla\int\rd\bm x^{\prime\prime} \,
\frac{1}{|\bm x - \bm x^{\prime\prime}|}
\,\rho_{\rm ex}(\bm x^{\prime\prime})\right]
\nonumber\\[10 pt]&=&
-\frac{e}{4\pi\epsilon_0}\int\rd\bm x\,
\phi^\dagger(\bm x)\,\phi(\bm x)
\int\rd\bm x^{\prime}\,\frac{1}{|\bm x - \bm x^\prime|}
\,\rho_{\rm ex}(\bm x^{\prime})
\nonumber\\[10 pt]&=&
-e\int\rd\bm x\,\phi^\dagger(\bm x){\it\Phi}_{\rm ex}(\bm x)\phi(\bm x) \, ,
\label{eq:ebl}
\end{eqnarray}
where
\begin{eqnarray}
{\it\Phi}_{\rm ex}(\bm x) &=& 
\frac{1}{4\pi\epsilon_0}\int\rd\bm x^\prime\,
\frac{1}{|\bm x - \bm x^\prime|}
\,\rho_{\rm ex}(\bm x^{\prime}) \, .
\end{eqnarray}
Substitution of the interaction energy given by
Eq.~(\ref{eq:ebl}) into
Eq.~(\ref{eq:freee}) gives
\begin{eqnarray}
E_0 + U_E^{\,\rm I}
 &=& \int\rd\bm x \,\phi^\dagger(\bm x)
\nonumber\\[10 pt]&&\bm\times
\left[c\,\bm\alpha\bm\cdot\bm p + \beta \, m_{\rm e} c^2
-e\,{\it\Phi}_{\rm ex}(\bm x)\right]\phi(\bm x)  .\qquad
\label{eq:dpot}
\end{eqnarray}
Because
\begin{eqnarray}
E_0 = \int\rd\bm x\,\phi^\dagger(\bm x)\,c\,p^0\phi(\bm x)  ,
\end{eqnarray}
the external potential term in Eq.~(\ref{eq:dpot}) could also be
inserted by the substitution
\begin{eqnarray}
p^0\rightarrow 
p^0 + e\,A^0(\bm x) =
p^0 +\frac{e}{c}\,{\it \Phi}_{\rm ex}(\bm x) \, ,
\end{eqnarray}
as in Eq.~(\ref{eq:mcs}).

\subsection{External magnetic field}
\label{ssec:emf}

\subsubsection{Transverse magnetic field}
\label{sssec:tmf}

The charge current density corresponding to the state $\phi$ is
\begin{eqnarray}
\bm j_\phi(\bm x) &=&
 -e c\,\phi^\dagger(\bm x) \,\bm\alpha\, \phi(\bm x) \, ,
\label{eq:cc}
\end{eqnarray}
which is transverse, because
\begin{eqnarray}
\bm\nabla\bm\cdot 
\phi^\dagger(\bm x) \,
\bm \alpha \,
 \phi(\bm x) &=& 
\phi^\dagger(\bm x)
\left[ \bm \alpha\bm\cdot \overleftarrow{\bm\nabla}
+\bm\alpha\bm\cdot \bm \nabla
\right] \phi(\bm x) 
\nonumber\\[10 pt]
&=& 0  ,
\label{eq:treq}
\end{eqnarray}
which follows from the difference between the expression
\begin{eqnarray}
\phi^\dagger(\bm x)
\left[ -\ri\hbar c\,\bm \alpha\bm\cdot \bm \nabla 
 +\beta\,m_{\rm e}c^2 -e{\it\Phi}_{\rm ex}(\bm x) \right]\phi(\bm x)
\end{eqnarray}
and its adjoint
\begin{eqnarray}
\phi^\dagger(\bm x)
\left[ \ri\hbar c\,\bm \alpha\bm\cdot \overleftarrow{\bm\nabla} 
 + \beta\, m_{\rm e}c^2  -e{\it\Phi}_{\rm ex}(\bm x) \right]
\phi(\bm x)
\end{eqnarray}
which vanishes because they are equal.  The current is the source of a
transverse magnetic field that satisfies [see Eq.~(\ref{eq:me2})]
\begin{eqnarray}
\bm\nabla\bm\times\bm B_\phi(\bm x) = -e\mu_0c\,
\phi^\dagger(\bm x) \,\bm\alpha\, \phi(\bm x)  ,
\end{eqnarray} 
so that [see Eq.~(\ref{eq:trans})]
\begin{eqnarray}
\bm B_\phi^{\rm T}(\bm x) = -\frac{e\mu_0c}{4\pi}
\,\bm\nabla\bm\times
\int\rd\bm x^\prime\,\frac{1}{|\bm x - \bm x^\prime|} \,
\phi^\dagger(\bm x^\prime)\,\bm\alpha\,\phi(\bm x^\prime)  .
\nonumber\\
\quad
\end{eqnarray}
An external current density $\bm j_{\rm ex}(\bm x)$ with an associated
field $\bm B_{\rm ex}(\bm x)$ are related by
\begin{eqnarray}
\bm\nabla\bm\times\bm B_{\rm ex}(\bm x) &=& 
\mu_0 \, \bm j_{\rm ex}(\bm x)
\end{eqnarray}
and
\begin{eqnarray}
\bm B_{\rm ex}^{\rm T}(\bm x) &=& \frac{\mu_0}{4\pi} \,
\bm\nabla\bm\times
\int\rd\bm x^\prime\, \frac{1}{|\bm x - \bm x^\prime|}\,
\bm j_{\rm ex}(\bm x^\prime)  .
\end{eqnarray}
According to the Poynting theorem, the energy density associated with
both magnetic fields is
\begin{eqnarray}
u_{\rm B^{\rm T}}(\bm x) &=& 
\frac{\epsilon_0c^2}{2}\left|\bm B_\phi^{\rm T}(\bm x) +
\bm B_{\rm ex}^{\rm T}(\bm x) \right|^2  ,
\end{eqnarray}
where the interaction term is
\begin{eqnarray}
u_{B\rT}^{\,\rm I}(\bm x)
= \epsilon_0c^2\bm B_\phi^{\rm T}(\bm x)
\bm\cdot\bm B_{\rm ex}^{\rm T}(\bm x)  ,
\label{eq:bbint}
\end{eqnarray}
and the interaction energy is
\begin{eqnarray}
&&U_{B\rT}^{\rm I} 
\nonumber\\[10 pt]&&=
-\frac{e\mu_0c}{(4\pi)^2}\int\rd\bm x
\int\rd\bm x^\prime\int\rd\bm x^{\prime\prime}\,
\left[\bm\nabla\bm\times
\frac{\phi^\dagger(\bm x^\prime)\,\bm\alpha\,\phi(\bm x^\prime)}
{|\bm x - \bm x^\prime|}\right] 
\nonumber\\[10 pt]&&\qquad\qquad
\bm\cdot
\left[\bm\nabla\bm\times\frac{\bm j_{\rm ex}(\bm x^{\prime\prime})}
{|\bm x - \bm x^{\prime\prime}|}\right]
\nonumber\\[10 pt]&&=
-\frac{e\mu_0c}{(4\pi)^2}\int\rd\bm x
\int\rd\bm x^\prime\int\rd\bm x^{\prime\prime}\,
\frac{\bm j_{\rm ex}(\bm x^{\prime\prime})}
{|\bm x - \bm x^{\prime\prime}|}\bm\times\bm\nabla
\nonumber\\[10 pt]&&\qquad\qquad\bm\cdot
\bm\nabla\bm\times
\frac{\phi^\dagger(\bm x^\prime)\,\bm\alpha\,\phi(\bm x^\prime)}
{|\bm x - \bm x^\prime|} \, .
\label{eq:tint}
\end{eqnarray}
In Eq.~(\ref{eq:tint}), on the second line, the differentiations
act only on the terms that follow on the right within the square
brackets, and the third and fourth lines follow from integration
by parts.  From the identity
\begin{eqnarray}
(\bm a\bm\times\bm b)\bm\cdot(\bm c\bm\times\bm d) = (\bm a\bm\cdot\bm c)(\bm b\bm\cdot
\bm d) - (\bm a\bm\cdot\bm d)(\bm b\bm\cdot\bm c) \, ,
\label{eq:identity}
\end{eqnarray}
we have
\begin{eqnarray}
&&\frac{\bm j_{\rm ex}(\bm x^{\prime\prime})}
{|\bm x - \bm x^{\prime\prime}|}\bm\times\bm\nabla\bm\cdot
\bm\nabla\bm\times
\frac{\phi^\dagger(\bm x^\prime)\,\bm\alpha\,\phi(\bm x^\prime)}
{|\bm x - \bm x^\prime|}
\nonumber\\[10 pt]
&&\qquad=
\frac{\bm j_{\rm ex}(\bm x^{\prime\prime})}
{|\bm x - \bm x^{\prime\prime}|}\bm\cdot\bm\nabla \,
\bm\nabla\bm\cdot\frac{\phi^\dagger(\bm x^\prime)\,\bm\alpha\,\phi(\bm x^\prime)}
{|\bm x - \bm x^\prime|}
\nonumber\\[10 pt]&&\qquad\qquad-
\frac{\bm j_{\rm ex}(\bm x^{\prime\prime})}
{|\bm x - \bm x^{\prime\prime}|}\bm\cdot\bm\nabla^2
\,\frac{\phi^\dagger(\bm x^\prime)\,\bm\alpha\,\phi(\bm x^\prime)}
{|\bm x - \bm x^\prime|}
\nonumber\\[10 pt]&&\qquad\rightarrow
4\pi\,\frac{\bm j_{\rm ex}(\bm x^{\prime\prime})}
{|\bm x - \bm x^{\prime\prime}|} \bm\cdot
\phi^\dagger(\bm x^\prime)\,\bm\alpha\,\phi(\bm x^\prime) \,
\delta(\bm x - \bm x^\prime)  .
\qquad
\label{eq:exa}
\end{eqnarray}
The first term on the right-hand side of Eq.~(\ref{eq:exa}) vanishes
when integrated over $\bm x^\prime$:
\begin{eqnarray}
&&\int\rd\bm x^\prime \, \bm\nabla\bm\cdot
\frac{\phi^\dagger(\bm x^\prime)\,\bm\alpha\,\phi(\bm x^\prime)}
{|\bm x - \bm x^\prime|}
\nonumber\\[10 pt]
&&\qquad=-\int\rd\bm x^\prime
\phi^\dagger(\bm x^\prime)\,\bm\alpha\,\phi(\bm x^\prime)
\bm\cdot\bm\nabla^\prime\,\frac{1}{|\bm x - \bm x^\prime|}
\nonumber\\[10 pt]&&\qquad=
\int\rd\bm x^\prime\,\frac{1}{|\bm x - \bm x^\prime|}\,
\bm\nabla^\prime \bm\cdot
\phi^\dagger(\bm x^\prime)\,\bm\alpha\,\phi(\bm x^\prime)
\nonumber\\[10 pt]&&\qquad =0 ,
\end{eqnarray}
according to Eq.~(\ref{eq:treq}).  We thus have
\begin{eqnarray}
U_{B\rT}^{\,\rm I} &=&
-\frac{e\mu_0c}{4\pi}\int\rd\bm x
\int\rd\bm x^{\prime}\,
\frac{\bm j_{\rm ex}(\bm x^{\prime})}
{|\bm x - \bm x^{\prime}|}
\bm\cdot
\phi^\dagger(\bm x)\,\bm\alpha\,\phi(\bm x)  .
\nonumber\\\label{eq:dmagint}
\end{eqnarray}
If we write this as
\begin{eqnarray}
U_{B\rT}^{\,\rm I} &=&
-ec\int\rd\bm x\,
\phi^\dagger(\bm x)\,\bm\alpha\bm\cdot
\bm A_{\rm ex}(\bm x)\,\phi(\bm x)  ,
\label{eq:bpt}
\end{eqnarray}
where
\begin{eqnarray}
\bm A_{\rm ex}(\bm x) &=& \frac{\mu_0}{4\pi}
\int\rd\bm x^{\prime}\,
\frac{\bm j_{\rm ex}(\bm x^{\prime})}
{|\bm x - \bm x^{\prime}|}  ,
\label{eq:exvecpot}
\end{eqnarray}
then this corresponds to a perturbation given by 
\begin{eqnarray}
-ec \,\bm\alpha\bm\cdot\bm A_{\rm ex}(\bm x)
\label{eq:depot}
\end{eqnarray}
in Eq.~(\ref{eq:dpot}), where the expression in Eq.~(\ref{eq:exvecpot})
is the same as the vector potential associated with the external current
$\bm j_{\rm ex}$.  However the sign of the perturbation in
Eq.~(\ref{eq:depot}) is the opposite of that given by the minimal
substitution in Eq.~(\ref{eq:mcs}), which gives a perturbation of
\begin{eqnarray}
+ec \,\bm\alpha\bm\cdot\bm A(\bm x)  .
\end{eqnarray}
Thus, in order to arrive at the correct Dirac equation that includes
external fields, we write
\begin{eqnarray}
&&E_0 + U_E^{\,\rm I} - U_{B\rT}^{\,\rm I}
= \int\rd\bm x \,\phi^\dagger(\bm x)
\big[c\,\bm\alpha\bm\cdot\bm p + \beta \, m_{\rm e} c^2
\nonumber\\[10 pt]&&\qquad\qquad
-e\,{\it\Phi}_{\rm ex}(\bm x) + ec\bm\alpha\bm\cdot\bm A_{\rm ex}(\bm x)
\big]\phi(\bm x)  . \qquad
\end{eqnarray}
This sign difference is related to the sign difference in
Sec.~\ref{sssec:tmdi} and is discussed in Sec.~\ref{sec:iiq}.

\subsubsection{Longitudinal magnetic field}
\label{sssec:lmf}

According to the extended Maxwell equations, a longitudinal magnetic
field associated with the Dirac equation is
defined by [see Eq.~(\ref{eq:rgme4})]
\begin{eqnarray}
\bm\nabla\bm\cdot c\bm B_\phi^{\rm L}(\bm x) 
= \frac{\sigma_\phi(\bm x)}{\epsilon_0}  ,
\end{eqnarray}
so that [see Eq.~(\ref{eq:long})]
\begin{eqnarray}
c\bm B_\phi^{\rm L}(\bm x) &=& -\frac{1}{4\pi\epsilon_0}
\,\bm\nabla\int\rd\bm x^\prime
\,\frac{\sigma_\phi(\bm x^\prime)}{|\bm x - \bm x^\prime|}  ,
\end{eqnarray}
where the source $\sigma_\phi$ is given below.
Similarly, a magnetic moment density $\sigma_{\rm ex}$ is associated
with a longitudinal external field $\bm B_{\rm ex}^{\rm L}$ by the
relation
\begin{eqnarray}
\bm\nabla\bm\cdot c\bm B_{\rm ex}^{\rm L}(\bm x) 
=\frac{\sigma_{\rm ex}(\bm x)} {\epsilon_0}
\end{eqnarray}
or
\begin{eqnarray}
c\bm B_{\rm ex}^{\rm L}(\bm x)
&=& -\frac{1}{4\pi\epsilon_0}\,\bm\nabla\int\rd\bm x^\prime \,
\frac{\sigma_{\rm ex}(\bm x^\prime)}{|\bm x - \bm x^\prime|}  .
\end{eqnarray}
The energy density of the combined fields is
\begin{eqnarray}
u_{B^{\rm L}}(\bm x) &=& \frac{\epsilon_0}{2}\,\left|c\bm B_\phi^{\rm L}(\bm x)
+c\bm B_{\rm ex}^{\rm L}(\bm x)\right|^2   ,
\end{eqnarray}
the interaction energy density is
\begin{eqnarray}
u_{B^{\rm L}}^{\,\rm I}(\bm x) &=& \epsilon_0c^2 \,
\bm B_\phi^{\rm L}(\bm x)\bm\cdot\bm B_{\rm ex}^{\rm L}(\bm x)  ,
\end{eqnarray}
and so the total interaction energy is
\begin{eqnarray}
U_{B^{\rm L}}^{\,\rm I} &=& \frac{1}{(4\pi)^2\epsilon_0}\int\rd\bm x\,
\left[\bm\nabla\int\rd\bm x^\prime \,
\frac{1}{|\bm x - \bm x^\prime|}
\,\sigma_\phi(\bm x^\prime) \right]
\nonumber\\[10 pt]&&\qquad
\bm\cdot\left[\bm\nabla\int\rd\bm x^{\prime\prime} \,
\frac{1}{|\bm x - \bm x^{\prime\prime}|}
\,\sigma_{\rm ex}(\bm x^{\prime\prime})\right]
\\[10 pt]&=&
\frac{1}{4\pi\epsilon_0}\int\rd\bm x \, \sigma_\phi(\bm x)
\int\rd\bm x^{\prime}\,\frac{1}{|\bm x - \bm x^\prime|}
\,\sigma_{\rm ex}(\bm x^{\prime})  .
\nonumber
\label{eq:blen}
\end{eqnarray}

Here, we consider the dipole interaction with an external magnetic
moment $\bm m_{\rm ex}$, located at $\bm x_0$, given by
\begin{eqnarray}
\sigma_{\rm ex}(\bm x) &=& -\frac{1}{c}\,
\bm m_{\rm ex}\bm\cdot\bm\nabla\delta(\bm x-\bm x_0)
\nonumber\\[10 pt]
&=&\frac{1}{c}\,\delta(\bm x-\bm x_0)\,
\bm m_{\rm ex}\bm\cdot\bm\nabla  ,
\end{eqnarray}
where the second equality indicates integration by parts, and
\begin{eqnarray}
\int\rd\bm x^{\prime}\,\frac{1}{|\bm x - \bm x^\prime|}
\,\sigma_{\rm ex}(\bm x^{\prime})
&=& \frac{1}{c}\,\bm m_{\rm ex}\bm\cdot\bm\nabla \,
\frac{1}{|\bm x - \bm x_0|} 
\nonumber\\[10 pt]&=& 
-\frac{1}{c}\,\bm m_{\rm ex}\bm\cdot\frac{\bm x - \bm x_0}
{|\bm x - \bm x_0|^3}  .\qquad
\end{eqnarray}
We suggest the convention
\begin{eqnarray}
\sigma_\phi(\bm x) &=& e\,\bm\nabla\bm\cdot\phi^\dagger(\bm x)\, 
\bm x\bm\times\bm\alpha\, \phi(\bm x) 
\nonumber\\[10 pt] &=&
-e\, \phi^\dagger(\bm x)\,
\bm x \bm\times \bm\alpha\, \phi(\bm x)\bm\cdot\bm\nabla 
\nonumber\\[10 pt] &=&
\frac{1}{c}\,\bm m_\phi(\bm x)
\bm\cdot\bm\nabla  ,
\label{eq:momdef}
\end{eqnarray}
where
\begin{eqnarray}
\bm m_\phi(\bm x) &=& ec\,\phi^\dagger(\bm x)\,
\bm x \bm\times \bm\alpha\, \phi(\bm x) 
\end{eqnarray}
is the magnetic moment density.  This is a plausible assignment which
has the nonrelativistic form given by (see Appendix \ref{app:nade})
\begin{eqnarray}
\bm m_\phi(\bm x) &\rightarrow& \frac{e}{m_{\rm e}} 
\,\varphi^\dagger(\bm x) 
\left(\bm L + 2\bm S\right)\varphi(\bm x)  ,
\end{eqnarray}
where
\begin{eqnarray}
\bm L = \bm x \bm\times \bm p;\qquad 
\bm S = \frac{\hbar}{2}\, \bm\sigma  ,
\end{eqnarray}
based on
\begin{eqnarray}
&&\int\rd\bm x\,\phi^\dagger(\bm x)\left(\bm x\bm\times\bm\alpha\right)
\phi(\bm x)
\rightarrow
\frac{1}{2m_{\rm e}c}\int\rd\bm x\,
\varphi^\dagger(\bm x) \qquad
\nonumber\\[10 pt]&&\qquad\bm\times
\left[\left(\bm x\bm\times\bm \sigma\right)
\, \bm \sigma\bm\cdot\bm p
+\bm \sigma\bm\cdot\bm p \left(\bm x\bm\times\bm \sigma\right)\right]
\varphi(\bm x)
\end{eqnarray}
and the identity
\begin{eqnarray}
\left(\bm x\bm\times\bm \sigma\right)
\, \bm \sigma\bm\cdot\bm p
+\bm \sigma\bm\cdot\bm p \left(\bm x\bm\times\bm \sigma\right)
&=& 2\left(\bm L + 2\bm S\right)  .\qquad
\end{eqnarray}

We thus have
\begin{eqnarray}
U_{B^{\rm L}}^{\,\rm I} &=& 
\frac{1}{4\pi\epsilon_0c^2}\int\rd\bm x\,
\bm m_\phi(\bm x)
\bm\cdot\bm\nabla \,
\bm m_{\rm ex}\bm\cdot\frac{\bm x - \bm x_0}
{|\bm x - \bm x_0|^3} 
\nonumber\\[10 pt] &=&
\frac{\mu_0}{4\pi}\int\rd\bm x\,
\bigg[\frac{\bm m_\phi(\bm x)\bm\cdot\bm m_{\rm ex}}
{|\bm x - \bm x_0|^3} 
\nonumber\\[10 pt]&&
-3\,\frac{\bm m_\phi(\bm x)\bm\cdot
(\bm x - \bm x_0)\,\bm m_{\rm ex}\bm\cdot(\bm x - \bm x_0)}
{|\bm x - \bm x_0|^5} \bigg] . \qquad
\end{eqnarray}
Inclusion of this interaction energy in the Dirac equation gives
\begin{eqnarray}
E_0 + U_{B^{\rm L}}^{\,\rm I} &=& \int\rd\bm x \,\phi^\dagger(\bm x)
\big[c\,\bm\alpha\bm\cdot\bm p + \beta \, m_{\rm e} c^2
\nonumber\\[10 pt]&&\qquad
+ec\,\bm x \bm\times\bm\alpha\bm\cdot\bm O(\bm x)\big]\phi(\bm x)  ,
\qquad
\end{eqnarray}
where
\begin{eqnarray}
\bm O(\bm x) &=& 
\frac{\mu_0}{4\pi}\bigg[\frac{\bm m_{\rm ex}}{|\bm x - \bm x_0|^3} 
\nonumber\\[10 pt]&&
-3\,\frac{(\bm x - \bm x_0)\,\bm m_{\rm ex}\bm\cdot(\bm x - \bm x_0)}
{|\bm x - \bm x_0|^5} \bigg]  .
\label{eq:longhfs}
\end{eqnarray}

\subsection{Hyperfine structure}
\label{ssec:hfs}

An example of an interaction in the Dirac equation is the hyperfine
structure correction~(\citet{1930001}).  This well-known example is
included to illustrate the source of the contact interaction in the
nonrelativistic approximation.  The correction arises from the
interaction of the bound electron with the magnetic moment $\bm m_{\rm
N}$ of the nucleus where
\begin{eqnarray}
\bm m_{\rm N} &=& g_{\rm N}\mu_\rN\bm I  .
\end{eqnarray}
Here, $g_{\rm N}$ is the $g$-factor of the nucleus, $\bm I$ is its
angular momentum, and $\mu_{\rm N}=e\hbar/2m_{\rm p}$ is the nuclear
magneton, with the proton mass $m_{\rm p}$.

The conventional Hamiltonian for the transverse hyperfine interaction is
\begin{eqnarray}
H_{\rm hfs}\rT(\bm x) &=& e c \,\bm\alpha\bm\cdot\bm A_{\rm hfs}(\bm x)
\end{eqnarray}
where
\begin{eqnarray}
\bm A_{\rm hfs}(\bm x) &=& \frac{\mu_0}{4\pi}\,
\frac{\bm m_{\rm N} \bm\times \bm x} {|\bm x|^3}
\nonumber\\[10 pt]&=&
 - \frac{\mu_0}{4\pi}\,\bm m_{\rm N}
\bm\times \bm\nabla\,\frac{1}{|\bm x|}
\end{eqnarray}
and
\begin{eqnarray}
\bm B_{\bm m_{\rm N}}^{\rm T}(\bm x) &=&
\bm\nabla\bm\times\bm A_{\rm hfs}(\bm x) 
\nonumber\\[10 pt]&=& 
\frac{\mu_0}{4\pi}\left(
\bm m_{\rm N}\bm\cdot\bm\nabla\,\bm\nabla
-\bm m_{\rm N}\bm\nabla^2
\right)\frac{1}{|\bm x|}  ,
\end{eqnarray}
in agreement with Eq.~(\ref{eq:clm}).
We thus have
\begin{eqnarray}
H_{\rm hfs}^{\rm T}(\bm x) &=& \frac{e\mu_0c}{4\pi}\,
\frac{\bm m_{\rm N}\bm\cdot
\left(\bm x\bm\times\bm\alpha\right)}{|\bm x|^3}  .
\end{eqnarray}

For the longitudinal hyperfine interaction, the external particle is
located at the origin $\bm x_0 = 0$ and $\bm m_{\rm ex} = \bm m_{\rm
N}$ in Eq.~(\ref{eq:longhfs}), which gives the interaction Hamiltonian
\begin{eqnarray}
H_{\rm hfs}^{\rm L}(\bm x) &=& \frac{e\mu_0c}{4\pi} \,
\frac{\bm m_{\rm N}\bm\cdot(\bm x \bm\times \bm \alpha)}{|\bm x|^3}\, .
\end{eqnarray}
The second term in Eq.~(\ref{eq:longhfs}) does not contribute when $\bm
x_0 = 0$ due to the orthogonality to $\bm x\bm\times\bm\alpha$.  Evidently,
the extended Poynting theorem result based on the field energy can give
the same result as the conventional treatment for the hyperfine
correction.  However, in this case, we have
\begin{eqnarray}
\bm B_{\bm m_{\rm N}}^{\rm L}(\bm x) &=&
 -\frac{1}{4\pi\epsilon_0c^2}\,\bm\nabla\int\rd\bm x^\prime \,
\delta\left(\bm x^\prime\right)\bm m_{\rm N}\bm\cdot\bm\nabla^\prime
\,\frac{1}{|\bm x - \bm x^\prime|}
\nonumber\\&=&
\,\frac{\mu_0}{4\pi}\,\bm m_{\rm N}\bm\cdot\bm\nabla \, \bm\nabla
\,\frac{1}{|\bm x|}  ,
\end{eqnarray}
in agreement with Eq.~(\ref{eq:gsder}).

An unperturbed eigenfunction of the Dirac equation with an external
spherically symmetric binding field, as in Eq.~(\ref{eq:dpot}), can be
written as (see for example \citet{1998004})
\begin{eqnarray}
\phi_{n\kappa\mu}(\bm x) &=& 
\left(\begin{array}{c}
f_1(x) \chi_\kappa^\mu(\bm{\hat x}) \\
\ri f_2(x) \chi_{-\kappa}^\mu(\bm{\hat x}) \end{array}\right)  ,
\label{eq:dwf}
\end{eqnarray}
where $f_1$ and $f_2$ are radial wavefunctions with $x = |\bm x|$ and
$\chi_\kappa^\mu(\bm{\hat x})$ is the two-component Dirac spin-angle
function, with the property that
\begin{eqnarray}
\bm\sigma\bm\cdot\bm{\hat x}\,\chi_\kappa^\mu(\bm{\hat x})
&=& -\chi_{-\kappa}^\mu(\bm{\hat x})  .
\end{eqnarray}
In Eq.~(\ref{eq:dwf}), $\kappa$ is the Dirac angular-momentum-parity
quantum number with angular-momentum quantum number $j = |\kappa|-1/2$,
and $\mu$ is the $z$ projection of the angular momentum.  Thus the
hyperfine matrix element is given by
\begin{eqnarray}
&&\phi_{n\kappa\mu}^\dagger(\bm x)\,
\frac{\bm x\bm\times\bm\alpha}{x^3} \,
\phi_{n\kappa\mu^\prime}(\bm x)
\nonumber\\[10 pt]&&=
\ri \, \frac{f_1(x)f_2(x)}{x^3}
\big[
\chi_\kappa^{\mu\dagger}(\bm{\hat x}) \,
\bm x\bm\times\bm\sigma \,
\chi_{-\kappa}^{\mu^\prime}(\bm{\hat x}) 
\nonumber\\[10 pt]&&\qquad -
\chi_{-\kappa}^{\mu\dagger}(\bm{\hat x}) \,
\bm x\bm\times\bm\sigma \,
\chi_{\kappa}^{\mu^\prime}(\bm{\hat x}) 
\big]
\nonumber\\[10 pt]&&=
-2 \, \frac{f_1(x)f_2(x)}{x^2} \,
\chi_{\kappa}^{\mu\dagger}(\bm{\hat x}) \,
\left(\bm\sigma - \bm{\hat x}\,\bm\sigma\bm\cdot\bm{\hat x}\,
\right)
\chi_{\kappa}^{\mu^\prime}(\bm{\hat x})  ,\qquad
\end{eqnarray}
which follows from the identity
\begin{eqnarray}
\bm{\hat x}\bm\times\bm\sigma \,\bm\sigma\bm\cdot\bm{\hat x} -
\bm\sigma\bm\cdot\bm{\hat x} \,\bm{\hat x}\bm\times\bm\sigma
&=&
-2\,\ri \left(\bm\sigma - \bm{\hat
x}\,\bm\sigma\bm\cdot\bm{\hat x}\right)  , \qquad
\end{eqnarray}
and
\begin{eqnarray}
&&\left<n\kappa\mu\left|H_{\rm hfs}\right|n\kappa\mu^\prime\right>
= -\frac{e\mu_0c}{2\pi}
\int\rd\bm x\,\frac{f_1(x)f_2(x)}{x^2} 
\nonumber\\[10 pt]&&\qquad\qquad\times
\bm m_{\rm N}\bm\cdot\chi_\kappa^{\mu\dagger}(\bm x)
\left(\bm\sigma - \bm{\hat
x}\,\bm\sigma\bm\cdot\bm{\hat x}\right)
\chi_\kappa^{\mu^\prime}(\bm x)  . \qquad
\end{eqnarray}

For the 1S state ($\kappa = -1$, $\mu = \pm \fr{1}{2}$\,),
\begin{eqnarray}
\chi_{-1}^\mu(\bm{\hat x}) &=& 
\frac{1}{\sqrt{4\pi}}\,\left|\mu\right>
= \frac{1} {\sqrt{4\pi}}\left(\begin{array}{l}\fr{1}{2}+\mu \\ 
\fr{1}{2}-\mu \end{array}\right)  ,
\end{eqnarray}
so that
\begin{eqnarray}
\int\rd\iO \,
\chi_{-1}^{\mu\dagger}(\bm{\hat x}) \,
\left(\bm\sigma - \bm{\hat x}\,\bm\sigma\bm\cdot\bm{\hat x}\,
\right)
\chi_{-1}^{\mu^\prime}(\bm{\hat x})
=\frac{2}{3}\left<\mu\right|\bm\sigma\left|\mu^\prime\right> .
\quad
\end{eqnarray}
With $\bm s = \bm\sigma/2$, and the nucleus is the proton with spin
$\fr{1}{2}$, this gives
\begin{eqnarray}
\left< H_{\rm hfs} \right> &=& -
\frac{4\alpha g_{\rm p}\hbar^2}{3m_{\rm p}}
\int_0^\infty\rd x\, f_1(x)f_2(x)
\left<\bm I\bm\cdot\bm s\right>  . \quad
\end{eqnarray}
Since $\bm F = \bm I + \bm s$, we have $\bm I\bm\cdot\bm s =
\fr{1}{2}\left(\bm F^2 - \bm I^2 - \bm s^2\right)$ 
so that
\begin{eqnarray}
\left<\bm I\bm\cdot\bm s \right>
&=& \left\{\begin{array}{cc} \frac{1}{4}
& \quad \mbox{for } F = 1 \\
 -\frac{3}{4} & \quad \mbox{for } F = 0 \end{array}\right.
\end{eqnarray}
and for the splitting
\begin{eqnarray}
\Delta E_{\rm hfs} &=& - 
\frac{4\alpha g_{\rm p}\hbar^2}{3m_{\rm p}}
\int_0^\infty\rd x\, f_1(x)f_2(x)  .
\end{eqnarray}
The integral in this expression can be evaluated exactly with the result
(see Appendix~\ref{app:crhi})
\begin{eqnarray}
\int_0^\infty\rd x\, f_1(x)f_2(x) &=&
 -\frac{(Z\alpha)^3}{a(2a-1)\lbar^2}
 \nonumber\\[10 pt]&
\rightarrow& - \frac{(Z\alpha)^3}{\lbar^2}  ,
\label{eq:exact}
\end{eqnarray}
where $a = \sqrt{1-(Z\alpha)^2}$.  Although $Z=1$ for the proton, we
retain the charge number to show the $Z$ dependence.  This is not a
contact interaction, but rather the result of a direct calculation which
gives the nonrelativistic limit as the leading term as
$Z\alpha\rightarrow0$.  On the other hand, the same result may be viewed
a contact interaction in the nonrelativistic limit.  In this case, the
contact term arises as a surface term in the integral.  In the
nonrelativistic limit (see Appendix \ref{app:nade}\,)
\begin{eqnarray}
f_1(x) &\rightarrow& f(x) \, ,
\\[10 pt]
f_2(x) &\rightarrow& \frac{\lbar_{\rm e}}{2}\,
\frac{\partial}{\partial x}\,f(x)  ,
\label{eq:nrsc}
\end{eqnarray}
so that
\begin{eqnarray}
\int_0^\infty\rd x\, f_1(x)f_2(x) &\rightarrow& 
\frac{\lbar_{\rm e}}{2}\int_0^\infty\rd x\, f(x)
\frac{\partial}{\partial x}\,f(x)
\nonumber\\[10 pt]&=&
\frac{\lbar_{\rm e}}{4}\int_0^\infty\rd x\,
\frac{\partial}{\partial x}\,f^2(x)
\nonumber\\[10 pt]&=&
- \frac{\lbar_{\rm e}}{4}
f^2(0) 
\nonumber\\[10 pt]&=&
 -\frac{(Z\alpha)^3}{\lbar^2}
\end{eqnarray}
which follows from Eq.~(\ref{eq:fsq}) and agrees with
Eq.~(\ref{eq:exact}).  We also have
\begin{eqnarray}
\varphi(\bm x) &=& f(x)\chi_{-1}^\mu(\bm{\hat x}) , 
\\[10 pt]
\left|\varphi(\bm x)\right|^2 
&=&
\frac{1}{4\pi}\,\left|f(x)\right|^2
\end{eqnarray}
and
\begin{eqnarray}
\int_0^\infty\rd x\, f_1(x)f_2(x) &\rightarrow&
- \pi\lbar_{\rm e}|\varphi(0)|^2  .
\end{eqnarray}
Thus
\begin{eqnarray}
\Delta E_{\rm hfs} &=& 
\frac{4\pi\alpha}{3}\,\frac{g_{\rm p}\hbar^2}{m_{\rm p}}
\,\lbar_{\rm e}\left|\varphi(0)\right|^2  ,
\label{eq:contact}
\end{eqnarray}
which gives as the leading 1S-state splitting in hydrogen
\begin{eqnarray}
\Delta E_{\rm hfs} &=&
\frac{4\alpha^4}{3}\,\frac{g\rp m_{\rm e}}{m\rp}
\,m_{\rm e}c^2 = 5.877\bm\times10^{-6}\mbox{ eV}  , \quad
\end{eqnarray}
corresponding to a wavelength of 21 cm.

Because the contact term in Eq.~(\ref{eq:contact}) can be
written as the surface term in the nonrelativistic reduction of
the Dirac hyperfine expression involving an integral over all
space, there is no reason to ascribe physical significance to
the contact interaction.  Moreover, the Dirac wavefunctions near
the origin include a negative power of the radial coordinate
(see Appendix \ref{app:crhi}), so the relativistic expression
does not allow for a contact interaction.

\section{The Poynting theorem and quantum electrodynamics (QED)}
\label{sec:iiq}

In the Furry picture of bound-state quantum electrodynamics, the
electron field is expressed in terms of creation and annihilation
operators for eigenstates of the electron in a static binding
field~\cite{1951015,schweber,ja}.  In this formulation, the interaction
between the particle current and the potential associated with the
electromagnetic field is given by
\begin{eqnarray}
{\cal H}_{\rm I}(\bm x) &=& j_\mu(\bm x) A^\mu(\bm x) 
\nonumber\\[10 pt]&=&
 -ec\,\phi^\dagger(\bm x)\alpha_\mu\phi(\bm x) A^\mu(\bm x) ,
\label{eq:qedint}
\end{eqnarray}
which is a local interaction at the point $\bm x$.  This interaction is
based on the external field Dirac equation, which may be written as 
\begin{eqnarray}
&&\phi^\dagger(\bm x)
\big[cp^0 - c\,\bm\alpha\bm\cdot\bm p - \beta \, m_{\rm e} c^2
+e{\it\Phi}_{\rm ex}(\bm x) 
\nonumber\\[10 pt]&&\qquad
- ec\bm\alpha\bm\cdot\bm A_{\rm ex}(\bm x)
\big]\phi(\bm x) = 0 \, ,
\end{eqnarray}
where the external field interactions are based in turn on the minimal
coupling substitution in Eq.~(\ref{eq:mcs}).

Alternatively, the interaction may be described by considering
the energy of the combined fields produced by the current and by
the fields corresponding to $A^\mu$, as suggested by the
Poynting theorem.  However, as shown below, when expressed in
terms of the fields, the interaction energy density corresponds
to $|\bm E|^2-|c\bm B|^2$.  Obviously, this is in disagreement
with the interaction energy density corresponding to $|\bm
E|^2+|c\bm B|^2$ suggested by the extended Poynting theorem.
The source of this difference is linked to the fact that the
magnetic interactions in QED are based on transverse fields.
This may be seen in the discussion of Sec.~\ref{sssec:tmdi}
where the sign difference already appears in the classical
interactions of particles with magnetic moments.  It also
appears in magnetic field interactions in the Dirac equation in
Sec.~\ref{sssec:tmf}.  This means that in order to calculate
magnetic interactions of particles by integration of transverse
field energies, the magnetic interaction energy must be taken to
be $-\epsilon_0|c\bm B|^2$.

\subsection{External field interaction}
\label{ssec:exfi}

Here we consider the example in which fields produced by an electron
interact with fields produced by an external source.

\subsubsection{Electric interactions}
\label{sssec:ei}

Electric interactions are given by the $\mu=0$ term in
Eq.~(\ref{eq:qedint}), where
\begin{eqnarray}
j_0(\bm x) &=& -ec\,\phi^\dagger(\bm x)\,\phi(\bm x) 
= c\,\rho_\phi(\bm x) \, ,
\\[10 pt]
A_{\rm ex}^0(\bm x) &=& \frac{1}{c}\,{\it\Phi}_{\rm ex}(\bm x) =
\frac{1}{4\pi\epsilon_0c}\int\rd\bm x^\prime\,\frac{\rho_{\rm ex}(\bm
x^\prime)}{|\bm x-\bm x^\prime|}  . \quad
\end{eqnarray}
These expressions yield
\begin{eqnarray}
&&\int\rd\bm x\,j_0(\bm x)\,A_{\rm ex}^0(\bm x)
\nonumber\\[3 pt]&&\quad=
\frac{1}{4\pi\epsilon_0}
\int\rd\bm x\int\rd\bm x^\prime\,\rho_\phi(\bm x)
\,\frac{1}{|\bm x-\bm x^\prime|}\,\rho_{\rm ex}(\bm
x^\prime) 
\nonumber\\[3 pt]&&\quad=
\frac{1}{4\pi\epsilon_0}\int\rd\bm x
\int\rd\bm x^\prime\int\rd\bm x^{\prime\prime}
\nonumber\\[3 pt]&&\qquad\times
\rho_\phi(\bm
x^{\prime\prime})
\,\frac{\delta(\bm x-\bm x^{\prime\prime})}
{|\bm x-\bm x^\prime|}
\,\rho_{\rm ex}(\bm x^\prime) \quad
\nonumber\\[3 pt]&&\quad=-\frac{1}{(4\pi)^2\epsilon_0}
\int\rd\bm x
\int\rd\bm x^\prime\int\rd\bm x^{\prime\prime}\,
\rho_\phi(\bm x^{\prime\prime})
\nonumber\\[3 pt]&&\qquad\times
\left(\bm\nabla^2\,
\frac{1}{|\bm x-\bm x^{\prime\prime}|}\right)
\,\frac{1}{|\bm x-\bm x^\prime|}
\,\rho_{\rm ex}(\bm x^\prime)
\nonumber\\[3 pt]&&\quad=\frac{1}{(4\pi)^2\epsilon_0}
\int\rd\bm x \int\rd\bm x^\prime\int\rd\bm x^{\prime\prime}\,
\nonumber\\[10 pt]&&\qquad\times
\rho_\phi(\bm x^{\prime\prime})
\frac{1}{|\bm x-\bm x^{\prime\prime}|}\,
\overleftarrow{\bm\nabla}
\bm\cdot\bm\nabla \,
\frac{1}{|\bm x-\bm x^\prime|}
\,\rho_{\rm ex}(\bm x^\prime)
\nonumber\\[3 pt]&&\quad=
\epsilon_0\int\rd\bm x \,
\bm E_\phi(\bm x)\bm\cdot\bm E_{\rm ex}(\bm x)  .
\label{eq:qedex}
\end{eqnarray}

\subsubsection{Magnetic interactions}
\label{sssec:mi}

The magnetic interaction is given by the three-vector terms in
Eq.~(\ref{eq:qedint}) where
\begin{eqnarray}
\bm j(\bm x) &=& -ec\,\phi^\dagger(\bm x)\,\bm\alpha\,\phi(\bm x)
\, ,
\\[3 pt]
\bm A_{\rm ex}(\bm x) &=& \frac{\mu_0}{4\pi}\int\rd\bm x^\prime \,
\frac{\bm j_{\rm ex}(\bm x^\prime)}
{|\bm x-\bm x^\prime|}  ,
\end{eqnarray}
and
\begin{eqnarray}
-\int\rd\bm x\,\bm j(\bm x)\bm\cdot
\bm A_{\rm ex}(\bm x)
= - \frac{e\mu_0c}{4\pi}
\int\rd\bm x \int\rd\bm x^\prime
\nonumber\\[3 pt]\times
\phi^\dagger(\bm x) \,\bm\alpha\,\phi(\bm x)
\bm\cdot \frac{1}{|\bm x-\bm x^\prime|} \,
\bm j_{\rm ex}(\bm x^\prime)  .
\label{eq:exmagint}
\end{eqnarray}
This is just Eq.~(\ref{eq:dmagint}), so by reversing the steps
to Eq.~(\ref{eq:bbint}), we conclude that
\begin{eqnarray}
-\int\rd \bm x\,\bm j(\bm x)\bm\cdot\bm
A_{\rm ex}(\bm x)
= -\epsilon_0 \int\rd\bm x 
\nonumber\\[3 pt]\times
c\bm B_\phi^{\rm T}(\bm x)\bm\cdot 
c\bm B_{\rm ex}^{\rm T}(\bm x)  .
\label{eq:exmagbb}
\end{eqnarray}
We thus have
\begin{eqnarray}
&&\int\rd\bm x\left[j_0(\bm x)A^0_{\rm ex}(\bm x) 
- \bm j(\bm x)\bm\cdot\bm
A_{\rm ex}(\bm x)\right]
= \epsilon_0 \int\rd\bm x
\nonumber\\[3 pt]&&\qquad\times
\big[\bm E_\phi(\bm x)\bm\cdot\bm E_{\rm ex}(\bm x) 
-
c\bm B_\phi^{\rm T}(\bm x)\bm\cdot 
c\bm B_{\rm ex}^{\rm T}(\bm x)\big]  ,\qquad
\end{eqnarray}
which is the interaction part of 
\begin{eqnarray}
\frac{\epsilon_0}{2}\int\rd\bm x
\left[|\bm E_\phi(\bm x)+\bm E_{\rm ex}(\bm x)|^2 -
|c\bm B_\phi^{\rm T}(\bm x)
+c\bm B_{\rm ex}^{\rm T}(\bm x)|^2\right]
\nonumber\\[5 pt]\qquad
=\frac{\epsilon_0}{2}\int\rd\bm x
\left[|\bm E(\bm x)|^2 -
|c\bm B^{\rm T}(\bm x)|^2\right]  . \qquad
\end{eqnarray}

\subsection{One-photon exchange}
\label{ssec:ope}

An example of an exact QED expression is the one-photon interaction
between two electrons bound in an atom.  With a highly-charged nucleus,
as a first approximation the two electrons may be taken to be hydrogenic
product states where the binding to the charged nucleus is much stronger
than the electron-electron interaction.  The one-photon interaction
takes the same form as the interaction in the previous section when
expressed in terms of electric and magnetic fields.

The relevant expression from QED in the Furry picture is \cite{1951015,
1985050}
\begin{eqnarray}
E_{\rm d} &=& \alpha\hbar c \int\rd\bm x_2\int\rd\bm x_1 \,
\nonumber\\[3 pt]&&\times
\Psi^\dagger(\bm x_2,\bm x_1) \, \frac{\alpha_\mu^{(2)}\alpha^{\mu(1)}}
{|\bm x_2-\bm x_1|}\,\Psi(\bm x_2,\bm x_1)  ,
\label{eq:photex}
\end{eqnarray}
where the exchange term is omitted (as would be the case for an
electron-muon interaction).  In Eq.~(\ref{eq:photex}), the wavefunction
is a sum of products of hydrogenic wavefunctions
\begin{eqnarray}
\Psi(\bm x_2,\bm x_1) &=&
\sum_{\sigma\sigma^\prime}D_{\sigma\sigma^\prime}
\phi_{\beta_2}^\sigma(\bm x_2)
\phi_{\beta_1}^{\sigma^\prime}(\bm x_1)  ,
\end{eqnarray}
where $\beta_i$ denotes the subset of quantum numbers $\{n,l,j\}$ and
$\sigma$ denotes the remaining quantum number $\{m\}$.  The coefficients
$D_{\sigma\sigma^\prime}$ produce eigenfunctions of angular momentum.
It is sufficient to consider a single term in the sum, which is
\begin{widetext}
\begin{eqnarray}
E_{\rm d}^{\sigma\sigma^\prime\nu\nu^\prime} &=&
\alpha\hbar c \int\rd\bm x_2\int\rd\bm x_1 
\phi_{\beta_2}^{\sigma\dagger}(\bm x_2)
\phi_{\beta_1}^{\sigma^\prime\dagger}(\bm x_1)
\, \frac{\alpha_\mu^{(2)}\alpha^{\mu(1)}}
{|\bm x_2-\bm x_1|}
\phi_{\beta_2}^\nu(\bm x_2)
\phi_{\beta_1}^{\nu^\prime}(\bm x_1)
\nonumber\\[10 pt]&=&
\frac{e^2}{4\pi\epsilon_0} \int\rd\bm x_2\int\rd\bm x_1
\phi_{\beta_2}^{\sigma\dagger}(\bm x_2)
\alpha_\mu
\phi_{\beta_2}^\nu(\bm x_2)
\, \frac{1}
{|\bm x_2-\bm x_1|}\,
\phi_{\beta_1}^{\sigma^\prime\dagger}(\bm x_1)
\alpha^\mu
\phi_{\beta_1}^{\nu^\prime}(\bm x_1)
\nonumber\\[10 pt]&=&
\frac{e^2}{4\pi\epsilon_0} \int\rd\bm x_2\int\rd\bm x_1 \,
\bigg[\phi_{\beta_2}^{\sigma\dagger}(\bm x_2)
\phi_{\beta_2}^\nu(\bm x_2)
\, \frac{1}
{|\bm x_2-\bm x_1|}\,
\phi_{\beta_1}^{\sigma^\prime\dagger}(\bm x_1)
\phi_{\beta_1}^{\nu^\prime}(\bm x_1) 
\nonumber\\[10 pt]&&\qquad\qquad\qquad
-\phi_{\beta_2}^{\sigma\dagger}(\bm x_2)
\bm\alpha
\phi_{\beta_2}^\nu(\bm x_2)
\bm\cdot \frac{1}
{|\bm x_2-\bm x_1|}\,
\phi_{\beta_1}^{\sigma^\prime\dagger}(\bm x_1)
\bm\alpha
\phi_{\beta_1}^{\nu^\prime}(\bm x_1) \bigg]
\nonumber\\[10 pt]&=&
\frac{1}{4\pi\epsilon_0} \int\rd\bm x_2\int\rd\bm x_1 \,
\bigg[\rho_2(\bm x_2)
\, \frac{1}
{|\bm x_2-\bm x_1|}\,
\rho_1(\bm x_1)
-
\frac{1}{c^2}\,\bm j_2(\bm x_2)
\bm\cdot \frac{1}
{|\bm x_2-\bm x_1|}\,
\bm j_1(\bm x_1)\bigg] ,\qquad
\label{eq:relope}
\end{eqnarray}
where
\begin{eqnarray}
\rho_1(\bm x_1) &=& -e\,
\phi_{\beta_1}^{\sigma^\prime\dagger}(\bm x_1)
\phi_{\beta_1}^{\nu^\prime}(\bm x_1)  ,
\label{eq:rho1}
\\[0 pt]
\rho_2(\bm x_2) &=& -e\,
\phi_{\beta_2}^{\sigma\dagger}(\bm x_2)
\phi_{\beta_2}^\nu(\bm x_2)  ,
\label{eq:rho2}
\\[0 pt]
\bm j_1(\bm x_1) &=& -ec\,
\phi_{\beta_1}^{\sigma^\prime\dagger}(\bm x_1)
\bm\alpha
\phi_{\beta_1}^{\nu^\prime}(\bm x_1)  , 
\label{eq:j1}
\\[0 pt]
\bm j_2(\bm x_2) &=& -ec\,
\phi_{\beta_2}^{\sigma\dagger}(\bm x_2)
\bm\alpha
\phi_{\beta_2}^\nu(\bm x_2)  .
\label{eq:j2}
\end{eqnarray}
Following the derivations in Eqs.~(\ref{eq:qedex}) and
(\ref{eq:exmagint})-(\ref{eq:exmagbb}), we find
\begin{eqnarray}
E_{\rm d}^{\sigma\sigma^\prime\nu\nu^\prime} &=&
\epsilon_0 \int\rd\bm x
\left[\bm E_2(\bm x)\bm\cdot\bm E_1(\bm x) -
c\bm B_2^{\rm T}(\bm x)\bm\cdot c\bm B_1^{\rm T}(\bm x)\right]  ,
\end{eqnarray}
where $\bm E_i$ and $\bm B_i^{\rm T}$ are the fields corresponding to
the sources in Eqs.~(\ref{eq:rho1}) to (\ref{eq:j2}).  As for the
external-field interaction correction, this is the interaction part of
\begin{eqnarray}
&&\frac{\epsilon_0}{2}\int\rd\bm x
\left[|\bm E_2(\bm x)+\bm E_1(\bm x)|^2 -
|c\bm B_2^{\rm T}(\bm x)+c\bm B_1^{\rm T}(\bm x)|^2\right]
\nonumber\\[5 pt]&&\qquad\qquad
=\frac{\epsilon_0}{2}\int\rd\bm x
\left[|\bm E(\bm x)|^2 -
|c\bm B^{\rm T}(\bm x)|^2\right]  .
\end{eqnarray}
If the replacements
\begin{eqnarray}
\rho_i(\bm x) \rightarrow q_i\,\delta(\bm x - \bm x_i)
\end{eqnarray}
are made in the first term of Eq.~(\ref{eq:relope}), it reproduces the
classical expression in Eq.~(\ref{eq:eeint}).

On the other hand, the interaction can be expressed in terms of the
extended Poynting theorem and longitudinal magnetic fields.  In this
case, we have the interaction energy density
\begin{eqnarray}
u_{B^{\rm L}}(\bm x) &=& \epsilon_0\left|c\bm B_{\phi_2}^{\rm L}(\bm x)
\bm\cdot c\bm B_{\phi_1}^{\rm L}(\bm x)\right|^2   ,
\end{eqnarray}
where
\begin{eqnarray}
c\bm B_{\phi_i}^{\rm L}(\bm x) &=& -\frac{1}{4\pi\epsilon_0}
\,\bm\nabla\int\rd\bm x_i
\,\frac{\sigma_{\phi_i}(\bm x_i)}{|\bm x - \bm x_i|}  ,
\end{eqnarray}
and
\begin{eqnarray}
U_{B^{\rm L}}(\bm x) &=& \frac{1}{(4\pi)^2\epsilon_0}
\int\rd\bm x
\left[\bm\nabla\int\rd\bm x_2
\,\frac{\sigma_{\phi_2}(\bm x_2)}{|\bm x - \bm x_2|}\right]
\bm\cdot
\left[\bm\nabla\int\rd\bm x_1
\,\frac{\sigma_{\phi_1}(\bm x_1)}{|\bm x - \bm x_1|}\right]
\nonumber\\[10 pt]
&=& -\frac{1}{(4\pi)^2\epsilon_0}
\int\rd\bm x
\int\rd\bm x_2
\,\frac{\sigma_{\phi_2}(\bm x_2)}{|\bm x - \bm x_2|}
\,\bm\nabla^2 \int\rd\bm x_1
\,\frac{\sigma_{\phi_1}(\bm x_1)}{|\bm x - \bm x_1|}
\nonumber\\[10 pt]
&=& \frac{1}{4\pi\epsilon_0}
\int\rd\bm x
\int\rd\bm x_2
\int\rd\bm x_1
\,\frac{\sigma_{\phi_2}(\bm x_2)}{|\bm x - \bm x_2|}
\,\delta(\bm x - \bm x_1)
\,\sigma_{\phi_1}(\bm x_1)
\nonumber\\[10 pt]
&=& \frac{1}{4\pi\epsilon_0}
\int\rd\bm x_2
\int\rd\bm x_1
\,\sigma_{\phi_2}(\bm x_2)\, \frac{1}{|\bm x_2 - \bm x_1|}
\,\sigma_{\phi_1}(\bm x_1)
\end{eqnarray}
\end{widetext}
where
\begin{eqnarray}
\sigma_{\phi_2}(\bm x_2) &=& 
e\,\bm\nabla_2 \bm\cdot
\phi_{\beta_2}^{\sigma\dagger}(\bm x_2) \,
\bm x_2\bm\times\bm\alpha_2\,
\phi_{\beta_2}^\nu(\bm x_2)
\nonumber\\&=&
-e\, \phi_{\beta_2}^{\sigma\dagger}(\bm x_2) \,
\bm x_2\bm\times\bm\alpha_2\,
\phi_{\beta_2}^\nu(\bm x_2) \bm\cdot \bm\nabla_2
\nonumber\\&=&
\frac{1}{c}\,\bm m_{\phi_2}(\bm x_2)\bm\cdot\bm\nabla_2  ,
\nonumber\\
\sigma_{\phi_1}(\bm x_1) &=&
e\,\bm\nabla_1 \bm\cdot
\phi_{\beta_1}^{\sigma^\prime\dagger}(\bm x_1) \,
\bm x_1\bm\times\bm\alpha_1\,
\phi_{\beta_1}^{\nu^\prime}(\bm x_1)
\nonumber\\&=&
-e\, \phi_{\beta_1}^{\sigma^\prime\dagger}(\bm x_1) \,
\bm x_1\bm\times\bm\alpha_1\,
\phi_{\beta_1}^{\nu^\prime}(\bm x_1) \bm\cdot \bm\nabla_1
\nonumber\\&=&
\frac{1}{c}\,\bm m_{\phi_1}(\bm x_1)
\bm\cdot\bm\nabla_1 \, ,
\end{eqnarray}
as in Eq.~(\ref{eq:momdef}).  We thus have
\begin{eqnarray}
&&U_{B^{\rm L}}(\bm x)
= \frac{\mu_0}{4\pi}
\int\rd\bm x_2
\int\rd\bm x_1 
\nonumber\\[10 pt]&&\times
\,\bm m_{\phi_2}(\bm x_2)\bm\cdot\bm\nabla_2 \, 
\,\bm m_{\phi_1}(\bm x_1) \bm\cdot\bm\nabla_1\,
\frac{1}{|\bm x_2 - \bm x_1|}  .
\label{eq:ope}
\end{eqnarray}
If the replacements
\begin{eqnarray}
\bm m_{\phi_i}(\bm x) \rightarrow 
\bm m_i\,\delta(\bm x - \bm x_i)
\end{eqnarray}
are made in Eq.~(\ref{eq:ope}), it becomes
\begin{eqnarray}
\frac{\mu_0}{4\pi}
\,\bm m_2\bm\cdot\bm\nabla_2 \, 
\,\bm m_1\bm\cdot\bm\nabla_1\,
\frac{1}{|\bm x_2 - \bm x_1|}  .
\end{eqnarray}
in agreement with Eq.~(\ref{eq:bsqint}).

\section{The extended Maxwell equations and electromagnetic radiation}
\label{sec:rd}

It is of interest to examine the consequences of the magnetic dipole
source terms for phenomena that involve electromagnetic radiation.
Propagation of electromagnetic waves in vacuum is not affected by the
source terms, but the interaction with atoms is.  This section examines
radiation from Dirac transition currents based on the extended Maxwell
equations.  This process is discussed from a similar point of view in
\citet{2010043}, however in that work, there is no lower-component
source term.\footnote{Here we do not include a factor of two in front of
$\bm j$ as was done in \citet{2010043}, and there is no factor of
one-half added to the radiation rate as in Eq.~(518) of \citet{2010043}.
These two changes result in a net factor of one-half for the transition
rate compared to the one in \citet{2010043}, which is explained below
Eq.~(\ref{eq:qedrate}).}

When a hydrogenlike atom in a volume $V$ makes a transition from state
$i$ to state $f$, the current associated with the rearrangement of the
charge and magnetic moment distributions of the atom are sources of
emitted radiation according to the Maxwell equations.  The change of
energy of the atom is the energy of the emitted radiation.  Energy
conservation for this process is described by the extended Poynting
theorem
\begin{eqnarray}
&&\int_{\rm V}\rd\bm x \,\frac{\partial u}{\partial t} 
+ \int_{\rm S}\rd A\,\bm{\hat
n}\cdot \bm S 
\nonumber\\[10 pt]&&\qquad = 
-\int_{\rm V}\rd\bm x\,\bm J\cdot\bm E -\int_{\rm
V}\rd\bm x\,\bm K\cdot c\bm B , 
\label{eq:xpt}
\end{eqnarray}
where the power $P$ of the radiation passing out of the volume is given
by the second term on the left-hand side the equation.  The power
radiated is the transition rate $A_{if}$, or the photon emission rate,
times the photon energy
\begin{eqnarray} 
P &=& A_{if} \hbar\omega  ,
\label{eq:rate} 
\end{eqnarray} 
where $\omega$ is the frequency of the emitted photon.  We have 
\begin{eqnarray} P &=& \int_{\rm S}\rd
A\,\bm{\hat n}\cdot \bm S \, , 
\end{eqnarray} where (see
Sec.~\ref{sec:rpt}) 
\begin{eqnarray} \bm S &=& \frac{1}{\mu_0}\,{\rm
Re}\,\bm E\times\bm B^* = \frac{1}{2c\mu_0}\,\overline{\iP}\bm
\gamma\iP .  
\end{eqnarray}

The Dirac extended source for the radiation is given by [see
Eq.~(\ref{eq:sources})]
\begin{eqnarray}
\iX_{\rm D}(x) &=& 
c\mu_0\left(\begin{array}{c} -\bm j^{if}_{\rm s}(x) \\
\ri\,\bm k^{if}_{\rm s}(x)
\end{array}\right) .
\label{eq:currents}
\end{eqnarray}
The upper component is the conventional current density
\begin{eqnarray}
\bm j^{if}(x) &=& 
- ec\, \phi_f^\dagger(x)\bm\alpha\phi_i(x)
\nonumber\\[0 pt]&=&
- ec\, \phi_f^\dagger(\bm x)\bm\alpha\phi_i(\bm x)
\,\re^{-\ri\omega_{if}t} \, ,
\end{eqnarray}
where $\phi_i$ and $\phi_f$ are the initial and final hydrogen-like atom
Dirac wave functions, here $\bm\alpha$ is the $4\times4$ Dirac matrix
[see Eq.~(\ref{eq:diracmat})], and
\begin{eqnarray}
\omega_{if} &=& \frac{E_i-E_f}{\hbar}
\end{eqnarray}
is the frequency corresponding to the energy difference of the
transition.  

Correspondence to a classical current is seen by taking the
nonrelativistic limit of the matrix element in the dipole approximation
$\bm k\cdot\bm x\rightarrow 0$ (see Appendix~\ref{app:nade})
\begin{eqnarray}
&&\int\rd\bm x\,\phi_f^\dagger(\bm x)\bm\alpha\,
\re^{-\ri\,\bm k\cdot\bm x} \phi_i(\bm x)
\nonumber\\[10 pt]
&&\qquad\rightarrow
\frac{1}{2m_{\rm e}c}\int\rd\bm x\,
\varphi_f^\dagger(\bm x)
\left(\bm \sigma\, \bm \sigma\cdot\bm p
+\bm \sigma\cdot\bm p \,\bm \sigma\right)
\varphi_i(\bm x)\qquad
\nonumber\\[10 pt]&&\qquad=
\frac{1}{m_{\rm e}c}\int\rd\bm x\,
\varphi_f^\dagger(\bm x)
\,\bm p \, \varphi_i(\bm x)
\end{eqnarray}
and
\begin{eqnarray}
\bm j^{if}(x) &\rightarrow& 
- \frac{e}{m_{\rm e}}\,
\varphi_f^\dagger(\bm x)\,\bm p\, \varphi_i(\bm x) 
\,\re^{-\ri\omega_{if}t} .
\end{eqnarray}
The spatial factor has the classical correspondence $\rho(\bm x)\,\bm
v$, which is $\bm J(\bm x)$ in Eq.~(\ref{eq:chcurrent}).

Although the charge current source, the upper component in
Eq.~(\ref{eq:currents}), is based on the conventional Dirac transition
current operator, there is no conventional magnetic-moment current
source, the lower component in Eq.~(\ref{eq:currents}), that we are
aware of.  Thus, we shall conjecture what the corresponding operator can
be, within the known constraints.  One constraint is that the operator
has even parity, in order to be consistent with the properties outlined
in Table~\ref{tab:pt}. Another constraint is that the current should
reduce to the classical limit $\bm K(\bm x) = \sigma(\bm x)\,\bm v$, as
given in Eq.~(\ref{eq:kdef}) with $\bm x_0 = 0$, where
\begin{eqnarray}
c\,\sigma(\bm x) &=& -\bm m\cdot\bm\nabla\,\delta(\bm x) \, .
\end{eqnarray}
The nonrelativistic correspondence is
\begin{eqnarray}
\bm v &\rightarrow& \frac{\bm p}{m_{\rm e}} \, ,
\nonumber \\[10 pt]
\bm m\cdot\bm \nabla &\rightarrow& \left(\frac{e\hbar}{2m_{\rm e}}
\,\bm\sigma\right)\cdot\left(\frac{\ri}{\hbar}\,\bm p\right)
=\frac{\ri\,e}{2m_{\rm e}}\,\bm\sigma\cdot\bm p \, .
\end{eqnarray}
We thus expect the nonrelativistic form to be
\begin{eqnarray}
\bm k^{if}(x) &\rightarrow& 
-\frac{\ri\,e}{2m_{\rm e}^2c}\,\varphi_f^\dagger(\bm x)\,\bm p\,
\bm\sigma\cdot\bm p \, \varphi_i(\bm x) 
\,\re^{-\ri\omega_{if}t} . \qquad
\end{eqnarray}
A simple candidate Dirac operator is
\begin{eqnarray}
\bm\omega &=& \frac{\bm p}{2m_{\rm e}c}\,\eta =
\frac{\bm p}{2m_{\rm e}c}\left(\begin{array}{c@{\quad}c}
\bm 0 & \bm I \\ \bm I & \bm 0 
\end{array}\right) \, ,
\end{eqnarray}
which has even parity, because a parity change gives
\begin{eqnarray}
\bm p &\rightarrow& -\bm p
\nonumber\\
\bm \omega &\rightarrow& - \beta\,\bm\omega\,\beta = \bm\omega \, ,
\end{eqnarray}
in contrast to the charge current operator $\bm\alpha$ which has odd
parity, because
\begin{eqnarray}
\bm \alpha &\rightarrow& \beta\,\bm\alpha\,\beta = -\bm \alpha \, .
\end{eqnarray}
We also have
\begin{eqnarray}
\bm\nabla\times\bm\omega = 0 \, ,
\end{eqnarray}
so the operator is longitudinal.  In the nonrelativistic dipole
approximation
\begin{eqnarray}
&&\int\rd\bm x \,\phi_f^\dagger(\bm x)\,\bm\omega\,
\re^{-\ri\,\bm k\cdot\bm x}\,
\phi_i(\bm x)\nonumber\\[10 pt]&&\qquad\rightarrow
\frac{1}{(2m_{\rm e}c)^2}
\int\rd\bm x \,
\varphi_f^\dagger(\bm x)
\left(\bm p \,\bm\sigma\cdot\bm p
+\bm\sigma\cdot\bm p \, \bm p\right)
\varphi_i(\bm x) 
\nonumber\\[10 pt]&&\qquad=
\frac{1}{2(m_{\rm e}c)^2}
\int\rd\bm x \,
\varphi_f^\dagger(\bm x)
\,\bm p \,\bm\sigma\cdot\bm p\,
\varphi_i(\bm x) 
\end{eqnarray}
as expected, so we assume
\begin{eqnarray}
\bm k^{if}(x) &=& 
-\ri\,ec\,\phi_f^\dagger(x)\,
\bm\omega \, \phi_i(x) \, .
\end{eqnarray}
We thus have the extended source
\begin{eqnarray}
\iX_{\rm D}(x) &=& 
\frac{e}{\epsilon_0}
\left(\begin{array}{c} 
\phi_f^\dagger(\bm x)\bm\alpha\rs\phi_i(\bm x)\re^{-\ri\omega_{if}t} \\
\phi_f^\dagger(\bm x)\bm \omega\rs\phi_i(\bm x)\re^{-\ri\omega_{if}t}
\end{array}\right) \, .
\end{eqnarray}

The electric and magnetic fields radiated by the source transition
currents are given by
\begin{eqnarray}
\iP_{\rm D}(x_2) &=& \int \rd^4 x_1 \,
\cD\rM(x_2 - x_1)\,\iX_{\rm D}(x_1),
\label{eq:radiation}
\end{eqnarray}
where, from Eq.~(499) in \citet{2010043},
\begin{eqnarray} \cD\rM(x_2 - x_1)
&=&
\frac{\ri}{(2\pi)^4} \, \int_{\rm C_F} \rd^4 k \
\frac{\re^{-\ri k\cdot (x_2-x_1)}} {\gamma^\mu k_\mu -
{\cal I}\epsilon}   \qquad
\label{eq:mprop}
\end{eqnarray}
is the covariant Green function for the Maxwell equation in
(\ref{eq:mmeq}).  Recall that here, the gamma matrices are the
$6\times6$ versions defined in Eq.~\eqref{eq:ggammas}.  In
Eq.~(\ref{eq:radiation}) $\rd^4 x_1 = c\,\rd t_1 \rd\bm x_1$ and in
Eq.~(\ref{eq:mprop}) the exponent is $-\ri\,k\cdot(x_2-x_1) =
-\ri\,k_0(ct_2-ct_1) + \ri\,\bm k\cdot(\bm x_2 -\bm x_1)$, ${\cal I}$ is
the $6\times6$ identity matrix, and $\epsilon$ is an infinitesimal to
avoid a zero in the denominator.  In Eq.~(\ref{eq:mprop}), ${\rm C_F}$
indicates that the contour of integration over $k_0$ is the Feynman
contour, which passes from $-\infty$ below the negative real axis,
through $0$, and above the positive real axis to $+\infty$; this can be
implemented by making the replacement $k_0\rightarrow k_0(1+\ri\delta)$
in the denominator and integrating $k_0$ along the real axis.
Integration over $t_1$ and $k_0$ yields
\begin{widetext}
\begin{eqnarray}
\int \rd k_0 \,f(k_0)\int \rd ct_1 \, 
\re^{-\ri\,k_0(ct_2-ct_1)}
\,\re^{-\ri\omega_{if}t_1}
&=&
2\pi\int\rd k_0 \,f(k_0)\,\re^{-\ri\,k_0ct_2}
\,\delta(k_0 - \omega_{if}/c)
= 2\pi f(\omega_{if}/c)\,\re^{-\ri\,\omega_{if}t_2}
\end{eqnarray}
and
\begin{eqnarray}
\iP_{\rm D}(x_2) &=& \frac{\ri\,e}{(2\pi)^3\epsilon_0}
\int\rd\bm x_1\int\rd\bm k \, 
\frac{\re^{\ri\bm k\cdot(\bm x_2-\bm x_1)}}
{\gamma^0k_0\left(1+\ri\delta\right)-\bm\gamma\cdot\bm k}
\left(\begin{array}{c} 
\phi_f^\dagger(\bm x_1)\bm\alpha\rs\phi_i(\bm x_1) \\
\phi_f^\dagger(\bm x_1)\bm \omega\rs\phi_i(\bm x_1)
\end{array}\right) 
\re^{-\ri \omega_{if}t_2} \, ,
\qquad
\end{eqnarray}
where $k_0 = \omega_{if}/c$.  The fields defined by $\iP_{\rm D}$ may be
separated into transverse $\iP_{\rm D}\rT$ and longitudinal $\iP_{\rm
D}\rL$ components by applying the projection operators defined in
Appendix~\ref{app:tl}.  We thus have
\begin{eqnarray}
\iP_{\rm D}(x_2) &=& \iP_{\rm D}\rT(x_2) + \iP_{\rm D}\rL(x_2) \, ,
\end{eqnarray}
where
\begin{eqnarray}
\iP_{\rm D}\rT(x_2) = \iPi\rT(\bm \nabla_2) \iP_{\rm D}(x_2) \, ,
&\qquad&
\iP_{\rm D}\rL(x_2) = \iPi\rL(\bm \nabla_2) \iP_{\rm D}(x_2) \, .
\end{eqnarray}

For the transverse component
\begin{eqnarray}
\iP_{\rm D}\rT(x_2) &=&
\frac{\ri\,e}{(2\pi)^3\epsilon_0}
\int\rd\bm x_1\int\rd\bm k \,
\iPi\rT(\bm \nabla_2) \, 
\frac{\re^{\ri\bm k\cdot(\bm x_2-\bm x_1)}}
{\gamma^0k_0\left(1+\ri\delta\right)-\bm\gamma\cdot\bm k}
\left(\begin{array}{c} 
\phi_f^\dagger(\bm x_1)\bm\alpha\rs\phi_i(\bm x_1) \\
\phi_f^\dagger(\bm x_1)\bm \omega\rs\phi_i(\bm x_1)
\end{array}\right) 
\re^{-\ri \omega_{if}t_2}
\nonumber\\[5 pt]&=&
\frac{\ri\,e}{(2\pi)^3\epsilon_0}
\int\rd\bm x_1\int\rd\bm k \,
\iPi\rT(\bm{\hat k}) \,
\frac{\re^{\ri\bm k\cdot(\bm x_2-\bm x_1)}}
{\gamma^0k_0\left(1+\ri\delta\right)-\bm\gamma\cdot\bm k}
\left(\begin{array}{c}
\phi_f^\dagger(\bm x_1)\bm\alpha\rs\phi_i(\bm x_1) \\
\phi_f^\dagger(\bm x_1)\bm \omega\rs\phi_i(\bm x_1)
\end{array}\right)
\re^{-\ri \omega_{if}t_2} ,
\end{eqnarray}
where
\begin{eqnarray}
\iPi\rT(\bm{\hat k})
\frac{1}{\gamma^0 k_0(1+\ri\delta) - \bm\gamma\cdot\bm k}
&=&
\iPi\rT(\bm{\hat k}) \,
\frac{\gamma^0 k_0 - \bm\gamma\cdot\bm k}
{k_0^2 - \bm k^2 + \ri \delta}
=\left(\begin{array}{cc}k_0\,(\bm\tau\cdot\bm{\hat k})^2 &
-\bm\tau\cdot\bm{k} \\ 
\bm\tau\cdot\bm{k} &
-k_0\,(\bm\tau\cdot\bm{\hat k})^2 \end{array}\right)
\frac{1}{k_0^2-\bm k^2 + \ri \delta} \, .
\label{eq:trgfid}
\end{eqnarray}
Note that $\iPi\rT(\bm{\hat k})(\bm\gamma\cdot\bm k)^2 = \bm k^2{\cal I}$,
although $(\bm\gamma\cdot\bm k)^2 \ne \bm k^2{\cal I}$.  Thus
\begin{eqnarray}
\int\rd\bm k \,\iPi\rT(\bm{\hat k})
\frac{\re^{\ri\bm k\cdot(\bm x_2-\bm x_1)}}
{\gamma^0 k_0(1+\ri\delta) - \bm\gamma\cdot\bm k}
&=&
\left(\begin{array}{cc}k_0\,\bm 
\iPi\rs\rT(\bm\nabla_2) &
\ri\,\bm\tau\cdot\bm\nabla_2 \\
-\ri\,\bm\tau\cdot\bm\nabla_2 &
-k_0\,\bm \iPi\rs\rT(\bm\nabla_2) \end{array}\right)
\int\rd\bm k \,
\frac{\re^{\ri\bm k\cdot(\bm x_2-\bm x_1)}}
{k_0^2 - \bm k^2 + \ri\,\delta}
\nonumber\\[10 pt]&=& 
 - 2\pi^2
\left(\begin{array}{cc}k_0\,\bm 
\iPi\rs\rT(\bm\nabla_2) &
\ri\,\bm\tau\cdot\bm\nabla_2 \\
-\ri\,\bm\tau\cdot\bm\nabla_2 &
-k_0\,\bm \iPi\rs\rT(\bm\nabla_2) \end{array}\right)
\frac{\re^{\ri\,k_0|\bm x_2-\bm x_1|}}
{|\bm x_2-\bm x_1|} ,
\end{eqnarray}
where ${\rm Im}\left(k_0^2+\ri\,\delta\right)^{1/2}>0$,
$\left(k_0^2+\ri\,\delta\right)^{1/2}\rightarrow|k_0| = \omega_{if}/c$,
and
\begin{eqnarray}
\iP_{\rm D}\rT(x_2) &=& -
\frac{\ri\,e}{4\pi\epsilon_0}
\left(\begin{array}{cc}k_0\,\bm 
\iPi\rs\rT(\bm\nabla_2) &
\ri\,\bm\tau\cdot\bm\nabla_2 \\
-\ri\,\bm\tau\cdot\bm\nabla_2 &
-k_0\,\bm \iPi\rs\rT(\bm\nabla_2) \end{array}\right)
\int\rd\bm x_1
\frac{\re^{\ri\,k_0(|\bm x_2-\bm x_1|-ct_2)}}
{|\bm x_2-\bm x_1|}
\left(\begin{array}{c} 
\phi_f^\dagger(\bm x_1)\bm\alpha\rs\phi_i(\bm x_1) \\
\phi_f^\dagger(\bm x_1)\bm \omega\rs\phi_i(\bm x_1)
\end{array}\right) \, .
\end{eqnarray}

For the longitudinal component
\begin{eqnarray}
\iPi\rL(\bm\nabla_2) \,
\int\rd\bm k \, 
\frac{\re^{\ri\bm k\cdot(\bm x_2-\bm x_1)}}
{\gamma^0k_0\left(1+\ri\delta\right)-\bm\gamma\cdot\bm k}
&=&
\int\rd\bm k \, 
\iPi\rL(\bm{\hat k}) \,
\frac{\re^{\ri\bm k\cdot(\bm x_2-\bm x_1)}}
{\gamma^0k_0\left(1+\ri\delta\right)-\bm\gamma\cdot\bm k}  ,
\qquad
\end{eqnarray}
where
\begin{eqnarray}
\iPi\rL(\bm{\hat k})
\frac{1}{\gamma^0 k_0(1+\ri\delta) - \bm\gamma\cdot\bm k}
&=&
\iPi\rL(\bm{\hat k}) \, \frac{1}{\gamma^0 k_0(1+\ri\delta)}
\rightarrow 
\iPi\rL(\bm{\hat k}) \, \frac{1}{k_0}\,\gamma^0 .
\label{eq:lld}
\end{eqnarray}
Thus
\begin{eqnarray}
\int\rd\bm k \,\iPi\rL(\bm{\hat k})
\frac{\re^{\ri\bm k\cdot(\bm x_2-\bm x_1)}}
{\gamma^0 k_0(1+\ri\delta) - \bm\gamma\cdot\bm k}
&=& \frac{1}{k_0}
\left(\begin{array}{cc}\bm \iPi\rs\rL(\bm\nabla_2) & 
\bm 0 \\ \bm 0 &
-\bm \iPi\rs\rL(\bm\nabla_2) \end{array}\right)
\int\rd\bm k \, \re^{\ri\bm k\cdot(\bm x_2-\bm x_1)}
\nonumber\\[10 pt]&=&
 - \frac{4\pi}{k_0}
\left(\begin{array}{cc}\bm \iPi\rs\rL(\bm\nabla_2) &
\bm 0 \\ \bm 0 &
-\bm \iPi\rs\rL(\bm\nabla_2) \end{array}\right)
\delta(\bm x_2 - \bm x_1) , \qquad
\end{eqnarray}
and
\begin{eqnarray}
\iP_{\rm D}\rL(x_2) &=& -
\frac{\ri\,e}{2\pi^2\epsilon_0k_0}
\left(\begin{array}{cc}\bm \iPi\rs\rL(\bm\nabla_2) & 
\bm 0 \\ \bm 0 &
-\bm \iPi\rs\rL(\bm\nabla_2) \end{array}\right)
\left(\begin{array}{c} 
\phi_f^\dagger(\bm x_2)\bm\alpha\rs\phi_i(\bm x_2) \\
\phi_f^\dagger(\bm x_2)\bm \omega\rs\phi_i(\bm x_2)
\end{array}\right) \re^{-\ri k_0 ct_2} .
\label{eq:longrad}
\end{eqnarray}

We calculate the radiated power by considering the radiation passing
outward through the surface of a sphere centered at the atom, with
radius $R$.  The relevant quantity is the radial power at the surface
for which $|\bm x_2| = R$, as $R\rightarrow\infty$.  In this limit, the
longitudinal component of the fields may be neglected because of the
exponential falloff of the atom's wavefunction, according to
Eq.~(\ref{eq:longrad}).  

For the transverse component, we have, with $\bm r = \bm x_2-\bm x_1$,
[see Appendix E in \citet{2010043}]
\begin{eqnarray}
\bm \iPi\rs\rT(\bm\nabla_2) \,\frac{\re^{\ri k_0r}}{r}
&=& (\bm\tau\cdot\bm{\hat r})^2\,\frac{\re^{\ri k_0r}}{r}
\left[1+{\cal O}\left(\frac{1}{k_0r}\right)\right]  ,
\nonumber\\[10 pt]
\bm\tau\cdot\bm\nabla_2\,\frac{\re^{\ri k_0r}}{r} &=& 
\ri\,k_0\,\bm\tau\cdot\bm{\hat r} \, \frac{\re^{\ri k_0r}}{r}
\left[1+{\cal O}\left(\frac{1}{k_0r}\right)\right] \, .
\end{eqnarray}
We here define a vector $\bm k = k_0 \bm{\hat x}_2$, so that
\begin{eqnarray}
k_0r = kx_2 - \bm k\cdot \bm x_1 + \dots
\end{eqnarray}
and thus
\begin{eqnarray}
\iP_{\rm D}\rT(x_2) &=& -
\frac{\ri ek}{4\pi\epsilon_0}
\,\re^{\ri (kx_2-\omega_{if}t_2)}
\int\rd\bm x_1 \, \frac{\re^{-\ri\,\bm k\cdot\bm x_1}}{r}
\left(\begin{array}{cc}
(\bm\tau\cdot\bm{\hat r})^2 &
-\bm\tau\cdot\bm{\hat r} \\
\bm\tau\cdot\bm{\hat r} &
-(\bm\tau\cdot\bm{\hat r})^2 \end{array}\right)
\left(\begin{array}{c} 
\phi_f^\dagger(\bm x_1)\bm\alpha\rs\phi_i(\bm x_1) \\
\phi_f^\dagger(\bm x_1)\bm \omega\rs\phi_i(\bm x_1)
\end{array}\right) + \dots \, .
\end{eqnarray}
With $ \bm r
\approx\bm x_2 = \bm k$, we have
\begin{eqnarray}
\iP_{\rm D}\rT(x_2) &=& -
\frac{\ri ek}{4\pi\epsilon_0}
\,\frac{\re^{\ri (kx_2-\omega_{if}t_2)}}{x_2}
\left(\begin{array}{cc}
(\bm\tau\cdot\bm{\hat k})^2 &
-\bm\tau\cdot\bm{\hat k} \\
\bm\tau\cdot\bm{\hat k} &
-(\bm\tau\cdot\bm{\hat k})^2 \end{array}\right)
\int\rd\bm x_1 \, \re^{-\ri\,\bm k\cdot\bm x_1}
\left(\begin{array}{c} 
\phi_f^\dagger(\bm x_1)\bm\alpha\rs\phi_i(\bm x_1) \\
\phi_f^\dagger(\bm x_1)\bm \omega\rs\phi_i(\bm x_1)
\end{array}\right) + \dots \, .
\label{eq:trad}
\end{eqnarray}
With $\rd A = x_2^2\,\rd\iO$, the radiated power is
\begin{eqnarray}
P = \int\rd\iO_k\,x_2^2\,
\bm{\hat k}\cdot \bm S(\bm x_2) \, ,
\end{eqnarray}
or
\begin{eqnarray}
&&\frac{\rd P}{\rd\iO_k} = \frac{x_2^2}{2c\mu_0}\,
\overline{\iP}_{\rm D}\rT(\bm x_2)\bm\gamma\cdot\bm{\hat k} \,
\iP_{\rm D}\rT(\bm x_2)
\nonumber\\[10 pt]&&\quad = \frac{\alpha\hbar\omega_{if}^2}{8\pi}
\int\rd\bm x_1\,\re^{\ri\,\bm k\cdot\bm x_1}
\begin{array}{cc}
\left(\phi_i^\dagger(\bm x_1) 
\bm\alpha\rs^\dagger\phi_f(\bm x_1) \right.&
\left.\phi_i^\dagger(\bm x_1) 
\bm\omega\rs^\dagger\phi_f(\bm x_1) \right)\\
& \end{array}
W(\bm{\hat k})
\int\rd\bm x_1^\prime \,\re^{-\ri\,\bm k\cdot\bm x_1^\prime}
\left(\begin{array}{c} 
\phi_f^\dagger(\bm x_1^\prime)\bm\alpha\rs\phi_i(\bm x_1^\prime) \\
\phi_f^\dagger(\bm x_1^\prime)\bm \omega\rs\phi_i(\bm x_1^\prime)
\end{array}\right) . \qquad
\end{eqnarray}
Taking into account $(\bm\tau\cdot\bm{\hat k})^4 =
(\bm\tau\cdot\bm{\hat k})^2$ and $(\bm\tau\cdot\bm{\hat k})^3 =
\bm\tau\cdot\bm{\hat k}$, we have
\begin{eqnarray}
W(\bm{\hat k}) &=& 
\left(\begin{array}{cc}
(\bm\tau\cdot\bm{\hat k})^2 &
\bm\tau\cdot\bm{\hat k} \\
-\bm\tau\cdot\bm{\hat k} &
-(\bm\tau\cdot\bm{\hat k})^2 \end{array}\right)
\left(\begin{array}{cc}
\bm I & \bm 0 \\
\bm 0 & -\bm I
\end{array}\right)
\left(\begin{array}{cc}
\bm 0 & \bm\tau\cdot\bm{\hat k} \\
-\bm\tau\cdot\bm{\hat k} &
\bm 0 \end{array}\right)
\left(\begin{array}{cc}
(\bm\tau\cdot\bm{\hat k})^2 &
-\bm\tau\cdot\bm{\hat k} \\
\bm\tau\cdot\bm{\hat k} &
-(\bm\tau\cdot\bm{\hat k})^2 \end{array}\right)
\nonumber\\[10 pt]&=& 
2 \left(\begin{array}{cc}
(\bm\tau\cdot\bm{\hat k})^2 &
-\bm\tau\cdot\bm{\hat k} \\
-\bm\tau\cdot\bm{\hat k} &
(\bm\tau\cdot\bm{\hat k})^2 \end{array}\right) \, .
\end{eqnarray}
Thus
\begin{eqnarray}
\frac{\rd P}{\rd\iO_k} &=&
\frac{\alpha\hbar\omega_{if}^2}{4\pi}
\bigg\{
\int\rd\bm x\,
\phi_i^\dagger(\bm x) 
\bm\alpha\rs^\dagger\,
\re^{\ri\,\bm k\cdot\bm x}
\phi_f(\bm x)
(\bm\tau\cdot\bm{\hat k})^2 
\int\rd\bm x^\prime \,
\phi_f^\dagger(\bm x^\prime)\bm\alpha\rs\,
\re^{-\ri\,\bm k\cdot\bm x^\prime}
\phi_i(\bm x^\prime)
\nonumber\\[5 pt]&&\qquad\quad
-\int\rd\bm x\,
\phi_i^\dagger(\bm x) 
\bm\alpha\rs^\dagger\,
\re^{\ri\,\bm k\cdot\bm x}
\phi_f(\bm x) \,
\bm\tau\cdot\bm{\hat k}
\int\rd\bm x^\prime \,
\phi_f^\dagger(\bm x^\prime)\bm\omega\rs\,
\re^{-\ri\,\bm k\cdot\bm x^\prime}
\phi_i(\bm x^\prime)
\nonumber\\[5 pt]&&\qquad\quad
-\int\rd\bm x\,
\phi_i^\dagger(\bm x) 
\bm\omega\rs^\dagger\,
\re^{\ri\,\bm k\cdot\bm x}
\phi_f(\bm x) \,
\bm\tau\cdot\bm{\hat k}
\int\rd\bm x^\prime \,
\phi_f^\dagger(\bm x^\prime)\bm\alpha\rs\,
\re^{-\ri\,\bm k\cdot\bm x^\prime}
\phi_i(\bm x^\prime)
\nonumber\\[5 pt]&&\qquad\quad
+\int\rd\bm x\,
\phi_i^\dagger(\bm x) 
\bm\omega\rs^\dagger\,
\re^{\ri\,\bm k\cdot\bm x}
\phi_f(\bm x)
(\bm\tau\cdot\bm{\hat k})^2 
\int\rd\bm x^\prime \,
\phi_f^\dagger(\bm x^\prime)\bm\omega\rs\,
\re^{-\ri\,\bm k\cdot\bm x^\prime}
\phi_i(\bm x^\prime)
\bigg\} \, . \qquad
\label{eq:total}
\end{eqnarray}
The first term is radiation from the current $\bm j$ [see
Appendix~\ref{app:pv}]
\begin{eqnarray}
\frac{\rd P_{\bm j}}{\rd\iO_k} &=&
\frac{\alpha\hbar\omega_{if}^2}{4\pi} \sum_{\lambda=1}^2
\int\rd\bm x\,
\phi_i^\dagger(\bm x) \,
\bm\alpha_{\rm s}^\dagger\cdot\bm{\hat\epsilon}_\lambda(\bm{\hat k})\,
\re^{\ri\,\bm k\cdot\bm x}
\,\phi_f(\bm x)
\int\rd\bm x^\prime \,
\phi_f^\dagger(\bm x^\prime)
\,\bm{\hat\epsilon}_\lambda^\dagger(\bm{\hat k})\cdot\bm\alpha_{\rm s} \,
\re^{-\ri\,\bm k\cdot\bm x^\prime}
\phi_i(\bm x^\prime) \qquad
\end{eqnarray}
corresponding to a transition rate of
\begin{eqnarray}
A_{if}^{(\bm j)} &=&
\frac{\alpha\hbar\omega_{if}}{4\pi} \sum_{\lambda=1}^2
\int\rd\bm x\,
\phi_i^\dagger(\bm x) \,
\bm\alpha_{\rm s}^\dagger\cdot\bm{\hat\epsilon}_\lambda(\bm{\hat k})\,
\re^{\ri\,\bm k\cdot\bm x}
\,\phi_f(\bm x)
\int\rd\bm x^\prime \,
\phi_f^\dagger(\bm x^\prime)
\,\bm{\hat\epsilon}_\lambda^\dagger(\bm{\hat k})\cdot\bm\alpha_{\rm s} \,
\re^{-\ri\,\bm k\cdot\bm x^\prime}
\phi_i(\bm x^\prime) \qquad
\label{eq:qedrate}
\end{eqnarray}
\end{widetext}
according to Eq.~(\ref{eq:rate}).  This is exactly the relativistic
transition rate predicted by QED [see, for example, \citet{1974003,
2010043}] which is ${\cal O}\big(\alpha(Z\alpha)^4\big)$~s$^{-1}$ for
allowed electric dipole transitions.  The factor of two difference
between Eq.~(\ref{eq:qedrate}) and Eq.~(F.5) in \citet{2010043} is due
to the fact that the latter rate is a sum over two final spin-direction
states of the electron, while Eq.~\eqref{eq:qedrate} is for a single
final state.  This was overlooked in \citet{2010043}.

The second and third terms in Eq.~(\ref{eq:total}), nominally ${\cal
O}\big(\alpha(Z\alpha)^5\big)$~s$^{-1}$, are suppressed by an additional
power of $Z\alpha$ because they vanish when integrated over $\rd\iO_k$
in the dipole approximation, which gives the leading contribution.  The
fourth term is also ${\cal O}\big(\alpha(Z\alpha)^6\big)$~s$^{-1}$.
Thus the magnetic source current contributes at the level of
relativistic corrections, or of relative order $5\times10^{-5}$, which
is much smaller than the uncertainty of the measured rate for the
$2p\rightarrow1s$ transition~\cite{ak}.

\section{Summary}
\label{sec:s}

In Sec.~\ref{sec:pfi}, the exchange of energy between fields and a
particle in terms of the work done by the fields on the particle, and
vice versa, is described.  Sec.~\ref{sec:pt} states the conventional
Poynting theorem, including the role of the Poynting vector.  This
section also notes that the interaction of a magnetic moment with an
external magnetic field is not explicitly taken into account in the
theorem.  In Sec.~\ref{sec:mfpi}, the forces on a magnetic moment by
both transverse and longitudinal magnetic fields are considered.  The
definitions of these terms are reviewed in Appendix~\ref{app:tl}.  It is
shown that the force is the same in either case, although the
formulations are not the same.  An extension of the Poynting theorem to
take this force into account is provided.  Sec.~\ref{sec:eme} shows how
the Maxwell equations may be modified to be consistent with the extended
Poynting theorem.  The extended Maxwell equations contain a magnetic
moment source and the corresponding current.  In Sec.~\ref{sec:peme},
issues associated with the modifications described in
Secs.~\ref{sec:mfpi} and \ref{sec:eme} are expanded upon.
Sec.~\ref{sec:av} shows that the magnetic moment current in the Maxwell
equations associated with the magnetic moment source may be derived by
making a Lorentz transformation of the fields associated with the source
from its rest frame to a moving frame.  In Sec.~\ref{sec:mme}, the
matrix form of the Maxwell equations, as discussed by \citet{2010043},
is reviewed to provide for its use in Sec.~\ref{sec:li} where it
facilitates the otherwise tedious algebra to show that the extended
Maxwell equations are indeed Lorentz invariant.  That section also
reviews the Lorentz transformations of four vectors and electromagnetic
fields.  Sec.~\ref{sec:rpt} provides a relativistic derivation of the
extended Poynting theorem based on the extended Maxwell equations.  The
derivation takes a particularly simple form when the matrix formulation
of the Maxwell equations is employed.  Moreover, it implicitly shows
that the extended Poynting theorem is relativistically invariant because
it follows from the Maxwell equations which in turn are relativistically
invariant.  A comparison of two classical models of the magnetic moment
field is provided in Sec.~\ref{sec:cmdmm}.  The current loop (transverse
field) and the dual magnetic monopole (longitudinal field) models are
discussed.  Comparison of Eqs.~(\ref{eq:newlong}) and (\ref{eq:clm})
shows that the two models differ only by a delta function at the
location of the magnetic moment source.  This section examines the role
of electromagnetic field energy in classical mechanics.  The electric
and magnetic interactions of particles are derived in terms of the
energy of the combined fields of the particles.  The electric
interaction is just the conventional electric interaction of two
particles.  The longitudinal field interaction is the same as the
conventional interaction, except that the contact interaction term is
different from one derived for transverse fields.  It is shown in
Sec.~\ref{ssec:hfs} that this does not conflict with the hyperfine
contact term which arises from the nonrelativistic reduction of the
Dirac hyperfine expression.  The transverse classical magnetic
interaction calculated as a field energy reproduces the conventional
classical result, but with the opposite sign.  This is due to the $|\bm
E|^2 - |c\bm B|^2$ form of the energy for transverse magnetic field
interactions.  If the Poynting theorem is used, the associated $+ |c\bm
B|^2$ gives the opposite sign.  This section also considers the
classical self energy due to electromagnetic fields.  An interesting
result is that the electric self energy of a charged point source
outside of the classical Bohr radius is exactly the nonrelativistic
binding energy of an electron, including the correct dependence on the
principal quantum number.  Another interesting result is that the
magnetic self energy for the same cutoff is of the order of relativistic
corrections to the electron energy levels.  Moreover, with a lower
cutoff of $\lbar_{\rm C}/12$, the magnetic self energy is just the
energy equivalent of the electron mass.  The same is true for the muon
mass with a cutoff of $\lbar_{\rmssmu}/12$.  Sec.~\ref{sec:ptde} reviews
the Dirac equation and shows that the interaction terms for external
fields can be derived by considering only the field energy of the
electron and the external sources.  The terms obtained with the usual
minimal coupling substitution are reproduced in this way.  The same sign
issue associated with transverse magnetic fields appears here also.  A
longitudinal external field interaction is proposed and it takes a form
that differs from the transverse interaction, although both forms are
shown to reproduce the correct hyperfine structure interaction.  In
Sec.~\ref{sec:iiq} the QED interaction given by $j_\mu A^\mu$ is shown
to be equivalent to the field energy expression proportional to the
interaction part of $|\bm E|^2 - |c\bm B|^2$ for external fields in the
Dirac equation.  The QED one-photon exchange correction is also shown to
be of the same form.  In Sec.~\ref{sec:rd} the effect of the
magnetic-moment current on electromagnetic radiation is examined.  There
is no effect on the propagation of radiation in vacuum and the leading
term for semi-classical radiation for an E1 transition is exactly the
same as the result from QED.  This resolves the factor of two disparity
sometimes associated with such a comparison [see the footnote on page
407 of \citet{1998165}].  However, there is an effect on the rate that
is of the order of relativistic corrections, which is smaller than the
uncertainty of the measured transition rate.

\section{Conclusion}
\label{sec:c}

The role of magnetic moments in electrodynamics has been shown to
warrant scrutiny.  Interactions of the moments are described in the
context of conventional quantum electrodynamics where the magnetic
energy density enters as a negative quantity.  On the other hand, in the
context of the extended Poynting theorem and extended Maxwell equations,
the magnetic energy density is positive, in keeping with intuitive
expectations.  We have shown how magnetic moment effects are included in
either version of electrodynamics.

This work also shows how the interaction of a magnetic dipole moment may
be taken into account in the extended Poynting theorem, which in turn
corresponds to magnetic moment source terms in the Maxwell equations.
This addition to the Maxwell equations is shown to be consistent with
special relativity.  It is shown the the magnetic moment field can
contribute to a particle mass to a greater degree than the electric
field.  Also shown is that the extended Poynting theorem provides a
formulation of interaction terms in the Dirac equation, without the use
of potentials, which is contrary to expectations.  Interactions in QED
are shown to be expressable in terms of electromagnetic fields based on
the extended Poynting theorem.  It is also shown that the rate of
radiative decay of an atom based on a classical radiation calculation
from a Dirac source current gives exact agreement with the QED result.

On the other hand, this work does not uniquely identify an operator
corresponding to longitudinal magnetic dipole moments in the Dirac
equation, but rather postulates a candidate operator that is not
necessarily an exact result.  Also, application of the Poynting theorem
to formulate QED, while a promising prospect, is not yet shown to solve
the problem of infinities, a topic that remains to be investigated.

\appendix

\section{Transverse vs longitudinal fields}
\label{app:tl}

The separation of vector fields into transverse and longitudinal
components is based on the identity~\cite{1998165,j}
\begin{eqnarray}
\bm\nabla\bm\times[ \bm\nabla\bm\times\bm F(\bm x)] &=&
\bm\nabla[\bm\nabla\bm\cdot\bm F(\bm x)] 
-\bm\nabla^2\bm F(\bm x)  . \qquad
\end{eqnarray}
If we define the components
\begin{eqnarray}
\bm F^{\rm T}(\bm x) &=& \frac{1}{4\pi}\int\rd\bm x^\prime \,
\frac{1}{|\bm x - \bm x^\prime|}
\,\bm\nabla^\prime\bm\times[\bm\nabla^\prime\bm\times\bm F(\bm x^\prime)]  ,
\nonumber\\[10 pt]
 &=& \frac{1}{4\pi}\,\bm\nabla\bm\times\int\rd\bm x^\prime \,
\frac{1}{|\bm x - \bm x^\prime|}
\,[\bm\nabla^\prime\bm\times\bm F(\bm x^\prime)]  , \qquad
\label{eq:trans}
\\[10 pt]
\bm F^{\rm L}(\bm x) &=& -\frac{1}{4\pi}\int\rd\bm x^\prime \,
\frac{1}{|\bm x - \bm x^\prime|}
\,\bm\nabla^\prime[\bm\nabla^\prime\bm\cdot\bm F(\bm x^\prime)]  ,
\nonumber\\[10 pt]
&=& -\frac{1}{4\pi}\,\bm\nabla\int\rd\bm x^\prime \,
\frac{1}{|\bm x - \bm x^\prime|}
\,[\bm\nabla^\prime\bm\cdot\bm F(\bm x^\prime)]  ,
\label{eq:long}
\end{eqnarray}
where the two forms in each of Eqs.~(\ref{eq:trans}) and (\ref{eq:long})
are related through integration by parts, where the surface terms are
assumed to vanish.  The superscripts T and L denote transverse and
longitudinal, and
\begin{eqnarray}
\bm F^{\rm T}(\bm x) + \bm F^{\rm L}(\bm x) &=& 
-\frac{1}{4\pi}\int\rd\bm x^\prime \,
\frac{1}{|\bm x - \bm x^\prime|}
\,\bm\nabla^{\prime2}\, \bm F(\bm x^\prime)
 \nonumber\\[10 pt]&=&
 \bm F(\bm x) \, .
\end{eqnarray}
These components have the properties:
\begin{eqnarray}
\bm\nabla\bm\cdot\bm F^{\rm T}(\bm x) = 0 \, ,
&\qquad&
\bm\nabla\bm\times\bm F^{\rm L}(\bm x) = 0 \, ,
\label{eq:killt}
\end{eqnarray}
\begin{eqnarray}
\int\rd\bm x\,\bm F^{\rm T}(\bm x) \bm\cdot 
\bm F^{\rm L}(\bm x) = 0\, .
\label{eq:orth}
\end{eqnarray}

An equivalent separation can be made by defining $3\times3$ matrix
transverse and longitudinal Hermitian projection operators
$\bm{\iPi}\rs\rT(\bm \nabla)$ and $\bm{\iPi}\rs\rL(\bm \nabla)$ based on
the matrix representation described in Sec.~\ref{sec:mme} and in Sec.~5
of \citet{2010043}.  They are differential-integral operators acting on
coordinate-space functions given by
\begin{eqnarray}
\bm{\iPi}\rs\rT(\bm\nabla) = 
\frac{(\bm\tau\cdot\bm\nabla)^2}{\bm\nabla^2} \, ,
\label{eq:tpjop}
&\qquad&
\bm{\iPi}\rs\rL(\bm\nabla) =
\frac{\bm\nabla\rs\bm\nabla\rs^\dagger}{\bm\nabla^2} \, ,
\label{eq:lpjop}
\end{eqnarray}
which takes into account that fact that $\bm \nabla$ has real Cartesian
components (in the sense that they give real values when acting on a
real function).  The inverse Laplacian is defined by the relation
\begin{eqnarray}
\frac{1}{\bm\nabla^2}\,\bm F\rs(\bm x) &=& -\frac{1}{4\pi} \int{\rm d}\,\bm
x^\prime \, \frac{1}{|\bm x - \bm x^\prime|} \,\bm F\rs(\bm x^\prime),
\label{eq:invlap}
\end{eqnarray}
which yields
\begin{eqnarray}
\bm\nabla^2\,\frac{1}{\bm\nabla^2}\,\bm F\rs(\bm x) &=& -\frac{1}{4\pi} \int{\rm d}\,\bm
x^\prime \, \bm\nabla^2 \frac{1}{|\bm x - \bm x^\prime|} \,\bm F\rs(\bm x^\prime)
 \nonumber\\[10 pt]
 &=& \bm F\rs(\bm x) \, .
\label{eq:nabsnab}
\end{eqnarray}

Based on Eq.~(30) of \citet{2010043}, we have
\begin{eqnarray}
(\bm \tau\cdot\bm\nabla)^2
+ \bm\nabla\rs\bm\nabla\rs^\dagger 
=\bm I\bm\nabla^2 \, ,
\end{eqnarray}
where $I$ is the $3\times3$ identity matrix, so that
\begin{eqnarray}
\bm{\iPi}\rs\rT(\bm\nabla) +
\bm{\iPi}\rs\rL(\bm\nabla) &=& \bm I \, .
\end{eqnarray}
The projection operators have the following
properties:
\begin{eqnarray}
\begin{array}{l@{\ }l}
\left[\bm{\iPi}\rs\rT(\bm\nabla)\right]^2 =
\bm{\iPi}\rs\rT(\bm\nabla) ;
&
\left[\bm{\iPi}\rs\rL(\bm\nabla)\right]^2 =
\bm{\iPi}\rs\rL(\bm\nabla),  \quad
\\
\bm{\iPi}\rs\rT(\bm\nabla) \,
\bm{\iPi}\rs\rL(\bm\nabla) = \bm 0 ;
&
\bm{\iPi}\rs\rL(\bm\nabla) \,
\bm{\iPi}\rs\rT(\bm\nabla) = \bm 0 .
\end{array}
\end{eqnarray}
We also have
\begin{eqnarray}
\bm{\iPi}\rs\rT(\bm\nabla) \, \re^{\ri\,\bm k\cdot\bm x} &=& 
\bm{\iPi}\rs\rT(\bm{\hat k}) \, \re^{\ri\,\bm k\cdot\bm x}  ,
\\[10 pt]
\bm{\iPi}\rs\rL(\bm\nabla) \, \re^{\ri\,\bm k\cdot\bm x} &=& 
\bm{\iPi}\rs\rL(\bm{\hat k}) \, \re^{\ri\,\bm k\cdot\bm x}  .
\end{eqnarray}

Six-dimensional transverse and longitudinal projection operators
are defined by
\begin{eqnarray}
\iPi\rT(\bm \nabla) &=& \left(\begin{array}{ccc}
\bm{\iPi}\rs\rT(\bm\nabla) && \0 \\
\0 && \bm{\iPi}\rs\rT(\bm\nabla) \msp \end{array}\right),
\label{eq:ptop}
\\
\iPi\rL(\bm \nabla) &=& \left(\begin{array}{ccc}
\bm{\iPi}\rs\rL(\bm\nabla) && \0 \\
\0 && \bm{\iPi}\rs\rL(\bm\nabla) \msp \end{array}\right),
\label{eq:plop}
\end{eqnarray}
where $\0$ is the $3\times3$ matrix of zeros, and
\begin{eqnarray}
\iPi\rT(\bm \nabla)+\iPi\rL(\bm \nabla) = \cI \, ,
\end{eqnarray}
where $\cI$ is the $6\times6$ identity matrix.

\section{Lorentz transformation identities}
\label{app:lti}

This Appendix gives some details of the calculation of the identities in
Eqs.~(\ref{eq:firstid}) and (\ref{eq:secondid}).

\subsection{Equation~(\ref{eq:firstid})}
\label{app:lti1}

The product in Eq.~(\ref{eq:firstid}) is
\begin{eqnarray}
\gamma^\mu \partial^{\,\prime}_\mu \, {\cal V}(\bm v)
= \left(\begin{array}{cc} A_{11} & A_{12} \\[10 pt]
-A_{12} & -A_{11} \end{array}\right)  ,
\end{eqnarray}
where $A_{11}$ and $A_{12}$ are $3\bm\times3$ matrices.  Only two are needed
due to the repetition of terms in $\gamma^\mu \partial^{\,\prime}_\mu$
and ${\cal V}$.  The first is
\begin{eqnarray}
A_{11} 
&=&\ft{\partial}{\partial ct^\prime}\left[
\bm I\cosh{\zeta} -  \bm{\hat v}\rs \bm{\hat
  v}\rs^\dagger\left(\cosh{\zeta}-1\right)\right]
\nonumber\\[3 pt]&&\qquad
+\bm\tau\bm\cdot\bm\nabla^\prime 
\bm \tau \bm\cdot \bm{\hat v} \sinh{\zeta}
\nonumber\\[3 pt]&=&
\left[\ft{\partial}{\partial ct}
\cosh{\zeta} -\bm{\hat v}\bm\cdot\bm \nabla\sinh{\zeta}\right]
\nonumber\\[3 pt]&&\qquad\times
\left[
\bm I
  \cosh{\zeta} - \bm{\hat v}\rs \bm{\hat
  v}\rs^\dagger\left(\cosh{\zeta}-1\right)\right]
  \nonumber\\[3 pt] &&
+\bm \tau\bm\cdot\left[
\bm \nabla
 +\bm{\hat v}\bm{\hat v}\bm\cdot \bm \nabla\left(\cosh{\zeta}-1\right)
-\bm{\hat v} \ft{\partial}{\partial ct}\sinh{\zeta}\right]
\nonumber\\[3 pt]&&\qquad\times
\bm \tau \bm\cdot \bm{\hat v} \sinh{\zeta}
\nonumber\\[3 pt]&=&
\left[\bm I + \bm{\hat v}\rs \bm{\hat v}\rs^\dagger
\left(\cosh\zeta-1\right)\right]
 \ft{\partial}{\partial ct}
-\bm{\hat v}\rs\bm \nabla^\dagger\rs \sinh\zeta.
\nonumber\\
\end{eqnarray}
This result follows from expanding the products and taking into account
the relations: $(\bm\tau\bm\cdot\bm{\hat v})^2 = \bm I - \bm{\hat v}\rs
\bm{\hat v}\rs^\dagger$; $\bm\tau\bm\cdot\bm\nabla\,\bm\tau\bm\cdot\bm{\hat v}
= \bm{\hat v}\bm\cdot\bm\nabla\bm I-\bm{\hat v}\rs\bm\nabla\rs^\dagger$; and
$\cosh^2{\zeta} -\sinh^2{\zeta} = 1$.
The second matrix is
\begin{eqnarray}
A_{12} &=&
\bm \tau \bm\cdot \bm{\hat v}\,\ft{\partial}{\partial ct^\prime}
\sinh{\zeta} + \bm\tau\bm\cdot\bm\nabla^\prime 
\nonumber\\[3 pt]&&\qquad\times
\left[ \bm I \cosh{\zeta} - \bm{\hat v}\rs \bm{\hat
  v}\rs^\dagger\left(\cosh{\zeta}-1\right)\right]
\nonumber\\[3 pt]&=&
\bm \tau \bm\cdot \bm{\hat v}
\left[\ft{\partial}{\partial ct}
\cosh{\zeta} -\bm{\hat v}\bm\cdot\bm \nabla\sinh{\zeta}\right]
\sinh{\zeta} 
\nonumber\\[3 pt]&&
+\bm \tau\bm\cdot\left[
\bm \nabla
 +\bm{\hat v}\bm{\hat v}\bm\cdot \bm \nabla\left(\cosh{\zeta}-1\right)
-\bm{\hat v} \ft{\partial}{\partial ct}\sinh{\zeta}\right]
\nonumber\\[3 pt]&&\qquad\times
\left[ \bm I \cosh{\zeta} - \bm{\hat v}\rs \bm{\hat
  v}\rs^\dagger\left(\cosh{\zeta}-1\right)\right]
\nonumber\\[5 pt]&=&
\left[\bm I +\bm{\hat v}\rs\bm{\hat v}\rs^\dagger
(\cosh\zeta - 1)\right]
\bm \tau\bm\cdot \bm \nabla  ,
\end{eqnarray}
where the previously noted relations, together with
$\bm\tau\bm\cdot\bm{\hat v}\,\bm{\hat v}\rs = 0$ and
$\bm\tau\bm\cdot\bm\nabla\bm{\hat v}\rs = -\bm\tau\bm\cdot\bm{\hat
v}\,\bm\nabla\rs$\,, are taken into account.

\subsection{Equation~(\ref{eq:secondid})}
\label{app:lti2}

The product in Eq.~(\ref{eq:secondid}) is
\begin{eqnarray}
{\cal D}^\prime \, {\cal V}(\bm v)
= \left(\begin{array}{cc} B_{11} & B_{12} \\[10 pt]
-B_{12} & -B_{11} \end{array}\right)  ,
\end{eqnarray}
where $B_{11}$ and $B_{12}$ are $1\bm\times3$ matrices.
The first is
\begin{eqnarray}
B_{11}
&=&
-\bm\nabla\rs^{\prime\dagger}
\left[\bm I\cosh{\zeta} -
\bm{\hat v}\rs\bm{\hat v}\rs^\dagger(\cosh{\zeta}-1) \right]
\nonumber\\[3 pt]
&=&-\left[
\bm \nabla\rs^\dagger
 +\bm{\hat v}\bm\cdot \bm \nabla\bm{\hat v}\rs^\dagger
 \left(\cosh{\zeta}-1\right)
-\bm{\hat v}\rs^\dagger\, \ft{\partial}{\partial ct}\sinh{\zeta}
\right]
\nonumber\\[3 pt]&&\qquad\times
\left[\bm I\cosh{\zeta} -
\bm{\hat v}\rs\bm{\hat v}\rs^\dagger(\cosh{\zeta}-1) \right]
\nonumber\\[10 pt]&=&
-\bm \nabla\rs^\dagger \cosh\zeta
+\bm{\hat v}\rs^\dagger
\, \ft{\partial}{\partial ct}\sinh{\zeta} \, ,
\end{eqnarray}
and the second is
\begin{eqnarray}
B_{12} &=&
-\bm\nabla\rs^{\prime\dagger}\,
\bm \tau \bm\cdot \bm{\hat v} \,\sinh{\zeta}
\nonumber\\[3 pt]
&=&-\left[
\bm \nabla\rs^\dagger
 +\bm{\hat v}\bm\cdot \bm \nabla\bm{\hat v}\rs^\dagger
 \left(\cosh{\zeta}-1\right)
-\bm{\hat v}\rs^\dagger\, \ft{\partial}{\partial ct}\sinh{\zeta}
\right]
\nonumber\\[3 pt]&&\qquad\times
\bm \tau \bm\cdot \bm{\hat v} \,\sinh{\zeta}
\nonumber\\[3 pt]&=&
\bm{\hat v}\rs^\dagger\bm \tau \bm\cdot \bm\nabla \,\sinh{\zeta} \, ,
\end{eqnarray}
where the relation $\bm\nabla\rs^\dagger\bm\tau\bm\cdot\bm{\hat v} =
-\bm{\hat v}\rs^\dagger\bm\tau\bm\cdot\nabla$ is taken into account.

\section{Nonrelativistic approximation to the Dirac equation}
\label{app:nade}

The Dirac equation for an electron bound in a spherically symmetric
field is
\begin{eqnarray}
E_n \,\phi_n(\bm x) &=& 
\left[c\,\bm\alpha\bm\cdot\bm p + \beta\,m_{\rm e}c^2
+V(x)\right]\phi_n(\bm x)  , \qquad
\end{eqnarray}
where $x=|\bm x|$.  For a point nucleus with charge $Ze$, the potential
is
\begin{eqnarray}
V(x) &=& -\hbar c\,\frac{Z\alpha}{x} \, .
\end{eqnarray}
In terms of 2-component functions, this is
\begin{eqnarray}
E_n\left(\begin{array}{c}u_n(\bm x) \\ v_n(\bm x) \end{array}\right)
&=& \left(\begin{array}{c@{\qquad}c} m_{\rm e}c^2 + V(x)
& c\,\bm\sigma\bm\cdot\bm p \\
 c\,\bm\sigma\bm\cdot\bm p  & 
 -m_{\rm e}c^2 + V(x)
\end{array}\right)
\nonumber\\[3 pt]&&\times
\left(\begin{array}{c}u_n(\bm x) \\ v_n(\bm x) \end{array}\right)
\end{eqnarray}
The lower equation is 
\begin{eqnarray}
c\,\bm\sigma\bm\cdot\bm p \, u_n(\bm x) &=& 
\left[E_n + m_{\rm e}c^2-V(x)\right]v_n(\bm x)  .
\end{eqnarray}
In the nonrelativistic limit, $E_n\rightarrow m_{\rm e}c^2$ and
$V(x)\rightarrow0$, which gives
\begin{eqnarray}
v_n(\bm x) &\rightarrow& 
\frac{1}{2m_{\rm e}c}\,\bm\sigma\bm\cdot\bm p\,u_n(\bm x) \, .
\end{eqnarray}
The upper equation is
\begin{eqnarray}
c\,\bm\sigma\bm\cdot\bm p \, v_n(\bm x) &=&
\left[E_n - m_{\rm e}c^2-V(x)\right]u_n(\bm x)
 ,
\end{eqnarray}
where
\begin{eqnarray}
E_n - m_{\rm e}c^2 &\rightarrow& E_n^{\rm NR} 
\end{eqnarray}
is the nonrelativistic Schr\"odinger energy, and
\begin{eqnarray}
E_n^{\rm NR}\varphi_n(\bm x) &=& 
\left[\frac{1}{2m_{\rm e}}\,\bm p^2 + V(r)
\right]\varphi_n(\bm x)  ,
\end{eqnarray}
which is just the Pauli-Schr\"odinger equation with $u_n(\bm x)
\rightarrow \varphi_n(\bm x)$, the Pauli-Schr\"odinger wavefunction.
The leading terms are
\begin{eqnarray}
\phi_n(\bm x) &\rightarrow& \left(\begin{array}{c} \varphi_n(\bm x) \\
\frac{\bm\sigma\bm\cdot\bm p}{2m_{\rm e}c}\,\varphi_n(\bm x) 
\end{array}\right) ,
\\[10 pt]
\phi_n^\dagger(\bm x) &\rightarrow& \left(\begin{array}{c@{\quad}c} 
\varphi_n^\dagger(\bm x) &
\varphi_n^\dagger(\bm x)
\frac{\bm\sigma\bm\cdot{\bm p}}{2m_{\rm e}c}
\end{array}\right) .
\end{eqnarray}
For example, this gives
\begin{eqnarray}
\phi_i^\dagger(\bm x)\, \bm\alpha\,\phi_j(\bm x) &\rightarrow&
\frac{1}{2m_{\rm e}c}\,\varphi_i^\dagger(\bm x)\,\left(\bm\sigma\bm\cdot\bm p \,
\bm\sigma + \bm\sigma\bm\sigma\bm\cdot\bm p\right)\varphi_j(\bm x)
\nonumber\\ &=&
\varphi_i^\dagger(\bm x)\,\frac{\bm v}{c}\,\varphi_j(\bm x) \, .
\end{eqnarray}

For a spherically symmetric binding field, the Dirac wavefunction can be
written in the form [see, for example, \citet{1998004}]
\begin{eqnarray}
\phi_n(\bm x) = \left(\begin{array}{c}
                      f_1(x)\,\chi_\kappa^\mu(\hat{\bm x}) \\
            {\ri}   f_2(x)\,\chi_{-\kappa}^\mu(\hat{\bm x})
              \end{array} \right)  ,
\end{eqnarray}
where $f_1$ and $f_2$ are radial wave functions, and
$\chi_\kappa^\mu(\hat{\bm x})$ is the Dirac spin-angle function.  This
function is an eigenfunction of total angular momentum and parity with
the properties
\begin{eqnarray}
\left(\bm\sigma\bm\cdot\bm L + \hbar\bm I\right)
\chi_\kappa^\mu(\hat{\bm x}) &=& -\hbar\,\kappa \,
\chi_\kappa^\mu(\hat{\bm x}) ,
\\
\bm\sigma\bm\cdot\bm{\hat x}\,\chi_\kappa^\mu(\hat{\bm x})
&=&-\chi_{-\kappa}^\mu(\hat{\bm x})  ,
\end{eqnarray}
where $\bm L = \bm x\bm\times\bm p$ is the orbital angular momentum.  In
the nonrelativistic limit, the upper component becomes the Schr\"odinger
wave function
\begin{eqnarray}
f_1(x)\,\chi_\kappa^\mu(\hat{\bm x}) &\rightarrow& 
\varphi(\bm x) =
f(x)\,\chi_\kappa^\mu(\hat{\bm x})  ,
\end{eqnarray}
where $f$ is the Schr\"odinger radial wave function.
For the lower component, we have
\begin{eqnarray}
\ri\,f_2(x)\,\chi_{-\kappa}^\mu(\hat{\bm x}) &\rightarrow&
 \frac{1}{2m_{\rm e}c} \, \bm \sigma\bm\cdot\bm p \,
f(x)\,\chi_\kappa^\mu(\hat{\bm x})  ,
\end{eqnarray}
or
\begin{eqnarray}
f_2(x)\,\chi_\kappa^\mu(\hat{\bm x}) &\rightarrow&
- \frac{1}{2\,\ri\,m_{\rm e}c} \, 
\bm\sigma\bm\cdot\hat{\bm x} \,
\bm \sigma\bm\cdot\bm p \,
f(x)\,\chi_\kappa^\mu(\hat{\bm x})
\nonumber\\&=&
- \frac{1}{2\,\ri\,m_{\rm e}c} \, \left(
\hat{\bm x} \bm\cdot \bm p  + \ri \, \bm\sigma\bm\cdot \hat{\bm x}\bm\times\bm p \right)
f(x)\,\chi_\kappa^\mu(\hat{\bm x})
\nonumber\\&=&
 \frac{1}{2m_{\rm e}c} \, \left(
\hbar\,\frac{\partial}{\partial x} - \frac{\bm\sigma\bm\cdot \bm L}{x} \right)
f(x)\,\chi_\kappa^\mu(\hat{\bm x})
\nonumber\\&=&
 \frac{\hbar}{2m_{\rm e}c} \, \left(
\frac{\partial}{\partial x} + \frac{\kappa + 1}{x} \right)
f(x)\,\chi_\kappa^\mu(\hat{\bm x})  ,
\end{eqnarray}
so that
\begin{eqnarray}
f_2(x) &\rightarrow&
\frac{\lbar_{\rm e}}{2} \left(
\frac{\partial}{\partial x} + \frac{\kappa + 1}{x} \right) f(x)  .
\end{eqnarray}

\section{Hyperfine integral for the 1S state}
\label{app:crhi}

The radial hyperfine integral is
\begin{eqnarray}
I_{\rm hfs} = \int_0^\infty {\rm d}x\,f_1(x)f_2(x)  .
\end{eqnarray}
For the 1S state with nuclear charge $Z$
\begin{eqnarray}
f_1(x) &=& 2 N^\frac{1}{2} (m_{\rm e}c^2+E)^\frac{1}{2} 
(2\gamma x)^{a-1}\re^{-\gamma x}
\label{eq:f1} ,
\\
f_2(x) &=& -2 N^\frac{1}{2} (m_{\rm e}c^2-E)^\frac{1}{2} 
(2\gamma x)^{a-1}\re^{-\gamma x}  ,
\end{eqnarray}
where
\begin{eqnarray}
\gamma &=& \frac{Z\alpha}{\lbar_{\rm e}},
\\
E &=& \left[1-(Z\alpha)^2\right]^\frac{1}{2}\,m_{\rm e}c^2,
\\
a &=& \left[1-(Z\alpha)^2\right]^\frac{1}{2},
\\
N &=& \frac{\gamma^3}{\Gamma(2a+1)m_{\rm e}c^2},
\end{eqnarray}
so that as $Z\alpha\rightarrow0$
\begin{eqnarray}
I_{\rm hfs} &=& -\frac{(Z\alpha)^3}{a(2a-1)\lbar_{\rm e}^2}
\rightarrow - \frac{(Z\alpha)^3}{\lbar_{\rm e}^2}  ,
\\[10 pt]
f_1(x) &\rightarrow& f(x) = 2\gamma^{3/2}\re^{-\gamma x}  ,
\\[10 pt]
f^2(0) &=& 4\left(\frac{Z\alpha}{\lbar_{\rm e}}\right)^3  .
\label{eq:fsq}
\end{eqnarray}

\section{Polarization vectors}
\label{app:pv}

We first consider transverse fields, i.e., fields for
which the electric and magnetic fields are perpendicular to the
wave vector.  The polarization vector is a unit vector
proportional to the electric or magnetic fields, represented by
a three component, possibly complex, vector.

Two polarization vectors, both in the plane perpendicular to
$\bm{\hat k}$, are denoted by
\begin{eqnarray}
\bm{\hat \epsilon}_\lambda(\bm{\hat k}) \, ; \qquad \lambda =
1,2.
\label{eq:epst}
\end{eqnarray}
They have the orthonormality properties
\begin{eqnarray}
\bm{\hat \epsilon}_{\lambda_2}^\dagger(\bm{\hat k}) \,
\bm{\hat \epsilon}_{\lambda_1}(\bm{\hat k}) &=&
\delta_{{\lambda_2},{\lambda_1}},
\label{eq:epsorth}
\\
\bm{\hat k}_{\rm s}^\dagger\bm{\hat \epsilon}_\lambda(\bm{\hat
k})
&=& 0, 
\label{eq:epskorth}
\end{eqnarray}
and the completeness property
\begin{eqnarray}
\sum_{\lambda=1}^2
\bm{\hat \epsilon}_\lambda(\bm{\hat k}) \,
\bm{\hat \epsilon}_\lambda^\dagger(\bm{\hat k}) &=& \bm I
- \bm{\hat k}_{\rm s} \, \bm{\hat k}_{\rm s}^\dagger
= (\bm \tau \cdot \bm{\hat k})^2
 \nonumber\\[0 pt]&=&
 \bm \iPi\rs\rT(\bm{\hat k}).  
\label{eq:bewf}
\end{eqnarray}
From Eq.~(\ref{eq:epskorth}), we also have
\begin{eqnarray}
(\bm \tau \cdot \bm{\hat k})^2 \,
\bm{\hat \epsilon}_\lambda(\bm{\hat k})
=
\bm{\hat \epsilon}_\lambda(\bm{\hat k}).
\end{eqnarray}
Longitudinal fields are represented by a third polarization
state, labeled $\lambda = 0$, with the polarization vector taken
to be
\begin{eqnarray}
\bm{\hat \epsilon}_0(\bm{\hat k}) &=& \bm{\hat k}_{\rm s}.
\label{eq:epsl}
\end{eqnarray}
We thus have
\begin{eqnarray}
\sum_{\lambda=0}^2
\bm{\hat \epsilon}_\lambda(\bm{\hat k}) \,
\bm{\hat \epsilon}_\lambda^\dagger(\bm{\hat k}) &=& \bm I \, .
\end{eqnarray}

\end{document}